\documentclass[a4paper,fleqn,usenatbib]{mnras}
\usepackage{times}

\usepackage{mathtools}
\usepackage{aas_macros}

\usepackage{graphicx}
\usepackage{amstext}
\usepackage{enumerate}

\makeatletter
\newcommand{\mathleft}{\@fleqntrue\@mathmargin0pt}
\newcommand{\mathcenter}{\@fleqnfalse}
\makeatother

\title[R-method for core helium burning stars]{The treatment of mixing in core helium burning models -- II. Constraints from cluster star counts}

\author[Constantino et al.]{Thomas Constantino$^{1,2}$\thanks{E-mail: T.Constantino@exeter.ac.uk}, Simon W. Campbell$^{3,2}$, John C. Lattanzio$^{2}$, and \newauthor Adam van Duijneveldt$^{2}$\\
$^{1}$Physics and Astronomy, University of Exeter, Exeter, EX4 4QL, United Kingdom\\
$^{2}$Monash Centre for Astrophysics (MoCA), School of Physics and Astronomy, Monash University, Victoria, 3800, Australia\\
$^{3}$Max-Planck-Institut f\"{u}r Astrophysik, Karl-Schwarzschild-Stra{\ss}e 1, 85748 Garching bei M\"{u}nchen, Germany}

\begin{document}

\maketitle

\begin{abstract} 

The treatment of convective boundaries during core helium burning is a fundamental problem in stellar evolution calculations.  In Paper~I we showed that new asteroseismic observations of these stars imply they have either very large convective cores or semiconvection/partially mixed zones that trap g-modes.  We probe this mixing by inferring the relative lifetimes of asymptotic giant branch (AGB) and horizontal branch (HB) from $R_2$, the observed ratio of these stars in recent HST photometry of 48 Galactic globular clusters.  Our new determinations of $R_2$ are more self-consistent than those of previous studies and our overall calculation of $R_2 = 0.117 \pm 0.005$ is the most statistically robust now available.  We also establish that the luminosity difference between the HB and the AGB clump is $\Delta \log{L}_\text{HB}^\text{AGB} = 0.455 \pm 0.012$.  Our results accord with earlier findings that standard models predict a lower $R_2$ than is observed.  We demonstrate that the dominant sources of uncertainty in models are the prescription for mixing and the stochastic effects that can result from its numerical treatment.  The luminosity probability density functions that we derive from observations feature a sharp peak near the AGB clump.  This constitutes a strong new argument against core breathing pulses, which broaden the predicted width of the peak.  We conclude that the two mixing schemes that can match the asteroseismology are capable of matching globular cluster observations, but only if (i) core breathing pulses are avoided in models with a semiconvection/partially mixed zone, or (ii) that models with large convective cores have a particular depth of mixing beneath the Schwarzschild boundary during subsequent early-AGB `gravonuclear' convection.

\end{abstract}

\begin{keywords}
stars: evolution --- stars: horizontal-branch --- stars: interiors
\end{keywords}

\mathleft

\section{Introduction}
\label{sec:introduction}
In stellar evolution calculations the core helium burning (CHeB) phase is subject to considerable uncertainty.  The fundamental reason for this is that the amount of helium fuel that is brought into the convective core where it can burn, and hence the phase lifetime, is critically dependent on the treatment of convective boundaries.  Historically, star counts in Galactic globular clusters have been the most important empirical test for the efficiency of mixing in CHeB stars.  More recently, asteroseismology has provided complementary constraints on the structure and evolution of CHeB stars.  In this study we test insights gleaned from recent asteroseismology studies, particularly those in \citet[][hereinafter Paper~I]{2015MNRAS.452..123C}, against the wealth of high-quality photometry of globular clusters now available.

\subsection{CHeB: Key properties and uncertainties}
In low-mass CHeB models with convective overshoot, the position of the boundary of the convective core is volatile.  This is because carbon and oxygen, the products of helium burning, are more opaque than helium, so if any of the material is mixed across the formal convective boundary it tends to increase the opacity enough in the adjacent zone to make it unstable to convection according to the Schwarzschild criterion.  In such models, the feedback from this process generally precipitates the development of a large region with slow mixing that is approximately neutrally stable according to the Schwarzschild criterion.  This tends to happen regardless of whether there is a specific implimentation for semiconvection \citep[see e.g.][]{1986ApJ...311..708L,1990A&A...240..305C}.  The feedback from overshoot also causes the evolution of models with different treatments of convective boundaries to diverge significantly.  The prescription for overshoot can cause the total mass of helium consumed during CHeB, and therefore the duration of this phase, to vary by more than a factor of two \citep[see e.g.,][]{1971Ap&SS..10..355C,1986MmSAI..57..411B,2013ApJS..208....4P,2015MNRAS.452..123C}.

The uncertainty in the evolution worsens as CHeB progresses.  Depending on the mixing scheme, the phenomenon of `core breathing pulses' (CBP) may occur near the end of CHeB \citep{1973ASSL...36..221S,1985ApJ...296..204C}.  CBP are characterized by a rapid growth in the mass of the convective core when the central helium abundance is very low.  This process relies on feedback from the energy released by the $^{12}\text{C}(\alpha,\gamma)^{16}\text{O}$ reaction, which dominates when the helium abundance is low.  

Despite the considerable difference in the core structure produced by different treatments of mixing in models, there is little immediate effect on the conditions at the surface, e.g. luminosity, temperature, and composition.  Consequently, it is difficult to use observations to constrain the mixing treatment for stellar evolution calculations.  The most common method found in the literature makes use of star counts from globular clusters to infer the relative lifetimes of the CHeB and (shell helium burning) asymptotic giant branch (AGB) phases, because these strongly depend on the mixing prescription in models.  Recently, asteroseismology of white dwarfs \citep[e.g.][]{2002ApJ...573..803M}, red giants \citep[e.g.][]{2011Natur.471..608B,2012A&A...540A.143M,2014A&A...572L...5M} and sdB stars \citep[e.g.][]{2011MNRAS.414.2885R} has opened a new and much needed avenue for further constraining stellar models of the CHeB phase (see also Paper~I).

\subsection{Insights from asteroseismology}

Early attempts to constrain CHeB evolution from asteroseismology were indirect.  \citet{2002ApJ...573..803M} deduced that a higher $^{12}\text{C}(\alpha,\gamma)^{16}\text{O}$ reaction rate was needed to match the central C/O determination for a pulsating white dwarf.  \citet{2003ApJ...583..878S} then extended the study to the efficiency of mixing from overshoot.  While the former argued an increase in the $^{12}\text{C}(\alpha,\gamma)^{16}\text{O}$ rate was required to match the high central oxygen abundance, the latter found that the observations were consistent with the standard rate after accounting for semiconvection.  This highlights how inferences from asteroseismology can still suffer from some degeneracy caused by other uncertainties in stellar models, such as nuclear reaction rates.  

Following the detection of modes of mixed g- and p-character in the oscillations of red giants in the \textit{Kepler} field \citep{2011Natur.471..608B}, \citet{2012A&A...540A.143M,2014A&A...572L...5M} inferred the asymptotic g-mode period spacing $\Delta\Pi_1$ for hundreds of subgiant, red giant branch (RGB), and CHeB stars in the \textit{Kepler} field.  This is a particularly powerful probe because it is sensitive to the conditions deep in the core where g-modes propagate.  The high values of $\Delta\Pi_1$ typically inferred strongly contradict calculations from models with small convective cores \citep[see e.g.][]{2013ApJ...766..118M,2014IAUS..301..399C,2015MNRAS.452..123C,2015MNRAS.453.2290B}.  In Paper~I we also demonstrated that they are inconsistent with models with a semiconvection or partially mixed zone.  To match the range of $\Delta\Pi_1$ reported, we required models with our newly developed `maximal overshoot' scheme, which produces the largest possible convective core.  We also showed, however, that some modes can be strongly trapped at the boundary of the semiconvection or a partially mixed zone, raising the apparent $\Delta\Pi_1$ so that it is consistent with the determinations by \citet{2012A&A...540A.143M,2014A&A...572L...5M}.

\subsection{Globular cluster star counts and the AGB clump}

In Galactic globular clusters, which are composed of old, approximately coeval stars, the lifetime of each late phase of evolution is proportional to the number of stars observed in that phase.  This property is important in the current context because the mixing scheme strongly governs the amount of helium burned during CHeB.  It therefore also controls the respective longevity of the CHeB and early-AGB (subsequent helium-shell burning) phases.  

By considering the then available determinations of $R = n_\text{HB}/n_\text{RGB}$ and $R_1=n_\text{AGB}/n_\text{RGB}$ for three clusters (M15, M93, and NGC~5466), \citet{1978MNRAS.184..377C} concluded that the models with semiconvection by \citet{1971Ap&SS..10..355C}, that spend longer on the HB and more rapidly ascend the AGB, were a better match to the observations compared with the models without semiconvection from \citet{1970ApJ...161..587I} and \citet{1972ApJ...177..681R}.  This finding was then further supported by $R$ and $R_1$ determinations for 15 clusters \citep{1983A&A...128...94B}.  Although models with semiconvection were used for this comparison, the key finding is that CHeB models require a mechanism to transport additional helium into the convective core, prolonging the HB lifetime and speeding up the early-AGB evolution.

The ratio $R_2=n_\text{AGB}/n_\text{HB}$ is the most direct probe of the efficiency of mixing in CHeB globular cluster stars.  This was used by \citet{1989ApJ...340..241C}, who found that models with semiconvection, but without CBP (which decrease $R_2$), were consistent with observations of M5 from \citet{1981MNRAS.196..435B}, for which it was found that $R_2 = 0.18 \pm 0.04$.  Interestingly, this and the more precise value of $R_2 = 0.176 \pm 0.018$ determined from later observations by \citet{2004ApJ...611..323S}, are both higher than predicted from the models without CBP favoured by \citet{1989ApJ...340..241C}, which had $R_2=0.14$ and $R_2=0.15$, depending on the mass fraction of carbon in the core in the zero-age horizontal branch (ZAHB) models.  In contrast, \citet{1986MmSAI..57..411B} calculated relevant models with $R_2=0.16$ and $R_2 = 0.21$, depending on the overshoot parameter in their non-local overshoot treatment that produces large, fully mixed cores.  These examples demonstrate that $R_2$ has been used to both show the need for some kind of overshooting/semiconvection and also constrain the details of proposed mechanisms.

An inspection of the literature demonstrates why conclusions based on observations of $R_2$ ought to be revisited.  Studies that include star counts for several different clusters show a significant scatter in $R_2$.  In other studies, only a single cluster is used.  The nine clusters with more than 100 HB stars included in \citet{1983A&A...128...94B}, for example, span a range of $ 0.109 \leq R_2 \leq 0.215$.  Determinations of $R_2$, even for the same cluster, can vary considerably.  In the cluster NGC 6809 (M55), for example, \citet{1983A&A...128...94B}, \citet{2000MNRAS.313..571S}, and \citet{2007AJ....134..825V} determined $R_2 = 0.215$, $0.182$, and $0.156$, respectively, each with samples of more than 200 stars.  The disagreement can be even worse when fewer stars are observed: \citet{1983A&A...128...94B} found $R_2 = 0.133$ from 51 stars in NGC~6171 whereas \citet{2000MNRAS.313..571S} report $R_2 = 0.248$ from 146 stars.  The sizes of these differences suggest that inferences about stellar physics from $R_2$ may be bolstered by exploiting the more recent and numerous globular cluster photometry from HST.

Another related diagnostic is the magnitude (or luminosity) of the AGB clump, which is the observed clustering of early-AGB stars in the colour-magnitude diagram (CMD).  \citet{1977A&AS...27..367L} first noted that such a clump was visible 1.5\,mag above the HB level in globular clusters with clear AGB sequences.  In evolution calculations this coincides with core helium exhaustion and the subsequent slow luminosity change at the beginning of shell helium burning.  Importantly, the surface luminosity during this event is dependent on the mixing during the earlier CHeB phase.  \citet{2001A&A...366..578C} showed, for example, that artificially suppressing CBP increases the luminosity of the AGB clump.  This suppression also shortens the HB lifetime and increases the AGB lifetime, better matching the observed ratio $R_2$ in M5.

\subsection{Other uncertainties}
Despite Nature having provided us with a large sample of nearby globular clusters, the interpretation of the observations presents a number of challenges.  In order to identify the current stage of evolution of a star from photometry we require the AGB and the RGB stars to form distinct sequences.  To correctly infer lifetimes the photometry must be nearly complete, or not have a bias against one of the populations, e.g. by excluding hot HB stars.  Cluster membership should also be verified to avoid contamination.  

The ability to use star counts to explain specific physical phenomena is also dependent on our wider understanding of stellar evolution.  The determination of initial helium content from the ratio $R = n_\text{HB}/n_\text{RGB}$, for example, is sensitive to the $^{12}\text{C}(\alpha,\gamma)^{16}\text{O}$ reaction rate \citep{1998MNRAS.298..557B} and binary interaction which could affect that ratio by truncating the evolution before the HB.  Fortunately, binary interaction is unlikely to be problematic for inferring lifetimes from $R_2$ because those stars have already survived the RGB.  Instead, we must consider the possibility that $R_2$ is reduced because some HB stars, whose envelope is too small for there to be a second ascent of the giant branch, become `AGB-manque' stars \citep[see e.g.,][]{1974A&A....30...13S,1976ApJ...204..116G,1989A&A...221...27C,1990ApJ...364...35G,1993ApJ...419..596D}.  Recent spectroscopic evidence suggests this evolution may be more common than predicted from models \citep{2013Natur.498..198C}, which is an additional hazard for the interpretation of $R_2$.

\subsection{The current study: Revisiting the R-method}
The mixing in the cores of CHeB models is a fundamental uncertainty that has existed since it was first shown that a slowly mixing semiconvection zone could develop outside the convective core.  Efforts to understand this structure have been hampered by the absence of any immediately observable effects.  Lately however, asteroseismology of red giants, sdB stars, and white dwarfs has offered new insights into this phase of evolution.  In light of this recent progress, and with the aid of photometry superior in quality and quantity, we revisit the R-method for inferring the properties of CHeB evolution from observations of globular clusters.  Specifically, we compare the HB and AGB luminosity evolution implied from the populations in globular clusters with a suite of stellar models computed using different mixing schemes, composition, and input physics.

\section{Observational data}
\label{sec:methods}

\subsection{Photometry}
\label{sec:methods_photometry}

In this study we use HST photometry of 74 Galactic globular clusters from the Wide Field Planetary Camera 2 (WFPC2) in the $F439W$ and $F555W$ filters \citep{2002A&A...391..945P} and 65 clusters from the ACS Wide Field Channel (WFC) in the $F606W$ and $F814W$ filters \citep{2007AJ....133.1658S}.  Together, these samples contain photometry of 104 unique clusters.  These data are advantageous for two reasons: they comprise two large homogeneous samples, and the photometry is reasonably complete at the magnitudes relevant for this study (which we show below).  The two filter sets do not exactly match any other systems.  \citet{2005PASP..117.1049S} describes $F439W$ and $F555W$ bands as Johnson B and Johnson V, and the $F606W$ and $F814W$ bands as broad V and broad I, respectively.

In the ACS Globular Cluster Survey \citep{2007AJ....133.1658S}, the flux of each unsaturated star was determined by fitting a point spread function constructed for each exposure.  The technique from \citet{2004acs..rept...17G} was used to find the flux for stars with saturated pixels.  The reliability of the method for determining flux from saturated pixels was confirmed by comparing against unsaturated shorter exposures.  

Crowding is only problematic for stars in the centre of clusters with compact cores, such as NGC~2808.  In that case, the artificial star tests by \citet{2008AJ....135.2055A} predict a completeness of 60 per cent in the cluster centre for stars with the magnitude of the extreme HB (the faintest stars we are interested in).  The completeness then rapidly improves with increasing distance from the centre.  Fortunately, most clusters in the photometry sample are close to 100 per cent complete above the subgiant branch level \citep{2008AJ....135.2055A}.  The artificial star experiments performed by \citet{2002A&A...391..945P} for NGC~104 and NGC~6723 (clusters with high and low central density, respectively) showed that the photometry has high completeness to more than 2 magnitudes  fainter than the faintest HB stars.  Therefore, except for clusters with extreme HBs, we do not expect incompleteness to influence any of our findings.

We correct for reddening in the \citet{2002A&A...391..945P} data by using the corrections provided, which are originally from the \citet{1996AJ....112.1487H} catalogue.  We also use the \citet{1996AJ....112.1487H} catalogue (2010 edition) to correct for reddening in the \citet{2007AJ....133.1658S} photometry.  In that case we use the $E(B-V)$ correction because, according to the extinction law from \citet{1989ApJ...345..245C}, it corresponds almost exactly to the $E({F606W}-{F814W})$ correction required.

\subsection{Sample selection}
\label{sec:R2_star_counts}

We limit our analysis to clusters that have clearly defined HB and AGB sequences, which excludes photometry with obvious large photometric errors.  We note that this procedure could introduce a selection bias but that we still include 48 globular clusters that meet our requirements, out of 104, which represents a sizeable fraction of the 157 known in the Galaxy \citep{2010arXiv1012.3224H}. We do not expect our selection to impact on our conclusions because most of the reasons for excluding photometry are not related to stellar evolution during and after the CHeB phase.  Additionally, we have rejected clusters with only a very small number of (identifiable) AGB stars.  Selecting against clusters with few AGB stars could introduce a small bias because they could be a true reflection of the lower tail of real scatter in $R_2$.

\subsection{Comparison of $R_2$ between different data sets}
\label{sec:photometry_comparison}

In Table~\ref{table_R2_different_photometry} we present the results of our star counts for 48 Galactic globular clusters using data from \citet{2002A&A...391..945P} and \citet{2007AJ....133.1658S}.  We also include, for comparison, counts from \citet{2000MNRAS.313..571S}.  In our star counts we include all AGB stars that could be identified, but note that we later restrict the AGB count to stars no more than 10 times as luminous as the HB level.  Interestingly, there is significant variation in $R_2$, both between, and within each data set.  When considering the 15 clusters common to all three data sets, we find that $R_2$ has a standard deviation of 0.05, 0.03, and 0.05 for the respective \citet{2002A&A...391..945P}, \citet{2007AJ....133.1658S} and \citet{2000MNRAS.313..571S} samples.  The \citet{2002A&A...391..945P} and \citet{2007AJ....133.1658S} data sets are the most concordant pair; the average size of the discrepancy in $R_2$ between the two sources is 0.03.  Despite the differences between $R_2$ determinations for individual clusters, the overall average $R_2$ is consistent between the two HST data sets.  If the observations for the 15 common clusters are combined we find $R_2 = 0.121\pm 0.006$ and $0.127 \pm 0.005$ from the \citet{2002A&A...391..945P} and \citet{2007AJ....133.1658S} data respectively, compared with $R_2 = 0.148\pm 0.007$ from \citet{2000MNRAS.313..571S}, where the uncertainty is calculated from Equation~\ref{eq:R2_sigma}.  When this comparison is further restricted to the seven clusters that have more than $150$ (total AGB and HB) stars in each CMD, we find similar results except that the average $R_2$ from the \citet{2000MNRAS.313..571S} data decreases to 0.141, improving the agreement with the other data.

\begin{figure}
\includegraphics[width=\linewidth]{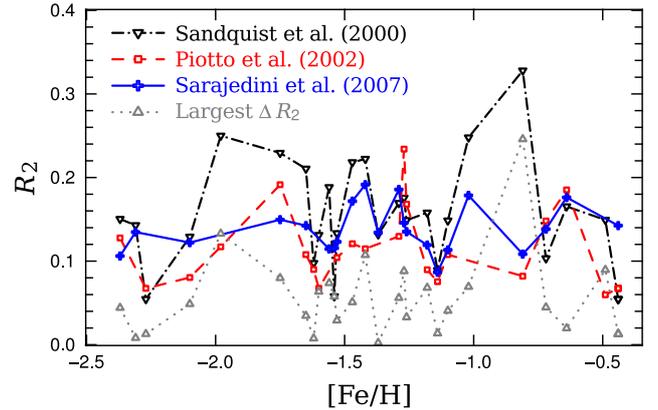}
  \caption{Comparison of $R_2$ for clusters shown in Table~\ref{table_R2_different_photometry}, limited to those with at least two different sources of photometry.  The $R_2$ determined from the \citet{2000MNRAS.313..571S}, \citet{2002A&A...391..945P}, and \citet{2007AJ....133.1658S} data are shown in black dash-dots, red dashes, and a blue solid line, respectively.  The dotted grey line shows the maximum difference between $R_2$ determinations from different photometry.}
  \label{figure_all_R2_comparison}
\end{figure}

In Figure~\ref{figure_all_R2_comparison} we plot $R_2$ for the 31 clusters with multiple sources of photometry.  The dotted grey line shows the largest difference between $R_2$ determinations for each cluster.  In several clusters this difference is more than 0.1, which is almost as large as the average $R_2$.  There is no obvious dependence of $R_2$, or its consistency between data sets, on metallicity.  We further discuss the statistics of this data in Section~\ref{sec:statistical_errors} and the effect of the metallicity and HB morphology in Section~\ref{sec:z_and_hb_morph}.

We have investigated the causes of some of the significant discrepancies between $R_2$ determinations from different photometry.  To this end, we have chosen three clusters with three independent counts: NGC~6093, NGC~6652, and NGC~7078.  In each of the three CMDs of NGC~6093 there are a total of more than 190 HB and AGB stars.  Despite the large sample size, $R_2$ is not consistent: we find $R_2 = 0.229$, $0.191$, and $0.150$, from \citet{2000MNRAS.313..571S}, \citet{2002A&A...391..945P}, and \citet{2007AJ....133.1658S}, respectively.  When examining the photometry from \citet{1998AJ....116.2415A}, which was used for the \citet{2000MNRAS.313..571S} count, the reason for the disagreement is obvious -- most of the blue HB, which is clear in the deeper \citet{2007AJ....133.1658S} photometry, is missing.  Excluding these ($\sim\,100$) stars from the \citet{2007AJ....133.1658S} count increases $R_2$ to about 0.22, consistent with the \citet{2000MNRAS.313..571S} result.  The disagreement for this cluster is also worsened to a lesser extent by the availability of $U$-band photometry in the \citet{1998AJ....116.2415A} photometry.  This better separates the luminous-AGB from the RGB, allowing more AGB stars to be identified, and increasing $R_2$.  

The most metal-poor cluster in our collection, NGC~7078 (M15), also has a blue HB and a similarly large number of stars in each CMD, but it appears that in this instance the difference in $R_2$, which ranges from $0.106$ to $0.150$, is primarily due to the difficulty of distinguishing AGB from RGB stars.  A gap is apparent between the blue and red parts of the HB in both the \citet{2002A&A...391..945P} and \citet{2007AJ....133.1658S} photometry.  The fraction of HB stars on the blue side of the gap is around 0.41, with agreement between the two sets to about 1 per cent, suggesting inconsistent completeness is not the problem.  Unlike the previous example, the blue HB also is well populated in the older photometry from \citet{1983A&AS...51...83B} that is included in \citet{2000MNRAS.313..571S}.  A similar issue seems to be at play for NGC~6752.  We find $R_2 = 0.116$ from the \citet{2007AJ....133.1658S} photometry which is exactly double the result in \citet{2000MNRAS.313..571S}.  Again, this difference appears to be due to the difficulty of distinguishing between AGB and RGB stars in the older $B V$ photometry from \citet{1986A&AS...66...79B}, rather than missing blue HB stars.  Our result is also consistent with $UBV$ photometry from Y. Momany (private communication) which has a very clear AGB sequence, and from which we find $R_2 = 0.104$.

There is a considerable spread in $R_2$ determinations for NGC~6652.  We find $R_2 = 0.082$ and $R_2 = 0.108$ from the \citet{2002A&A...391..945P} and \citet{2007AJ....133.1658S} photometry, respectively, while \citet{2000MNRAS.313..571S} reports $R_2 = 0.267$.  We attribute this variation to the small number of HB and AGB stars (there are fewer than a total of 100 in each CMD), and the difficulty in positively identifying AGB stars, especially for the ground-based photometry from \citet{1994A&A...286..444O} used by \citet{2000MNRAS.313..571S}.

\subsection{Colour transformations and bolometric corrections}
\label{sec:colour_and_bc}

In order to compare observations with theoretical models we must relate the observed magnitude to luminosity.  Throughout this study we do this by converting from the magnitude observed to luminosity.  This conveniently minimizes the importance of colour-temperature transformations and the MLT mixing length calibration (see Section~\ref{sec:MLT}).  To convert magnitude to luminosity we use the colour-temperature relations from \citet{2000AJ....119.2018O} and the bolometric corrections from \citet{2008PASP..120..583G} for the \citet{2002A&A...391..945P} photometry.  We also use bolometric corrections from \citet{2008PASP..120..583G} for the \citet{2007AJ....133.1658S} photometry, but in that case also use them to derive a colour-temperature relation.  We do not attempt to find the absolute luminosity of the observed stars, which would be subject to the uncertainties in the distance of each cluster.  We instead rescale the luminosity so that it is expressed relative to that of the HB, i.e. we use
\begin{equation}
\label{eq:delta_logL}
\Delta\log{L} = \log{L}-\log{L_\text{HB}},
\end{equation}
where $L$ is the derived luminosity and $\log{L_\text{HB}}$ is the mode of the $\log{L}$ distribution for a cluster, i.e. the typical luminosity for HB stars.

\subsection{Luminosity probability density functions}
\label{sec:obs_PDF}
We calculate $R_2$ and $\Delta\log{L_\text{HB}^\text{AGB}}$ (defined below) from the luminosity probability density function (PDF) determined from the observations.  Each luminosity PDF $P(\Delta\log{L})$ is constructed from a sample of $N$ stars by the addition of Gaussian functions so that
\begin{equation}
\label{eq:obs_PDF}
P(\Delta\log{L}) = \frac{1}{N}\sum\limits_{i=1}^N  \frac{1}{\sigma \sqrt{2\pi}}\exp{\left[-\frac{\left(\Delta \log{L}-\Delta \log{L_i}\right)^2}{2\sigma^2}\right]},
\end{equation}
where $i$ represents each star, $\Delta \log{L_i} = \log{L_i}-\log{L_\text{HB}}$, $L_i$ is the luminosity of each star, $L_\text{HB}$ is  the mode of the $\log{L}$ distribution, and $\sigma$ is the standard deviation which determines the smoothness of the resulting function.  The value of $\sigma$ chosen for each cluster depends on the number of stars but it is generally around $\sigma = 0.04$.  This is typically large enough for there to be a well defined peak from the AGB clump in the luminosity PDF.

The luminosity difference between the AGB clump and the HB is defined as
\begin{equation}
\Delta\log{L_\text{HB}^\text{AGB}} = \log{L_\text{AGB}}-\log{L_\text{HB}}.
\end{equation}
This is just the difference in $\Delta \log{L}$ between the two peaks in the luminosity PDF.  We can also caculate $R_2$ from the luminosity PDF.  We define the boundary between the HB and AGB to be the location of the minimum in the luminosity PDF between the HB and AGB clump peaks, $\Delta \log{L_\text{min}}$.  $R_2$ is thus the ratio of the integral of the luminosity PDF above and below this peak:
\begin{equation}
R_2 = \frac{\int_{\Lambda_\text{min}}^{\Lambda_\text{lim}} P(\Lambda) \text{d} \Lambda}{\int_{-\infty}^{\Lambda_\text{min}} P(\Lambda) \text{d} \Lambda} \simeq \frac{n_\text{AGB}}{n_\text{HB}}  ,
\end{equation}
where for convenience we have defined $\Lambda$ as $\Delta \log{L}$, and $\Lambda_\text{lim}$ is the luminosity cut-off for the AGB that we introduce in Section~\ref{sec:z_and_hb_morph}.

\subsection{Statistical errors}
\label{sec:statistical_errors}
We account for statistical errors in a manner similar to previous studies \citep[e.g.,][]{1971PASP...83..697I,1983A&A...128...94B,2004ApJ...611..323S}, i.e. by assuming that the ratio of AGB to HB stars follows a Poisson distribution.  This gives a variance
\begin{equation}
  \sigma^2 (n_\text{AGB}) = n_\text{HB} R_2,
\end{equation}
and therefore standard error
\begin{equation}
  \label{eq:R2_sigma}
  \sigma (R_2) = \frac{\sigma (n_\text{AGB})}{n_\text{HB}} = \sqrt{\frac{R_2}{n_\text{HB}}}.
\end{equation}
This is strictly true only if the observed ratio $R_2 = n_\text{AGB}/n_\text{HB}$ is the ratio expected from an infinitely large sample.  It is also larger, by a very small factor of $\sqrt{1 + R_2}$, than the result found by assuming that there is a binomial distribution of HB and AGB stars.  It is important to note that this error analysis is not exhaustive.  There are other sources of error that were raised in Section~\ref{sec:introduction}, but which are not easily quantifiable.

We have determined the statistical uncertainty in $R_2$ and $\Delta\log{L_\text{HB}^\text{AGB}}$ in our aggregate data sets (e.g. those in Section~\ref{sec:red_hb_clusters}) from Monte Carlo simulations.  In this method we use the observed luminosity probability density functions to randomly populate hundreds of artificial data sets from which we calculate $R_2$ and $\Delta\log{L_\text{HB}^\text{AGB}}$ using the same method as for observations.  

The error bars for $\Delta \log{L_\text{HB}^\text{AGB}}$ in Figure~\ref{figure_cluster_properties_z} and Figure~\ref{figure_cluster_properties_hb} are 1-$\sigma$ and generated by determining  $\Delta \log{L_\text{HB}^\text{AGB}}$ from Monte Carlo simulations using different sample sizes and the luminosity PDF of the data set comprising every cluster without a blue HB.  This also revealed a bias towards higher $\Delta \log{L_\text{HB}^\text{AGB}}$ for smaller data sets.  For example, we calculated the average $\Delta \log{L_\text{HB}^\text{AGB}}$ for samples of $100$ and $200$ stars to be $0.496$ and $0.474$, respectively, compared with the true value of $0.455$, This discrepancy, however, is well within the respective 1-$\sigma$ uncertainties ($0.096$ and $0.065$).  We have not corrected for this source of error in Figure~\ref{figure_cluster_properties_z} and Figure~\ref{figure_cluster_properties_hb}. 

We have also tested the effect of random photometric errors $\sigma_\text{phot}$ by allowing for them when populating the luminosity PDF in the Monte Carlo simulations.  With sample sizes of $N=100$ and $N=200$ the error in $R_2$ and $\Delta \log{L_\text{HB}^\text{AGB}}$ is unaffected if $\sigma_\text{phot}(\log{L}) \la 0.05$ and $\sigma_\text{phot}(\log{L}) \la 0.065$, respectively.  This increases to $\sigma_\text{phot}(\log{L}) \la 0.10$ for the combined data set 6366 stars from all clusters without a blue HB.  This allowance can accommodate the size of the estimated errors provided with the photometry, with the possible exception of blue HB stars.  The uncertainty for these stars does not, however, affect the determination of $R_2$ and $\Delta \log{L_\text{HB}^\text{AGB}}$.

It is still possible that photometric errors affect our results by causing the misidentification of stars, e.g. blending of the RGB and AGB sequences, or the blue HB and the main sequence.  The former is possible because the two are so close in the CMD while the latter is possible because the error is larger for fainter stars.  However, if it is assumed that the HB and AGB stars are properly identified then the photometric errors do not add to the uncertainty of either $R_2$ or $\Delta \log{L_\text{HB}^\text{AGB}}$.

We have tested whether, after accounting for the known statistical uncertainty, the values of $R_2$ are consistent with the overall weighted mean $R_2 = 0.113 \pm 0.002$.  In Figure~\ref{figure_R2_statistics} we show the distribution of the difference between $R_2$ for each cluster and the overall mean, expressed as a fraction of the 1-$\sigma$ standard error.  In both the \citet{2002A&A...391..945P} and \citet{2007AJ....133.1658S} cases, the distribution is wider than expected from the above hypothesis.  The standard deviations for the respective samples are $1.81$ and $1.40$ compared with an expected value of $1.0$.  This suggests that (i) we have underestimated the errors and/or (ii) $R_2$ is cluster dependent.  We already know that the first possibility is true, since we have only accounted for one of the possible sources of error (which were discussed in Section~\ref{sec:methods}).  In Section~\ref{sec:z_and_hb_morph} we investigate the second possibility by analysing the observations and in Section~\ref{composition_and_input_physics} we quantify how various factors, such as composition and stellar mass, affect theoretical predictions of $R_2$.

\subsection{Cluster metallicity and HB morphology} 
\label{sec:z_and_hb_morph}

\begin{figure}
\includegraphics[width=\linewidth]{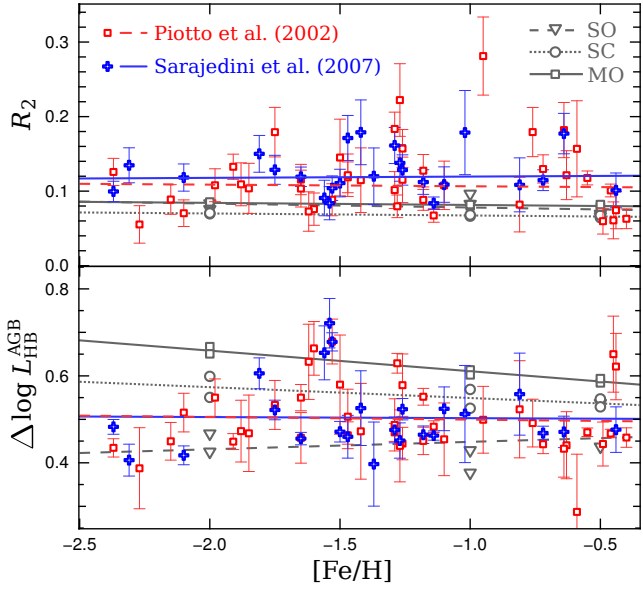}
  \caption{\textbf{Upper panel:} $R_2$ as a function of cluster metallicity (from the \citealt{1996AJ....112.1487H} catalogue; 2010 edition) for selected clusters from \citet{2002A&A...391..945P} photometry (red squares) and \citet{2007AJ....133.1658S}  photometry (blue crosses).  Error bars are 1-$\sigma$ according to Equation~\ref{eq:R2_sigma}.  \textbf{Lower panel:}  luminosity difference between the HB (defined as the peak of the luminosity probability density function) and the AGB clump (similarly defined) for the same clusters.  Error bars are 1-$\sigma$ according to the method in Section~\ref{sec:statistical_errors}.  The lines of best fit (dashed red and solid blue lines for the \citealt{2002A&A...391..945P} and \citealt{2007AJ....133.1658S} photometry, respectively) were constructed by weighting the clusters according to the reciprocal of the 1-$\sigma$ error.  Example results of theoretical evolution calculations with the standard-overshoot (SO), semiconvection (SC), and maximal-overshoot (MO) mixing schemes are denoted by grey triangles, circles, and squares with dotted, dashed, and solid trend lines, respectively (see Section~\ref{sec:overall_model_obs_comparison}).}
  \label{figure_cluster_properties_z}
\end{figure}

In Figure~\ref{figure_cluster_properties_z} we present $R_2$ and $\Delta \log{L_\text{HB}^\text{AGB}}$ for the 48 clusters in this study.  In this calculation of $R_2$, we limit the AGB count to $\log{L} < \log{L_\text{HB}}+1.0$, which is different from the method used for Figure~\ref{figure_all_R2_comparison} where all (identified) AGB stars are included.  We use this luminosity limit for the remainder of the paper because it enables a consistent comparison of $R_2$ between different clusters and between observations and predictions from models.  Figure~\ref{figure_all_R2_comparison} shows that we do not detect any significant trend in either $R_2$ or $\Delta \log{L_\text{HB}^\text{AGB}}$ with metallicity.  The lines of best fit (and those in subsequent figures) are constructed by weighting the data points according to the reciprocal of the 1-$\sigma$ error.  The weighted average $\Delta \log{L_\text{HB}^\text{AGB}} \approx 0.5$ is the same for both data sets (note that this is higher than $\Delta \log{L_\text{HB}^\text{AGB}}$ calculated in Section~\ref{sec:red_hb_clusters} from combined data sets).  When accounting for the uncertainties there is also agreement in the average $R_2$: we find $R_2 = 0.111 \pm 0.007$ from the \citet{2002A&A...391..945P} data and $R_2 = 0.127 \pm 0.009$ from the \citet{2007AJ....133.1658S} data.

\begin{figure}
\includegraphics[width=\linewidth]{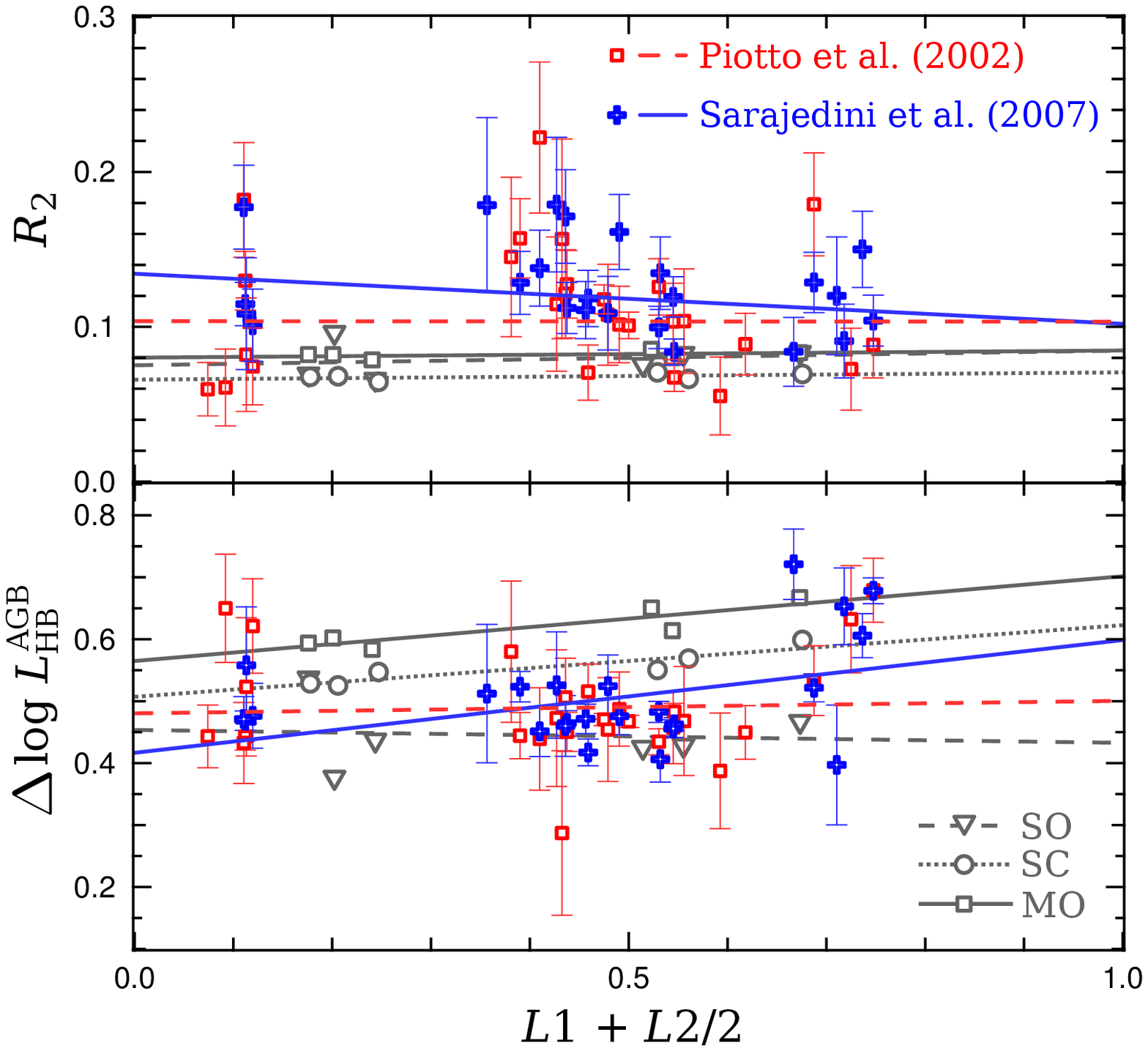}
  \caption{Same as Figure~\ref{figure_cluster_properties_z} except $R_2$ and $\Delta \log{L_\text{HB}^\text{AGB}}$ are plotted against $L1 + L2/2$, which is the colour ($F606W - F814W$) difference between the RGB and the middle of the HB determined by \citet{2014ApJ...785...21M}.  Note that this sample is restricted to those clusters in Figure~\ref{figure_cluster_properties_z} that have $L1$ and $L2$ determinations.}
  \label{figure_cluster_properties_hb}
\end{figure}

Figure~\ref{figure_cluster_properties_hb} is the same as Figure~\ref{figure_cluster_properties_z} except that $R_2$ and $\Delta \log{L_\text{HB}^\text{AGB}}$ are plotted against $L1 + L2/2$, the colour ($F606W - F814W$) difference between the middle of the HB and the RGB determined by \citet{2014ApJ...785...21M}, i.e. a measure of the `blueness' of the HB.  The 14 clusters without $L1$ and $L2$ determinations (i.e. those with only \citealt{2002A&A...391..945P} photometry) are not shown in this figure.  The clusters from both sets of data divide into three groups with distinct $L1 + L2/2$. In the middle group of clusters, with $L1 + L2/2 \approx 0.5$, there appears to be a strong negative correlation between $R_2$ and $L1 + L2/2$, but this trend is not preserved when the more red and the more blue HB groups are included.  When all of the clusters are considered, there appears to be at most a weak correlation between $R_2$ and $L1 + L2/2$, i.e. $R_2$ does not strongly depend on the stellar factors that control HB morphology, principally mass, metallicity, and helium content.   

We did not detect any dependence of $\Delta \log{L_\text{HB}^\text{AGB}}$ on $L1 + L2/2$ for the \citet{2002A&A...391..945P} observations shown in Figure~\ref{figure_cluster_properties_hb}.  Although the trend line for the \citet{2007AJ....133.1658S} sample shows a positive correlation between $\Delta \log{L_\text{HB}^\text{AGB}}$ and $L1 + L2/2$ (with a gradient of $0.18$) that is concordant with the example models, the trend in the observations is due entirely to the six clusters with the bluest HBs, and these have a large scatter.  We further discuss the dependence of $\Delta \log{L_\text{HB}^\text{AGB}}$ on HB morphology with reference to the stellar models in Section~\ref{sec:overall_model_obs_comparison}.

\begin{figure}
\includegraphics[width=\linewidth]{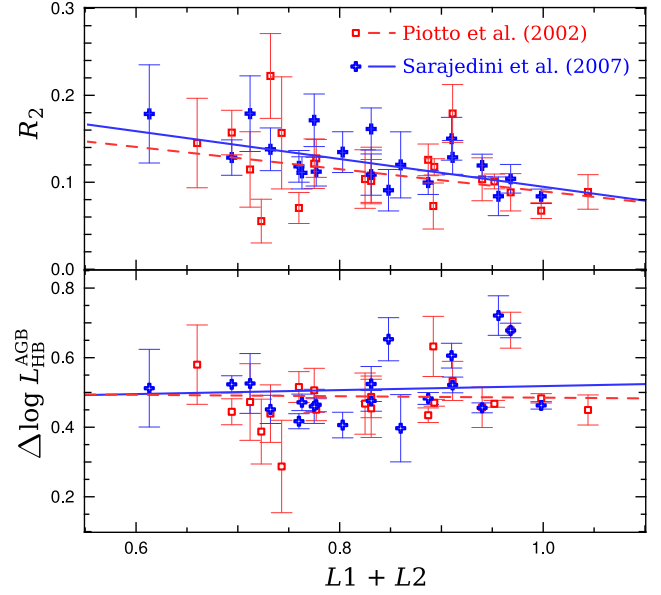}
  \caption{Same as Figure~\ref{figure_cluster_properties_z} except $R_2$ and $\Delta \log{L_\text{HB}^\text{AGB}}$ are plotted against $L1 + L2$, which is the colour ($F606W - F814W$) difference between the RGB and the fourth percentile of the HB population, determined by \citet{2014ApJ...785...21M}.  Note that this sample is restricted to those clusters in Figure~\ref{figure_cluster_properties_z} that have $L1$ and $L2$ determinations and a blue extension to the HB.}
  \label{figure_cluster_properties_blue_hb}
\end{figure}

\begin{figure}
\includegraphics[width=\linewidth]{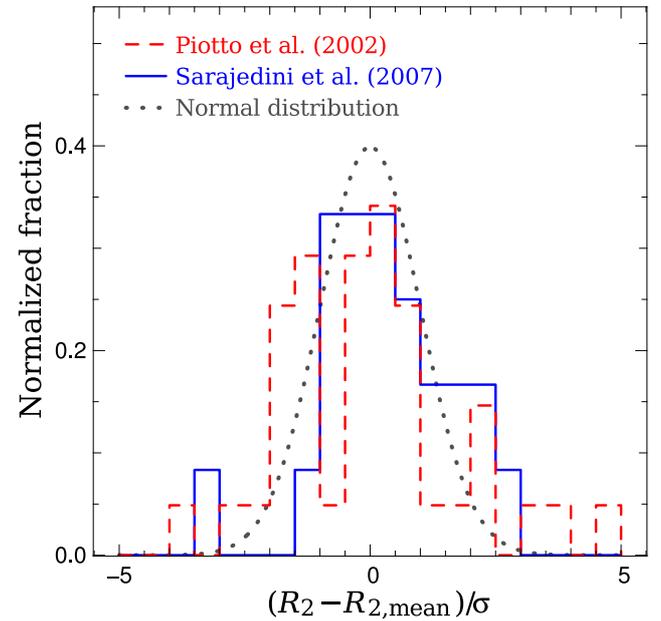}
  \caption{Histogram of the difference, as a fraction of the standard error (calculated from Equation~\ref{eq:R2_sigma}), between $R_2$ for each cluster shown in Figure~\ref{figure_cluster_properties_z} and the overall mean value of $R_2 = 0.113$ for the \citet{2002A&A...391..945P} photometry (dashed red line) and \citet{2007AJ....133.1658S} photometry (solid blue line).  The dotted grey curve is the standard normal distribution, i.e. with standard deviation $\sigma = 1$.}
  \label{figure_R2_statistics}
\end{figure}

It has been proposed that the lack of CN-strong \citep{1981ApJ...244..205N,1999AJ....118.1273I,2010MmSAI..81.1004C,2012ApJ...761L...2C} and sodium-rich \citep{2013Natur.498..198C,2013A&A...557L..17C,2014A&A...571A..81C,2015AJ....149...71J} AGB stars in globular clusters could be due to a sizeable fraction of the (low mass) blue HB stars not evolving to the AGB.  Our analysis of the observations reveals there is no dependence of $R_2$ on the colour of the midpoint of the HB (Figure~\ref{figure_cluster_properties_hb}), which would appear to contradict assertions that some blue HB stars do not reach the AGB.  The picture is changed, however, when the colour of the bluest extent, rather than the midpoint, of the HB is considered.  In Figure~\ref{figure_cluster_properties_blue_hb} we show $R_2$ and $\Delta\log{L_\text{HB}^\text{AGB}}$ for the clusters with a blue HB plotted against the colour difference between the blue end (fourth percentile) of the HB and the RGB ($L1+L2$).  It is clear from both the \citet{2002A&A...391..945P} and \citet{2007AJ....133.1658S} data sets that $R_2$ is lower in clusters with a bluer HB tail.  

It is conceivable that the dependence of $R_2$ on the extent of the blue HB results from the lower luminosity of blue HB stars: this would reduce the luminosity cut-off for the AGB and reduce the number of AGB stars included in the count, and therefore $R_2$.  There are two arguments against this though: (i) there is only a weak dependence of $\Delta \log{L_\text{HB}^\text{AGB}}$ on $L1 + L2$ (i.e. we are still including the same luminosity range of AGB stars, independent of the extent of the blue HB), and (ii) $R_2$ is not substantially lower for clusters in which the middle of the HB is blue (i.e. those with the highest $L1 +L2/2$), which would be the case if more AGB stars were excluded from our counts in clusters with a blue HB.  It thus appears from this sample that a considerable fraction of blue HB stars do not evolve to the AGB phase.

\subsection{General observed properties of red-HB clusters}
 \label{sec:red_hb_clusters}

\begin{figure}
\includegraphics[width=\linewidth]{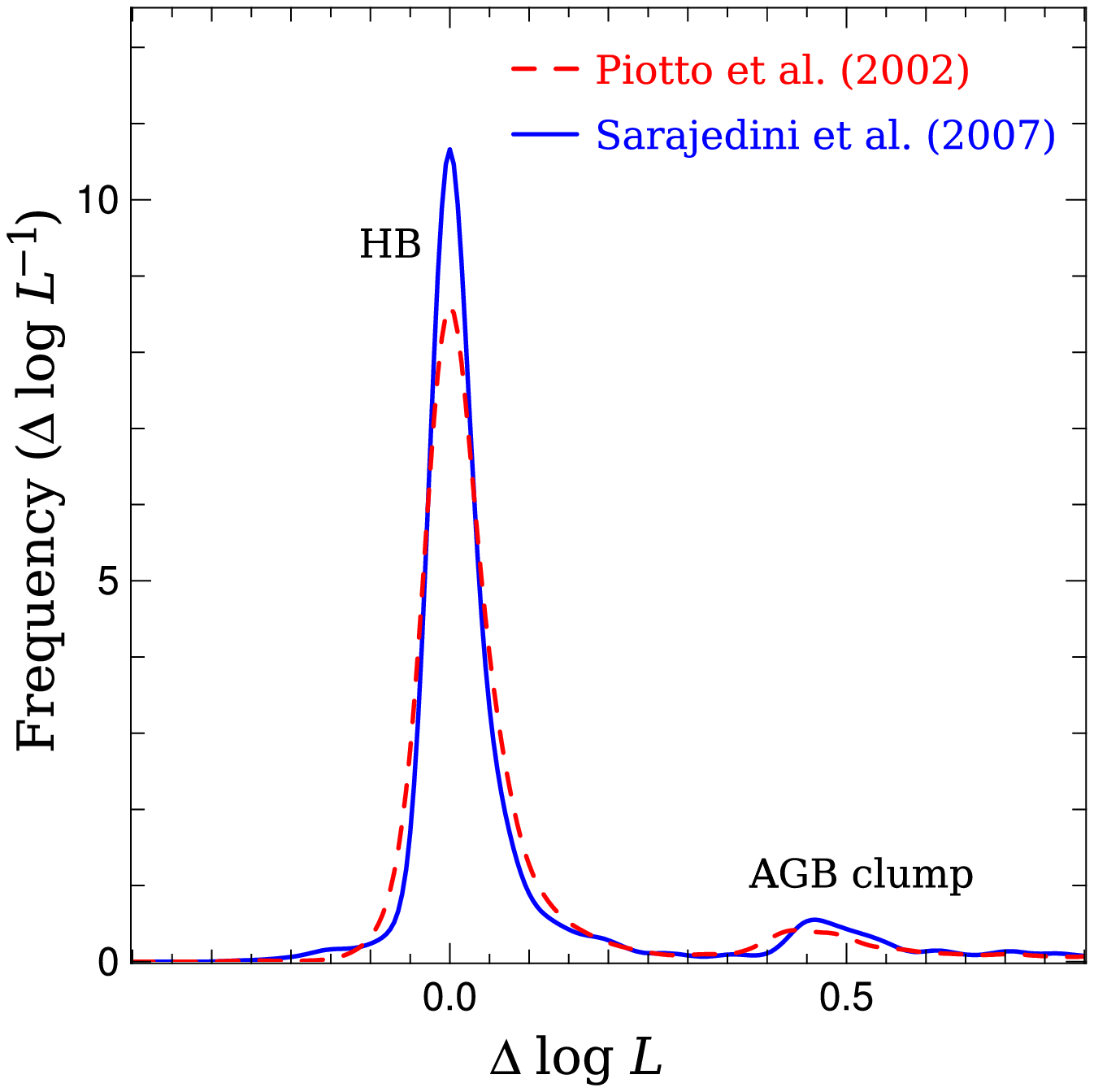}
  \caption{Observed probability density function of the luminosity of all HB and AGB stars in clusters without a blue extension of the HB (listed in Section~\ref{sec:photometry_comparison}).  The luminosity of the HB for each cluster (defined here as the peak of the distribution) has been rescaled so that $\log{L_\text{HB}} = 0$.  The sample has been truncated at $\Delta \log{L} = 1.0$.  The PDF for each cluster was constructed by adding a Gaussian function with $\sigma = 0.04$ for each star (see Section~\ref{sec:obs_PDF}).  These were then combined by weighting each cluster according to the number of stars counted.  Transformations from magnitude to luminosity are in accordance with the method in Section~\ref{sec:colour_and_bc}.}
  \label{figure_all_red_hb}
\end{figure}

In Figure~\ref{figure_all_red_hb} we present the luminosity PDFs from all of the HB and AGB stars in the 14 clusters in the \citet{2002A&A...391..945P} or \citet{2007AJ....133.1658S} data sets that do not have a blue extension to the HB.  Restricting the analysis to these clusters is beneficial for several reasons:
\begin{enumerate}[1.]
\item The total luminosity (and magnitude) range of the HB and the AGB is smaller, reducing the importance of any potential magnitude-dependent completeness function (such as that for the centrally dense cluster NGC~2808; \citealt{2008AJ....135.2055A}).
\item The colour range is smaller, reducing the effect of imperfect bolometric corrections.
\item The luminosity of the HB is unambiguous and therefore so is the cut-off for the AGB luminosity.  This also makes estimates of $\Delta \log{L_\text{HB}^\text{AGB}}$ more certain.
\item We expect all of the HB stars to be massive enough to ascend the AGB.
\end{enumerate}
The clusters NGC~104, NGC~362, NGC~1261, NGC~1851, NGC~6624, NGC~6637, and NGC~6652 are common to both samples while NGC~5927, NGC~6304, NGC~6356, NGC~6441, NGC~6539, and NGC~6569 are only in the \citet{2002A&A...391..945P} set and NGC~6171 is only in the \citet{2007AJ....133.1658S} set.  In the luminosity PDFs shown in Figure~\ref{figure_all_red_hb}, each cluster is weighted according to the total number of stars.  The agreement between the two consolidated data sets is remarkable.  The consistency between both $R_2$ and $\Delta \log{L}_\text{HB}^\text{AGB}$ provides a strong constraint for models.  The \citet{2002A&A...391..945P} sample gives $R_2 = 0.114 \pm 0.007 $ and $\Delta \log{L}_\text{HB}^\text{AGB} = 0.436 \pm 0.017$ compared with $R_2 = 0.127 \pm 0.009$ and $\Delta \log{L}_\text{HB}^\text{AGB} = 0.460 \pm 0.010$ for the \citet{2007AJ....133.1658S} sample, where the 1-$\sigma$ uncertainty is determined from the Monte Carlo method described in Section~\ref{sec:statistical_errors}.  When the two data sets are combined, we find $R_2 = 0.117 \pm 0.005$ and $\Delta \log{L}_\text{HB}^\text{AGB} = 0.455 \pm 0.012$.  We use these observational constraints -- the tightest yet -- in the following sections.

\setlength{\tabcolsep}{2pt}

\begin{table*}

  \caption{Comparison of horizontal branch (HB) and asymptotic giant branch (AGB) star counts from three different sources.  The photometry from \citet{2002A&A...391..945P} and \citet{2007AJ....133.1658S} for each cluster are included according to the criteria in Section~\ref{sec:R2_star_counts}.  The counts by \citet{2000MNRAS.313..571S} were performed on photometry available from various sources in the literature.  Metallicity [Fe/H] is from the \citet{1996AJ....112.1487H} catalogue (2010 edition) and the HB morphology parameters $L1$ and $L2$ are from \citet{2014ApJ...785...21M}.}
  \label{table_R2_different_photometry}
\begin{tabular}{rrrrrrrrrrrrr}
\hline
 & & & & \multicolumn{3}{c}{\hspace{0.2cm} \citet{2002A&A...391..945P} \hspace{0.2cm}}  & \multicolumn{3}{c}{\citet{2007AJ....133.1658S}} & \multicolumn{3}{c}{\hspace{0.2cm} \citet{2000MNRAS.313..571S} \hspace{0.2cm}} \\
\hspace{0.1cm} NGC & \hspace{0.2cm} [Fe/H] & $L1$ \hspace{0.1cm} & $L2$ \hspace{0.1cm} & \hspace{0.1cm} $n_\text{HB}$ \hspace{0.1cm} & $n_\text{AGB}$ &  $R_2$  \hspace{0.1cm}  & \hspace{0.2cm} $n_\text{HB}$  & $n_\text{AGB}$ & \hspace{0.1cm} $R_2$ \hspace{0.1cm} &  \hspace{0.2cm}$n_\text{HB}$ \hspace{0.1cm} & $n_\text{AGB}$ & \hspace{0.1cm} $R_2$ \hspace{0.1cm}\\
\hline
104 & -0.72 & 0.078 & 0.068 & 358 & 53 & 0.148 & 591 & 82 & 0.139 & 368 & 38 & 0.103 \\
362 & -1.26 & 0.086 & 0.608 & 238 & 40 & 0.168 & 318 & 43 & 0.135 & 94 & 14 & 0.149 \\
1261 & -1.27 & 0.088 & 0.644 & 94 & 22 & 0.234 & 233 & 34 & 0.146 & 148 & 26 & 0.176 \\
1851 & -1.18 & 0.098 & 0.679 & 272 & 37 & 0.136 & 411 & 49 & 0.119 & 209 & 24 & 0.115 \\
1904 & -1.60 & . . . & . . . & 163 & 11 & 0.067 & . . . & . . . & . . . & 122 & 16 & 0.131 \\
2419 & -2.15 & 0.192 & 0.852 & 225 & 22 & 0.098 & . . . & . . . & . . . & . . . & . . . & . . . \\
2808 & -1.14 & 0.094 & 0.904 & 809 & 61 & 0.075 & 1200 & 104 & 0.087 & 247 & 22 & 0.089 \\
4833 & -1.85 & 0.287 & 0.538 & 94 & 10 & 0.106 & . . . & . . . & . . . & . . . & . . . & . . . \\
5024 & -2.10 & 0.158 & 0.602 & 224 & 18 & 0.080 & 360 & 44 & 0.122 & 302 & 39 & 0.129 \\
5272 & -1.50 & 0.150 & 0.613 & . . . & . . . & . . . & 323 & 40 & 0.124 & 562 & 65 & 0.116 \\
5634 & -1.88 & . . . & . . . & 130 & 15 & 0.115 & . . . & . . . & . . . & . . . & . . . & . . . \\
5694 & -1.98 & . . . & . . . & 222 & 26 & 0.117 & . . . & . . . & . . . & 56 & 14 & 0.250 \\
5824 & -1.91 & . . . & . . . & 463 & 63 & 0.136 & . . . & . . . & . . . & . . . & . . . & . . . \\
5904 & -1.29 & 0.150 & 0.681 & 162 & 21 & 0.130 & 280 & 52 & 0.186 & 555 & 94 & 0.169 \\
5927 & -0.49 & 0.043 & 0.062 & 201 & 12 & 0.060 & . . . & . . . & . . . & 134 & 20 & 0.149 \\
6093 & -1.75 & 0.464 & 0.447 & 162 & 31 & 0.191 & 341 & 51 & 0.150 & 170 & 39 & 0.229 \\
6139 & -1.65 & . . . & . . . & 282 & 35 & 0.124 & . . . & . . . & . . . & 114 & 24 & 0.211 \\
6171 & -1.02 & 0.100 & 0.513 & . . . & . . . & . . . & 56 & 10 & 0.179 & 117 & 29 & 0.248 \\
6205 & -1.53 & 0.527 & 0.441 & 192 & 20 & 0.104 & 390 & 48 & 0.123 & 90 & 12 & 0.133 \\
6218 & -1.47 & 0.561 & 0.299 & . . . & . . . & . . . & 82 & 11 & 0.134 & 91 & 12 & 0.132 \\
6229 & -1.18 & . . . & . . . & 278 & 34 & 0.122 & . . . & . . . & . . . & 92 & 19 & 0.207 \\
6254 & -1.26 & 0.588 & 0.260 & . . . & . . . & . . . & 157 & 18 & 0.115 & 69 & 13 & 0.188 \\
6266 & -1.18 & . . . & . . . & 446 & 40 & 0.090 & . . . & . . . & . . . & 114 & 18 & 0.158 \\
6284 & -1.26 & . . . & . . . & 127 & 16 & 0.126 & . . . & . . . & . . . & . . . & . . . & . . . \\
6304 & -0.45 & 0.062 & 0.060 & 99 & 8 & 0.081 & . . . & . . . & . . . & . . . & . . . & . . . \\
6341 & -2.31 & 0.261 & 0.542 & . . . & . . . & . . . & 245 & 33 & 0.135 & 140 & 20 & 0.143 \\
6356 & -0.40 & . . . & . . . & 362 & 25 & 0.069 & . . . & . . . & . . . & . . . & . . . & . . . \\
6362 & -0.59 & 0.122 & 0.621 & 38 & 6 & 0.158 & . . . & . . . & . . . & . . . & . . . & . . . \\
6388 & -0.55 & 0.057 & 0.836 & 1347 & 176 & 0.131 & . . . & . . . & . . . & . . . & . . . & . . . \\
6402 & -1.28 & . . . & . . . & 349 & 29 & 0.083 & . . . & . . . & . . . & . . . & . . . & . . . \\
6441 & -0.46 & 0.048 & 0.904 & 1380 & 154 & 0.112 & . . . & . . . & . . . & . . . & . . . & . . . \\
6539 & -0.63 & . . . & . . . & 114 & 15 & 0.132 & . . . & . . . & . . . & . . . & . . . & . . . \\
6541 & -1.81 & 0.563 & 0.347 & . . . & . . . & . . . & 248 & 41 & 0.165 & . . . & . . . & . . . \\
6569 & -0.76 & . . . & . . . & 166 & 30 & 0.181 & . . . & . . . & . . . & . . . & . . . & . . . \\
6584 & -1.50 & 0.102 & 0.558 & 55 & 8 & 0.145 & . . . & . . . & . . . & . . . & . . . & . . . \\
6624 & -0.44 & 0.077 & 0.085 & 121 & 9 & 0.074 & 188 & 20 & 0.106 & 126 & 30 & 0.238 \\
6637 & -0.64 & 0.078 & 0.065 & 135 & 25 & 0.185 & 244 & 43 & 0.176 & 127 & 21 & 0.165 \\
6638 & -0.95 & . . . & . . . & 101 & 28 & 0.277 & . . . & . . . & . . . & . . . & . . . & . . . \\
6652 & -0.81 & 0.073 & 0.080 & 61 & 5 & 0.082 & 83 & 9 & 0.108 & 75 & 20 & 0.267 \\
6681 & -1.62 & 0.558 & 0.334 & 100 & 9 & 0.090 & . . . & . . . & . . . & 82 & 8 & 0.098 \\
6723 & -1.10 & 0.127 & 0.704 & 102 & 11 & 0.108 & 194 & 22 & 0.113 & 101 & 15 & 0.149 \\
6752 & -1.54 & 0.378 & 0.578 & . . . & . . . & . . . & 173 & 20 & 0.116 & 225 & 13 & 0.058 \\
6864 & -1.29 & . . . & . . . & 363 & 69 & 0.190 & . . . & . . . & . . . & 55 & 12 & 0.218 \\
6934 & -1.47 & 0.097 & 0.678 & 149 & 18 & 0.121 & 99 & 17 & 0.172 & . . . & . . . & . . . \\
6981 & -1.42 & 0.142 & 0.570 & 61 & 7 & 0.115 & 188 & 36 & 0.191 & 45 & 10 & 0.222 \\
7078 & -2.37 & 0.174 & 0.713 & 376 & 48 & 0.128 & 537 & 57 & 0.106 & 153 & 23 & 0.150 \\
7089 & -1.65 & 0.150 & 0.790 & 167 & 18 & 0.108 & 702 & 100 & 0.142 & . . . & . . . & . . . \\
7099 & -2.27 & 0.462 & 0.261 & 89 & 6 & 0.067 & . . . & . . . & . . . & 202 & 11 & 0.054 \\
\hline
\end{tabular}

\end{table*}

\section{Stellar models}
\label{sec:stellar_models_methods}
\label{sec:stellar_models}
\subsection{Description of models}
\label{sec:model_grid}

In this study we have computed a grid of stellar models that encompasses a range of values for the three parameters most important to the evolution of HB stars: stellar mass, helium abundance, and metallicity (and hence stellar age).  We have chosen to do this instead of calculating stellar models to specifically match each of the 48 clusters in our sample.  This would require at the very least running models with a suitable initial mass, metallicity, MLT mixing length parameter $\alpha_\text{MLT}$, and RGB mass loss rate (and then additional models with different initial mass and helium abundance to account for multiple populations) for each cluster.  

The stellar models were computed with the Monash University stellar evolution code {\sc monstar} \citep[which has been described previously, e.g.][]{2008A&A...490..769C,2014ApJ...784...56C}.  The models initially have a metal abundance in the solar ratio according to \citet{2009ARA&A..47..481A}, except with oxygen enhancement of $\text{[O/Fe]} = +0.4$ to mimic the $\alpha$-element enhancement observed in globular cluster stars \citep[see e.g.][]{2012A&ARv..20...50G}.  During the RGB evolution the models have the mass loss rate from \citet{1975MSRSL...8..369R} with $\eta = 0.4$.

The grid includes a total of 24 models with each of the four mixing schemes described in Section~\ref{sec:mixing_schemes} with every combination of $\text{[Fe/H]} = -2$, $-1$, $-0.5$, and $Y = 0.245$ and $0.284$.  The initial mass of each model was set so that the HB age is close to 13\,Gyr, consistent with that of the oldest Galactic globular clusters \citep[e.g.][]{2013ApJ...775..134V}.  Note that obtaining the correct age ZAHB age and RGB mass loss rate is not important here because the stellar structure is so well described by the aforementioned properties.  The difference in helium chosen, $\Delta Y = +0.039$, has an equivalent effect on lifetime to a 0.05\,$\text{M}_\odot$ decrease in initial mass for the $\text{[Fe/H]} = -1$ case, and is comparable to the modest limit on the spread of helium inferred for the majority of clusters (see Section~\ref{sec:helium} for references).  

In the following sections we quantify how the CHeB mixing scheme, initial mass, initial composition, and physical uncertainties affect the predictions for the observed quantities $R_2$ and $\Delta \log{L}_\text{HB}^\text{AGB}$.  When testing variables other than the mixing scheme we predominantly make use of models with the semiconvection and maximal overshoot prescriptions because, unlike the standard-overshoot sequences, their evolution is not strongly affected by the numerical treatment (see Section~\ref{sec:numerical_dependence}).  Unless stated otherwise, tests are carried out with models that have an initial mass of $M_\text{i} = 0.83\,\text{M}_\odot$, initial helium $Y=0.245$, and metallicity $\text{[Fe/H]} = -1$.

\subsection{Mixing schemes}
\label{sec:mixing_schemes}
In the CHeB phase we use the four different mixing schemes from Paper~I: (i) no overshoot, (ii) standard overshoot, (iii) semiconvection, and (iv) maximal overshoot.  In this paper we include only a brief summary of the mechanics and outcomes of these four mixing schemes: they are shown in more detail in Paper~I.  We also test the effects of uncertainties in the input physics, which are described in the relevant sections.

The \textbf{no-overshoot} models have the Schwarzschild criterion strictly applied.  That is, the location of the convective boundary is not found each time step by extrapolating $\nabla_\text{rad} - \nabla_\text{ad}$ across the prior position of the boundary to find the point of neutrality.  A zone may only become convective if the conditions ($T$, $p$, $\rho$, or composition from nuclear burning) change so that it becomes convectively unstable.  It cannot happen because of a change in composition from mixing or numerical diffusion (because this is not allowed).  

In the \textbf{semiconvection} scheme, slow mixing is allowed in regions that are formally stable according to the Schwarzschild criterion.  In this scheme, mixing is modelled as a diffusive process where the diffusion coefficient depends exponentially on how far $\nabla_\text{rad}/\nabla_\text{ad}$ is from unity.  The particular formulation and parameters used for this study are given in Paper~I.

The \textbf{standard overshoot} runs have overshooting at every convective boundary according to the scheme proposed by \citet{1997A&A...324L..81H} where there is an exponential decay of the diffusion coefficient that depends on the parameter $f_\text{OS}$ (see Section~\ref{sec:overshoot_dependence}).  Unless specified otherwise, we use $f_\text{OS} = 0.001$ in this study.  This value is consistent with the models in Paper~I, and in Section~ \ref{sec:overshoot_dependence} we also examine how $f_\text{OS}$ affects the evolution.

In the \textbf{maximal overshoot} scheme, convective overshoot is applied at the boundary of the convective core, and the core then allowed to grow, only if $\nabla_\text{rad}/\nabla_\text{ad} > 1 + \delta$ everywhere in the convection zone, where $\delta$ is a (small) parameter.  This ensures that mixing from overshoot does not reduce $\nabla_\text{rad}/\nabla_\text{ad}$ enough for part of the convection zone to become convectively stable (see Paper~I for details).  This scheme produces the largest convective core possible, i.e, there is a point in the convection zone (not necessarily at the boundary) that is close to convective neutrality, and would become stable if any more helium were mixed into the convection zone (Figure~3 in Paper~I).

The four mixing schemes in this study produce divergent internal structures.  The models without overshoot have the smallest possible convective core, i.e. the material adjacent to the boundary is close to convective neutrality according to the Schwarzschild criterion.  Because this boundary does not move during the evolution, conversion of helium to carbon and oxygen causes a large composition discontinuity to develop.  This is the only one of the four mixing schemes in which no helium is transported into the convective core, and it consequently has the shortest CHeB lifetime.  In calculations with standard overshoot, a large partially mixed region with a stepped composition profile develops around the convective core (Figure~2 in Paper~I).  This growth occurs via discrete mixing episodes and is driven by the higher opacity of the products of helium burning.  By the end of core helium burning, the mass of the partially mixed region can become comparable to that of the convective core beneath it.  The same effect is apparent in the semiconvection models except there is continuous, slow mixing, which results in a smooth composition gradient outside the convective core.  The maximal-overshoot sequences develop a structure that is similar to the models without overshoot, i.e. a large composition discontinuity at the core boundary, except that the mass enclosed by the convective core is much larger, and comparable to the total mass of the convective core plus partially mixed region in the standard overshoot models.

\subsection{Diagnostics for models}
We use the evolution sequences to generate theoretical $\log{L}$ PDFs $P(\Delta \log{L})$ which can then be compared with those derived from observations (Equation~\ref{eq:obs_PDF}).  The PDFs are constructed from models by iterating over the (post-core flash) evolution sequences and adding Gaussian functions so that
\begin{equation}
\label{eq:model_PDF}
\resizebox{.95\hsize}{!}{$
P(\Delta\log{L}) = \frac{1}{\tau}\sum\limits_{i=1}^n  \frac{\Delta t_i}{\sigma \sqrt{2\pi}}\exp{\left[-\frac{\left(\Delta \log{L}-\Delta \log{L_i}\right)^2}{2\sigma^2}\right]},
$}
\end{equation}
where $i$ represents each model in the sequence of $n$ models (where typically $n \approx 10^4$), $\Delta t_i$ is each time step, $\Delta\log{L_i} = \log{L_i}-\log{L_\text{HB}}$, $L_i$ is the luminosity of model $i$, $L_\text{HB}$ is the HB luminosity determined from the mode of the $\log{L}$ distribution, $\tau$ is the total time the model spends within the luminosity limits ($\log{L}<\log{L_\text{HB}} + 1.0$), and we have chosen $\sigma = 0.02$, which is sufficient to ensure that the theoretical luminosity PDFs are smooth.

\subsection{Overall comparison between models and observations}
\label{sec:overall_model_obs_comparison}

Along with the observations, Figure~\ref{figure_cluster_properties_z} and Figure~\ref{figure_cluster_properties_hb} also include models for comparison.  The trend lines (in grey) are each constructed from six models with combinations of $\text{[Fe/H]} = -2$, $-1$, and $-0.5$, and initial helium $Y = 0.245$ and $Y=0.284$.  

The predictions for $R_2$ from each of the these three schemes are below the observed average (by up to 3$\sigma$).  Although previous studies have shown examples of standard models with semiconvection zones predicting $R_2$ lower than that observed \citep[e.g.][]{1989ApJ...340..241C,2001A&A...366..578C,2003ApJ...588..862C,2007AJ....134..825V}, our tighter constraint on $R_2$ from two sets of homogeneous observations of a total of 48 clusters provides much stronger evidence that a discrepancy truly exists between the observations and standard models.  Importantly, however, the trend lines for $R_2$ in Figure~\ref{figure_cluster_properties_z} demonstrate that models and observations have the same insensitivity to stellar composition.  This implies that our conclusions about the validity of different mixing schemes are not weakened by uncertainty in the composition of the multiple populations of globular cluster stars.

In Section~\ref{sec:diff_mixing_results} we compare predictions from each mixing scheme to the observations and show how $\Delta\log{L_\text{HB}^\text{AGB}}$ depends on the stellar structure.  We specifically compare observations with standard-overshoot models in Section~\ref{sec:numerical_dependence} and \ref{sec:cbp}, and with maximal-overshoot models in Section~\ref{sec:gravonuclear_loops}.

The comparison between models and observations in Figure~\ref{figure_cluster_properties_hb} using the `blueness' of the HB ($L1 + L2/2$) shows the same offsets evident in Figure~\ref{figure_cluster_properties_z}.  The \citet{2007AJ....133.1658S} observations show a slight decrease of $R_2$ with an increase in $L1 + L2/2$ whereas the models and the \citet{2002A&A...391..945P} data show no trend.  The \citet{2007AJ....133.1658S} observations also show a dependence of $\Delta\log{L_\text{HB}^\text{AGB}}$ on $L1 + L2/2$ that is consistent with the example semiconvection and maximal-overshoot models (apart from the offset), especially given their large scatter (Figure~\ref{figure_cluster_properties_hb}).  In models, this slope is mostly due to the lower luminosity of bluer HB stars (luminosity is a strong function of envelope mass), rather than any affect on the luminosity of the AGB clump.  

\subsection{Effect of the mixing prescription}
\label{sec:diff_mixing_results}

\begin{figure}
\includegraphics[width=\linewidth]{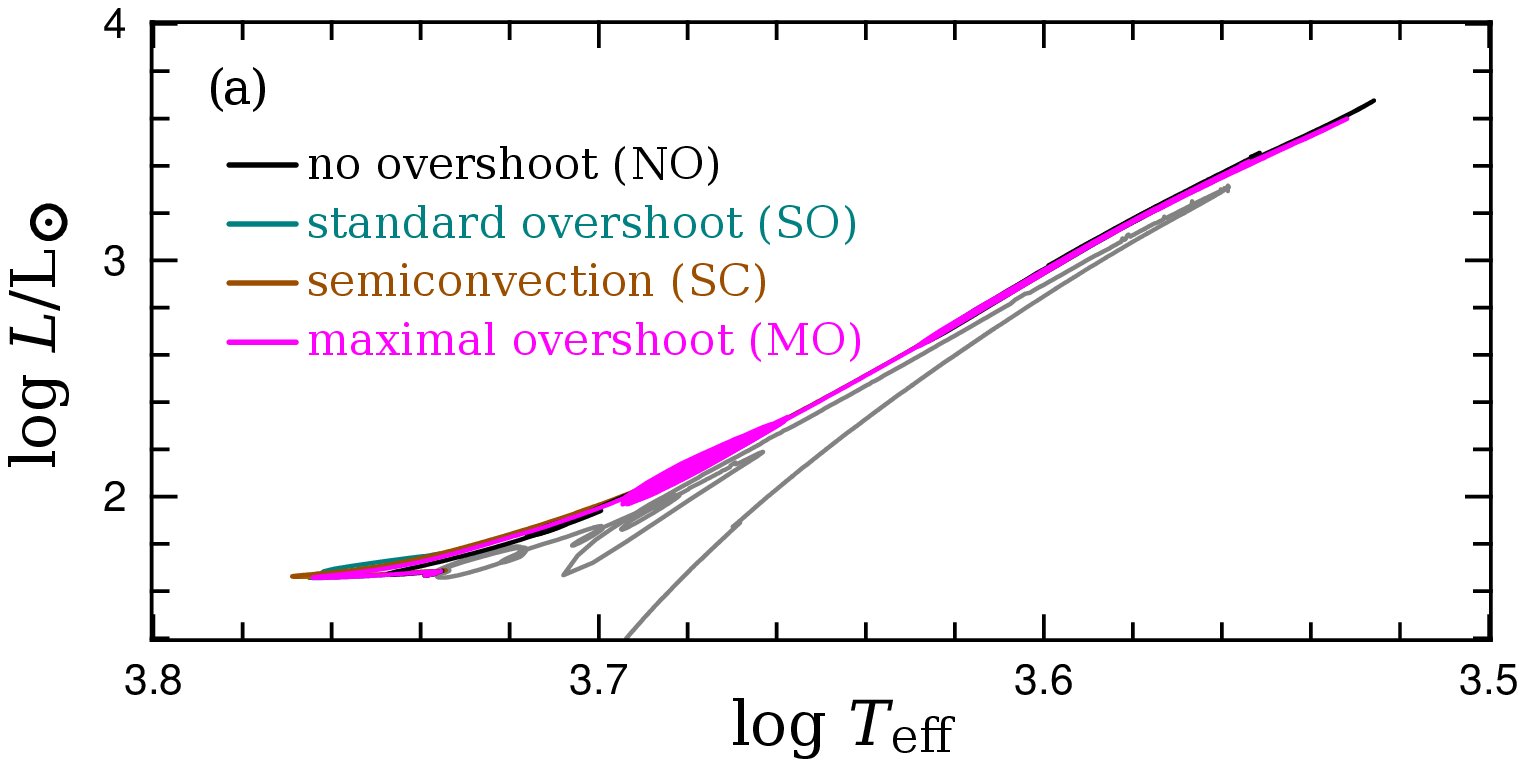}
\par
\vspace{0.35cm}
\includegraphics[width=\linewidth]{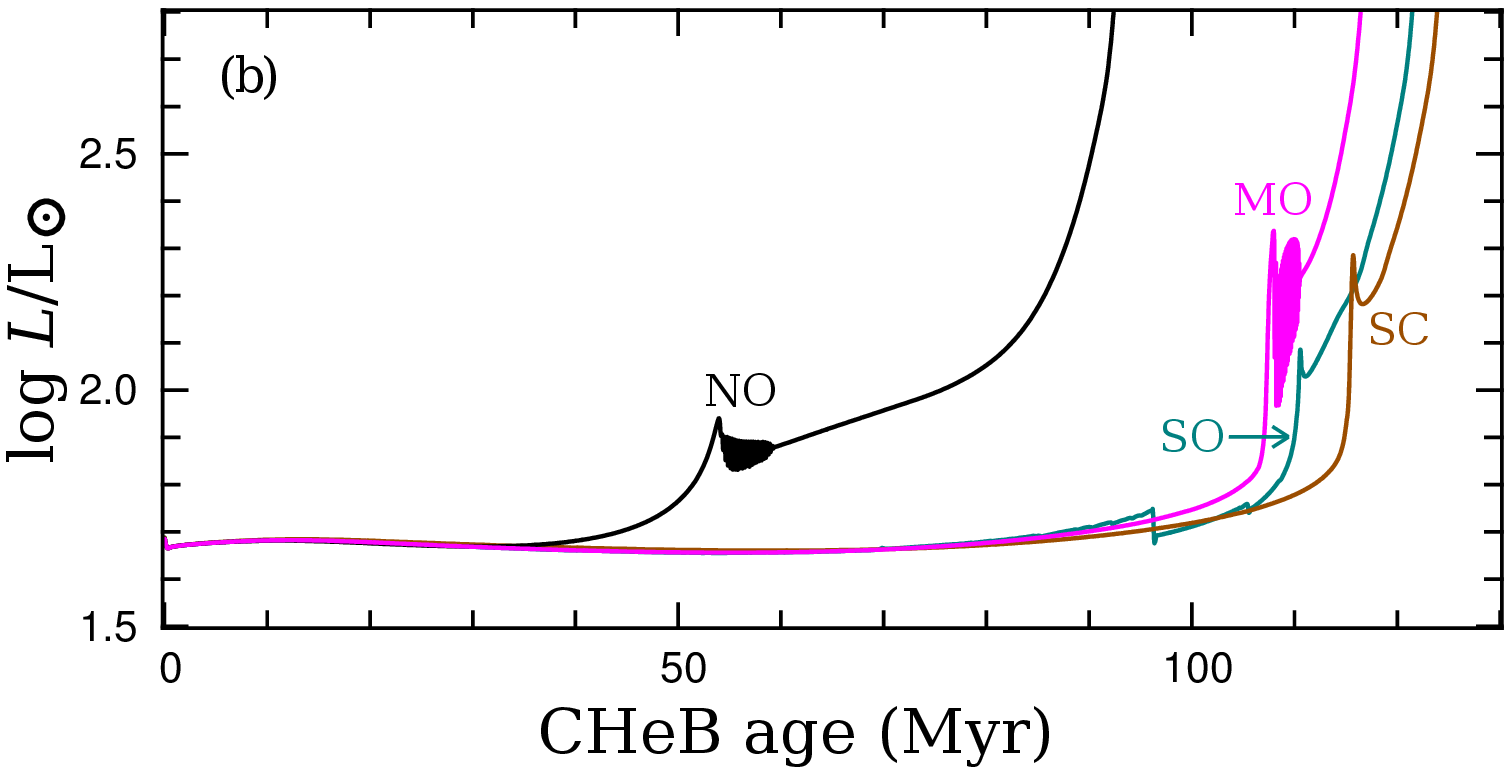}
\par
\vspace{0.35cm}
\includegraphics[width=\linewidth]{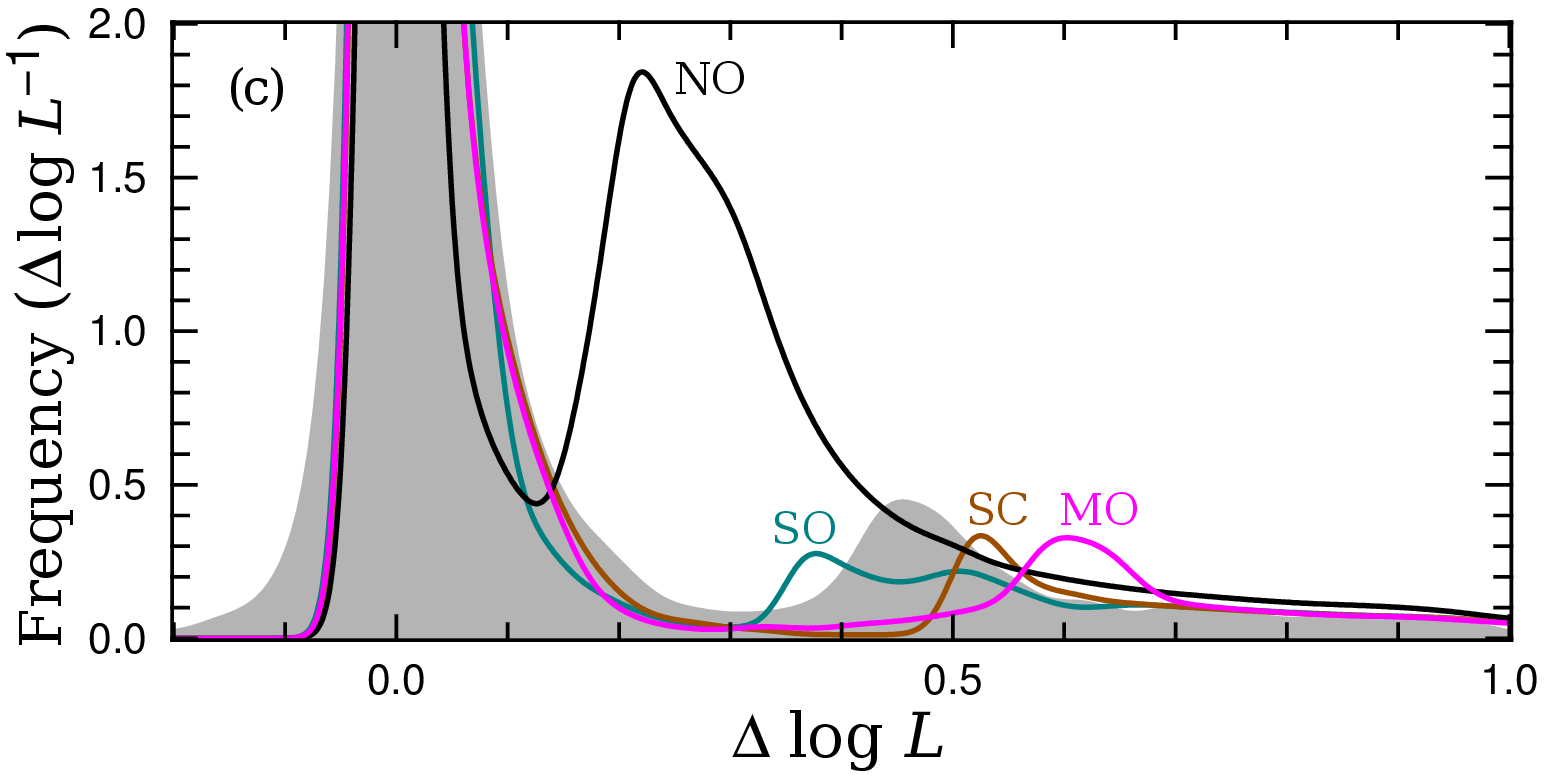}
  \caption{Comparison of models with different CHeB mixing schemes: standard overshoot, semiconvection, no overshoot, and maximal overshoot in cyan, orange, black, and magenta, respectively.  The models have initial mass $M_\text{i} = 0.83\,\text{M}_\odot$, metallicity $\text{[Fe/H]}=-1$, and initial helium $Y = 0.245$.  \textbf{Upper panel:} evolution tracks in the HR diagram.  \textbf{Middle panel:} surface luminosity evolution.  \textbf{Lower panel:} post-RGB luminosity probability density functions (PDF).  The shaded area is the observed PDF for all clusters without a blue HB, i.e. the combination of the two curves in Figure~\ref{figure_all_red_hb}.}
  \label{figure_mixing_comparison}
\end{figure}

In Figure~\ref{figure_mixing_comparison} we show the evolution of four models with different mixing schemes.  The resulting predictions of $R_2$ and $\Delta \log{L}_\text{HB}^\text{AGB}$ for these models are summarized in Table~\ref{table_model_predictions}.  It is evident from panel~(a) in Figure~\ref{figure_mixing_comparison} that each model follows the same path in the HR diagram.  The luminosity evolution of each sequence is nearly identical until they are close to exhausting helium in the core (Figure~\ref{figure_mixing_comparison}b).  The no-overshoot model is an obvious outlier because the lack of growth in the mass of the convective core restricts the fuel available and shortens the CHeB lifetime to less than half that of the others.  This increases the early-AGB lifetime and decreases $\Delta \log{L}_\text{HB}^\text{AGB}$ compared to the observations, producing a luminosity PDF (Figure~\ref{figure_mixing_comparison}c) that is starkly at odds with the observations shown in Figure~\ref{figure_all_red_hb}.  This result has been found previously \citep[e.g.][]{1983A&A...128...94B,1985A&A...145...97B,1986MmSAI..57..411B,1987ASSL..132..213C,1988ARA&A..26..199R,1989ApJ...340..241C,2001A&A...366..578C} and is consistent with the finding from asteroseismology that larger convective cores are preferred \citep{2013ApJ...766..118M,2015MNRAS.452..123C,2015MNRAS.453.2290B}.  It also has a strong theoretical basis because of the physical instability of the convective boundary.  We do not discuss the no-overshoot models further.

Among the other three models, the CHeB lifetime differs by less than 9\,Myr, which is only around 8 per cent.  The maximal-overshoot sequence has a larger $R_2$ and $\Delta \log{L}_\text{HB}^\text{AGB}$ than the semiconvection sequence; this is also true throughout this study, regardless of initial composition or input physics.  Of the three sequences in Figure~\ref{figure_mixing_comparison}, the one with standard overshoot has the lowest $\Delta \log{L}_\text{HB}^\text{AGB}$.  However, $\Delta \log{L}_\text{HB}^\text{AGB}$ and CHeB lifetime for the standard-overshoot sequences strongly depend on the time step constraints and the overshooting parameter $f_\text{OS}$.  These dependences are explored in Section~\ref{sec:numerical_dependence}.  Each of these three mixing schemes fails to match the average $R_2$ observed: the standard-overshoot, semiconvection, and maximal overshoot sequences have $R_2 = 0.096$, $0.068$, and $0.082$, respectively, compared with the observed average $R_2 = 0.117 \pm 0.005$.  

\setlength{\tabcolsep}{4pt}
\begin{table}
\centering
  \caption{Summary of observations and model predictions.  The models have initial mass $M_\text{i} = 0.83\,\text{M}_\odot$, metallicity $\text{[Fe/H]}=-1$, and initial helium $Y = 0.245$.  The observed values are derived from the 14 clusters without blue HBs (see Section~\ref{sec:red_hb_clusters}).  The uncertainty for the standard-overshoot models is the standard deviation from the results of the calculations using different $f_\text{OS}$ that are discussed in Section~\ref{sec:overshoot_dependence}.}
  \label{table_model_predictions}
\footnotesize
\begin{tabular}{lll}
\hline
 & $R_2$ & $\Delta \log{L}_\text{HB}^\text{AGB}$\\
\hline
Observations & $0.117 \pm 0.005$ & $0.455 \pm 0.012$ \\
No overshoot & 0.783 & 0.22 \\
Semiconvection & 0.068  & 0.53 \\
Standard overshoot & $0.075 \pm 0.025$ & $0.46 \pm 0.15$ \\
Maximal overshoot & 0.082 & 0.60 \\
\hline
\end{tabular}
\normalsize
\end{table}

In addition to the $R_2$ discrepancy, none of the models in Figure~\ref{figure_mixing_comparison} can match $\Delta \log{L}_\text{HB}^\text{AGB} = 0.455 \pm 0.012$ from observations.  The standard-overshoot, semiconvection, and maximal overshoot sequences have $\Delta \log{L}_\text{HB}^\text{AGB} = 0.38$, $0.53$, and $0.60$, respectively.  Contrary to the case for $R_2$, the observed $\Delta \log{L}_\text{HB}^\text{AGB}$ at least sits within the spread resulting from the three mixing schemes.  Figure~\ref{figure_mixing_comparison}(b) shows that the luminosity during CHeB is independent of the mixing scheme.  The broad range in $\Delta \log{L}_\text{HB}^\text{AGB}$ is due to the disparity in the masses of the helium-exhausted cores at the onset of shell helium burning.  In these sequences, shell helium burning begins with core masses of approximately 0.05\,M$_\odot$, 0.10\,M$_\odot$, and 0.14\,M$_\odot$, respectively.  The dependence of the AGB clump luminosity on the mass enclosed by the early-AGB helium-burning shell and strongly suggests that the CHeB partially mixed region extends too far in the semiconvection and maximal-overshoot models.

After CHeB, when helium burning moves to a shell, the no-overshoot and maximal-overshoot models both have chemical discontinuities at the convective core boundary.  This leads to `gravonuclear loops' \citep[see e.g.][]{1997ApJ...489..822B,1997ApJ...479..279B,2000LIACo..35..529S,2015ASSP...39...25B}, which cause an oscillation in surface luminosity that lasts for a few million years (near 55\,Myr and 110\,Myr for the respective sequences in Figure~\ref{figure_mixing_comparison}b).  These convection and burning episodes eventually end once a smooth helium composition profile has been established.  We discuss their effect on $R_2$ and $\Delta \log{L}_\text{HB}^\text{AGB}$ in the maximal-overshoot case in Section~\ref{sec:gravonuclear_loops}.  In addition, we show that convective overshoot during the early-AGB phase in these models can reduce the disagreement with observations.

\subsection{Effects of composition and other input physics}
\label{composition_and_input_physics}

\subsubsection{MLT mixing length}
\label{sec:MLT}
We have tested the importance of the choice of MLT mixing length parameter $\alpha_\text{MLT}$.  We did this by evolving models (beginning with the same model at the core flash) with three different values of $\alpha_\text{MLT}$: the solar-calibrated value ($\alpha_\text{MLT} = 1.53$) and increases of $\Delta \alpha_\text{MLT} = +0.2$ and $\Delta \alpha_\text{MLT} = +1.0$.  Each group of runs with different mixing schemes but the same $\alpha_\text{MLT}$ formed a distinct evolutionary track in the HR diagram (the effective temperature $T_\text{eff}$ increases with higher $\alpha_\text{MLT}$).  Models with the same mixing scheme but different $\alpha_\text{MLT}$ have exactly the same luminosity evolution.  The luminosity PDFs are similarly unaffected by changes to $\alpha_\text{MLT}$.  We may therefore safely proceed with our luminosity comparisons between models and observations without finding a suitable $\alpha_\text{MLT}$ to match the theoretical and observed $T_\text{eff}$ for each case.

\subsubsection{Effects of initial helium abundance}
\label{sec:helium}
It has been suggested that some globular clusters host helium-rich subpopulations.  Evidence for this includes the detection of multiple main sequences \citep[e.g.][]{2004ApJ...605L.125B,2004ApJ...612L..25N,2005ApJ...621..777P,2007ApJ...661L..53P,2013ApJ...767..120M,2015MNRAS.446.1672M}, HB morphology \citep[e.g.][]{2002A&A...395...69D,2005ApJ...621L..57L,2005A&A...435..987C,2007A&A...463..949C}, spectroscopy of hot HB stars \citep{2009A&A...499..755V,2012ApJ...748...62V,2014MNRAS.437.1609M}, and abundance patterns that point towards various scenarios of self-enrichment in the products of hydrogen burning \citep[e.g.][]{2002A&A...393..215V,2007A&A...464.1029D,2009A&A...507L...1D,2014MNRAS.437L..21D}.  The consequences of changing the initial helium abundance in models must therefore be considered.  

We tested the effect of increasing the initial helium abundance $Y$ by $\Delta Y = 0.039$.  This increase, while simultaneously decreasing the initial mass so that the age is unchanged, tends to decrease $R_2$.  In the metallicity range tested ($ -2 \leq \text{[Fe/H]} \leq -0.5$), however, the effect is small.  We find for the semiconvection and maximal-overshoot models that 
\begin{equation}
\frac{\partial R_2}{\partial Y} \approx -0.05.
\end{equation}
The important factors contributing to this trend are the decrease, with increasing helium, in the mass of both the envelope and the H-exhausted core.  The envelope mass is reduced because increasing the initial helium abundance, while keeping the age constant, reduces the initial stellar mass.  Both of the factors mentioned slow helium burning throughout CHeB.  This lengthens the CHeB phase and lowers the absolute luminosity of the AGB cut-off (see Section~\ref{sec:z_and_hb_morph}), which both decrease $R_2$.  The effects of changing only the H-exhausted core mass and only the initial mass are examined in Sections~\ref{sec:neutrino} and \ref{sec:initial_mass}, respectively.

The reduction in $R_2$ from increasing helium becomes more substantial when the metallicity is higher, whereas the effect on $\Delta \log{L}_\text{HB}^\text{AGB}$ becomes smaller with increasing metallicity.  There is also a significant difference between the effect on semiconvection and maximal-overshoot models.  Models with the former mixing scheme show a greater increase in $\Delta \log{L}_\text{HB}^\text{AGB}$ with increasing initial helium.  This difference appears to be due to the effect on the luminosity of the AGB clump.  Increasing helium by $\Delta Y =0.039$ decreases the HB luminosity by $\Delta \log{L_\text{HB}} = -0.015$ for both mixing schemes.  In contrast, it increases the luminosity of the AGB clump by nearly $\Delta \log{L_\text{HB}} = 0.03$ for the semiconvection run while having no effect on the maximal-overshoot sequence.  At its largest (for the $\text{[Fe/H]} = -2$ semiconvection model) we find that 
\begin{equation}
\frac{\partial \Delta \log{L}_\text{HB}^\text{AGB}}{\partial Y} \approx 1.25.
\end{equation}
It therefore appears that accounting for the small variation in initial helium allowed for most globular clusters could have a modest effect on $\Delta \log{L_\text{HB}^\text{AGB}}$ and a negligible effect on $R_2$.

\subsubsection{Metallicity}

\begin{figure}
\includegraphics[width=\linewidth]{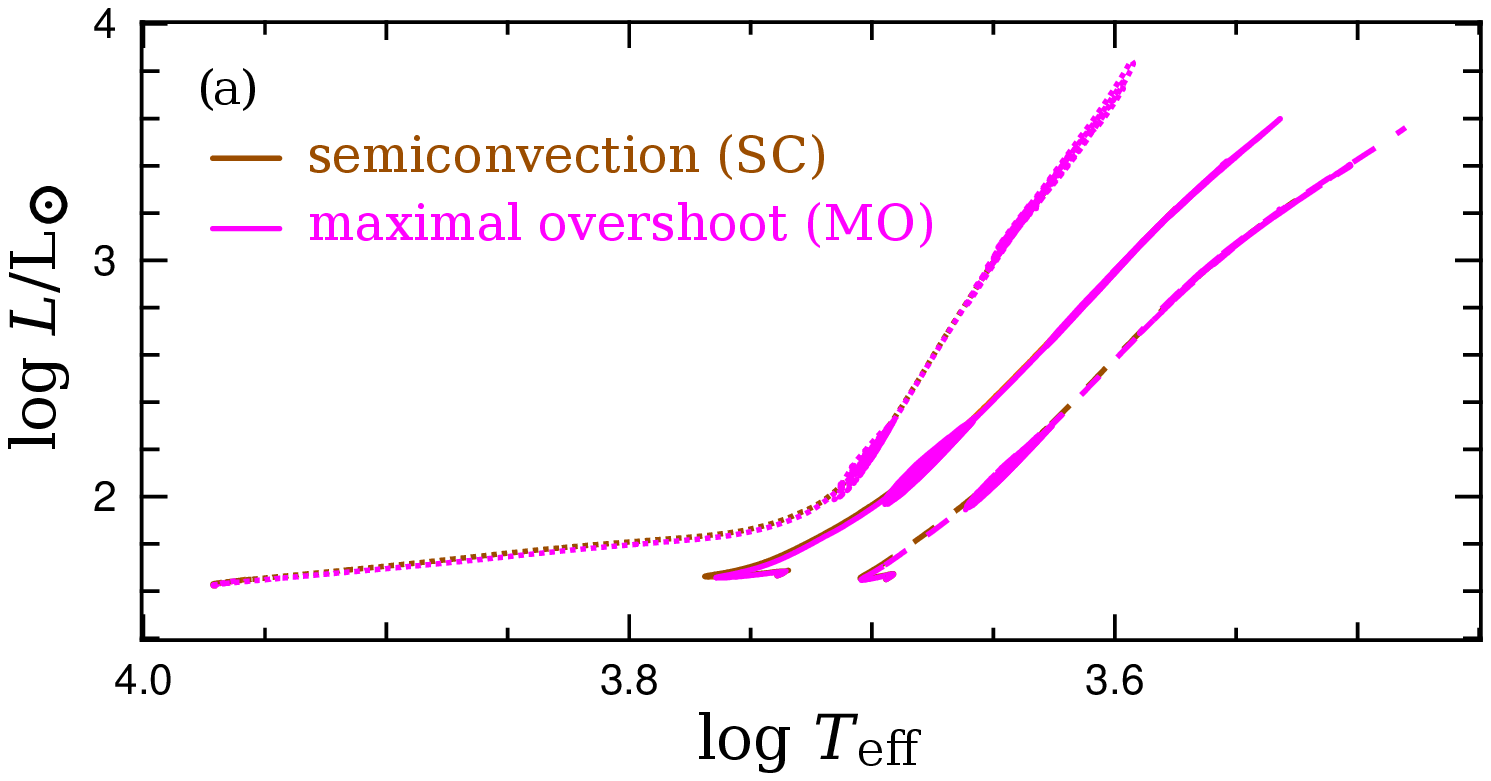}
\par
\vspace{0.35cm}
\includegraphics[width=\linewidth]{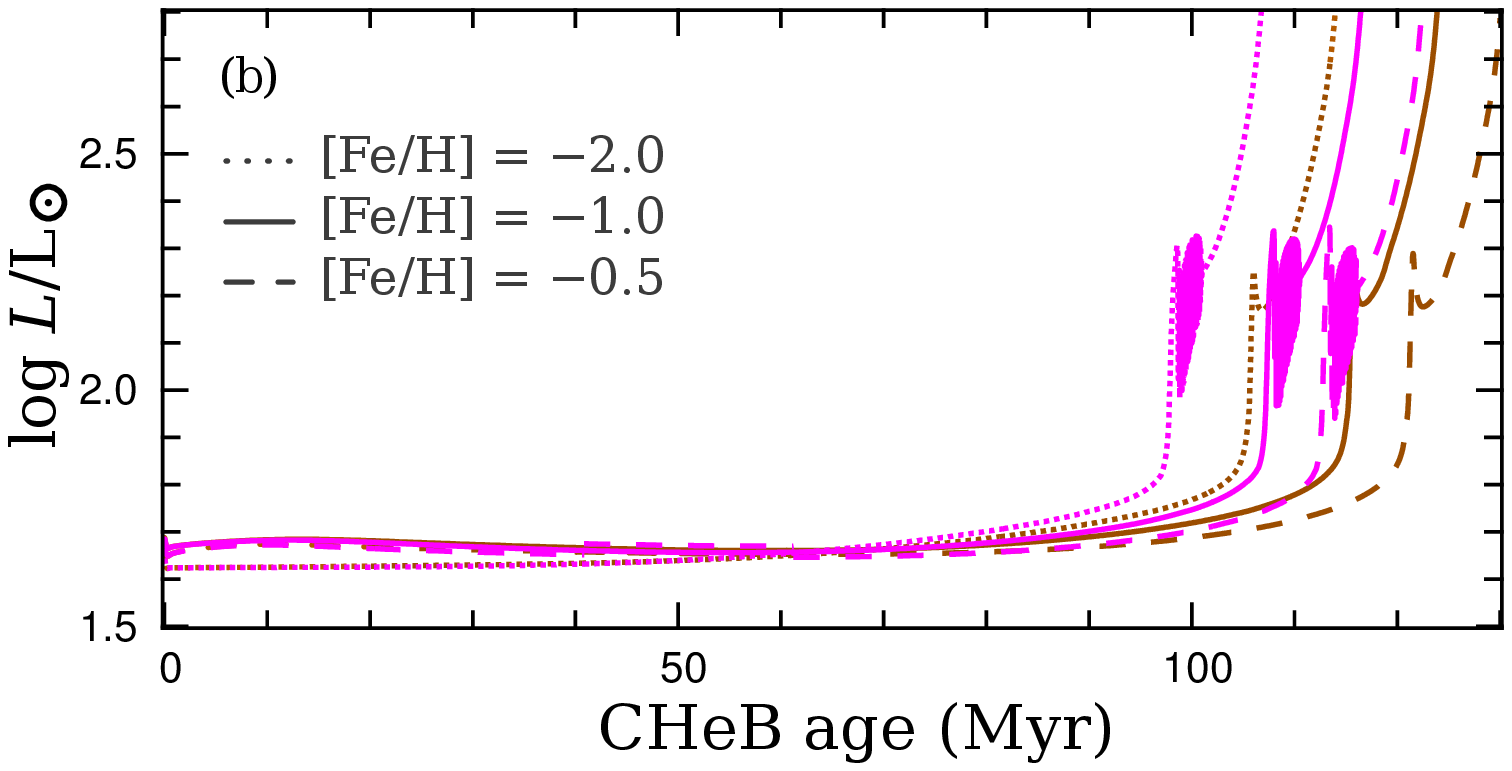}
\par
\vspace{0.35cm}
\includegraphics[width=\linewidth]{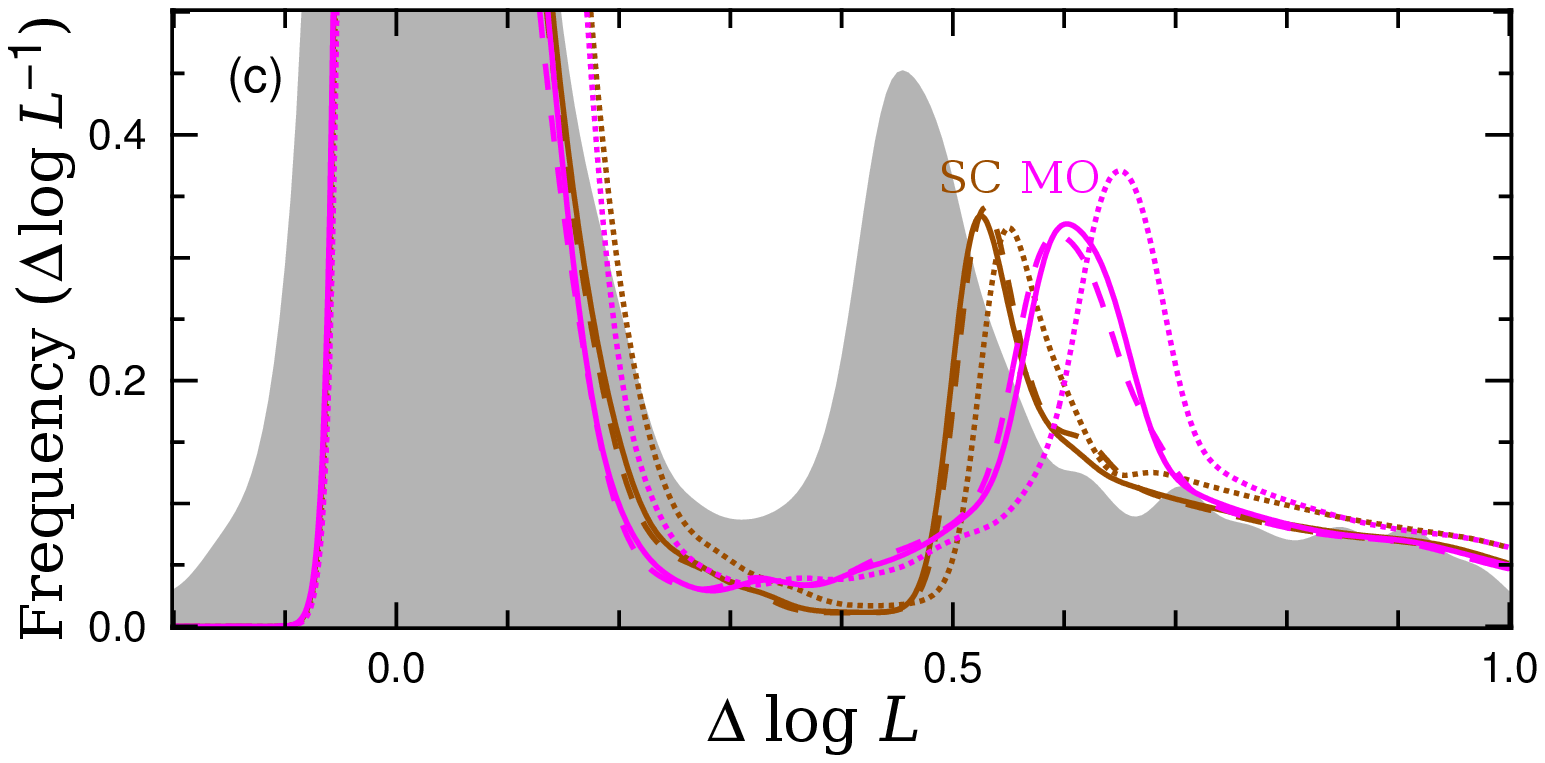}
  \caption{Comparison of semiconvection (orange) and maximal overshoot (magenta) models with different metallicities, $\text{[Fe/H]} = -2$, $-1$, and $-0.5$ (shown by dotted, solid, and dashed lines, respectively), and initial masses $M_\text{i}/\text{M}_\odot = 0.80$, $0.83$, and $0.89$ so that they are each 13\,Gyr old.  The panels are the same as Figure~\ref{figure_mixing_comparison}.}
  \label{figure_metallicity}
\end{figure}

In Figure~\ref{figure_metallicity} we present models with three different metallicities: $\text{[Fe/H]} = -2$, $-1$, and $-0.5$, which spans most of the metallicity range in our globular cluster sample (shown in Figure~\ref{figure_cluster_properties_z}).  Increasing the metallicity tends to very slightly decrease $R_2$ in models, where we find
\begin{equation}
\frac{\partial R_2}{\partial \text{[Fe/H]}} \approx -0.003.
\end{equation}
Similarly, $\Delta \log{L}_\text{HB}^\text{AGB}$ decreases with increasing metallicity according to
\begin{equation}
\frac{\partial \Delta \log{L}_\text{HB}^\text{AGB}}{\partial \text{[Fe/H]}} \approx -0.03.
\end{equation}
but this change is not consistent between models with different mixing schemes or composition.  Both of these trends are small enough to be consistent with the absence of a detectable metallicity trend in the observations, for which there is also considerable scatter (Figure~\ref{figure_cluster_properties_z}) and an unknown trend in helium abundance and cluster age which would also affect theoretical predictions (see Section~\ref{sec:helium} and \ref{sec:initial_mass}, respectively).

\subsubsection{Effects of helium burning reaction rates}

\begin{figure}
\includegraphics[width=\linewidth]{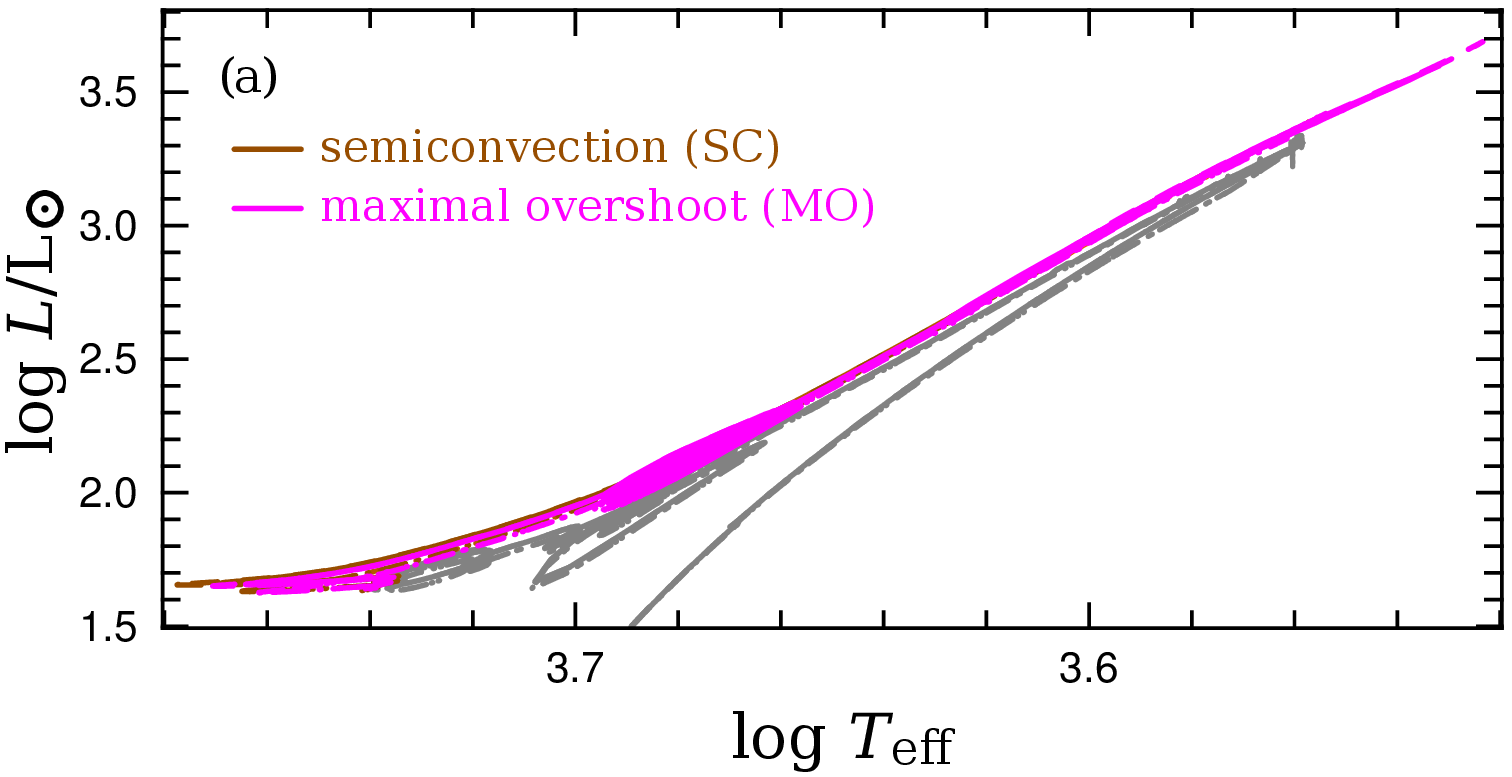}
\par
\vspace{0.35cm}
\includegraphics[width=\linewidth]{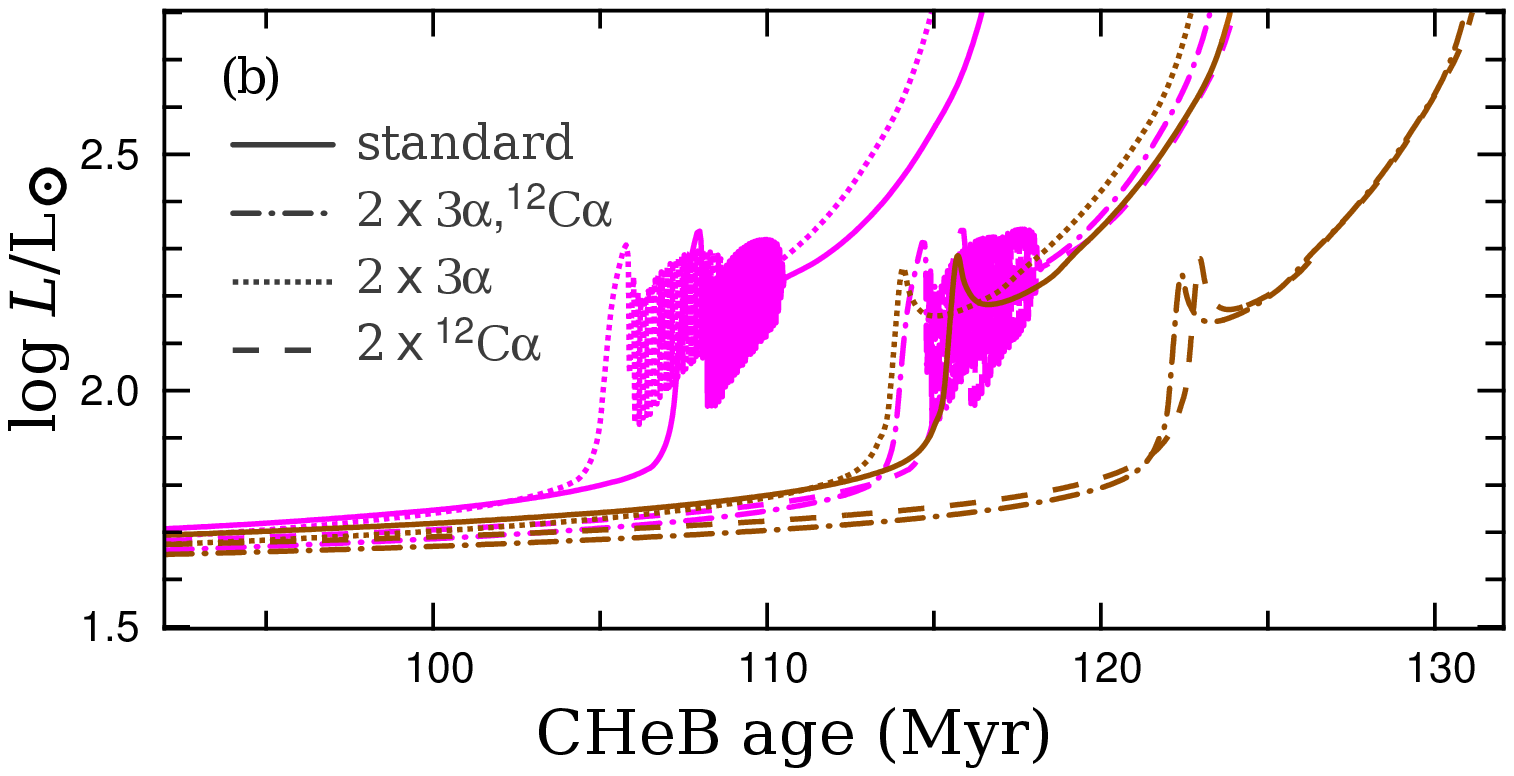}
\par
\vspace{0.35cm}
\includegraphics[width=\linewidth]{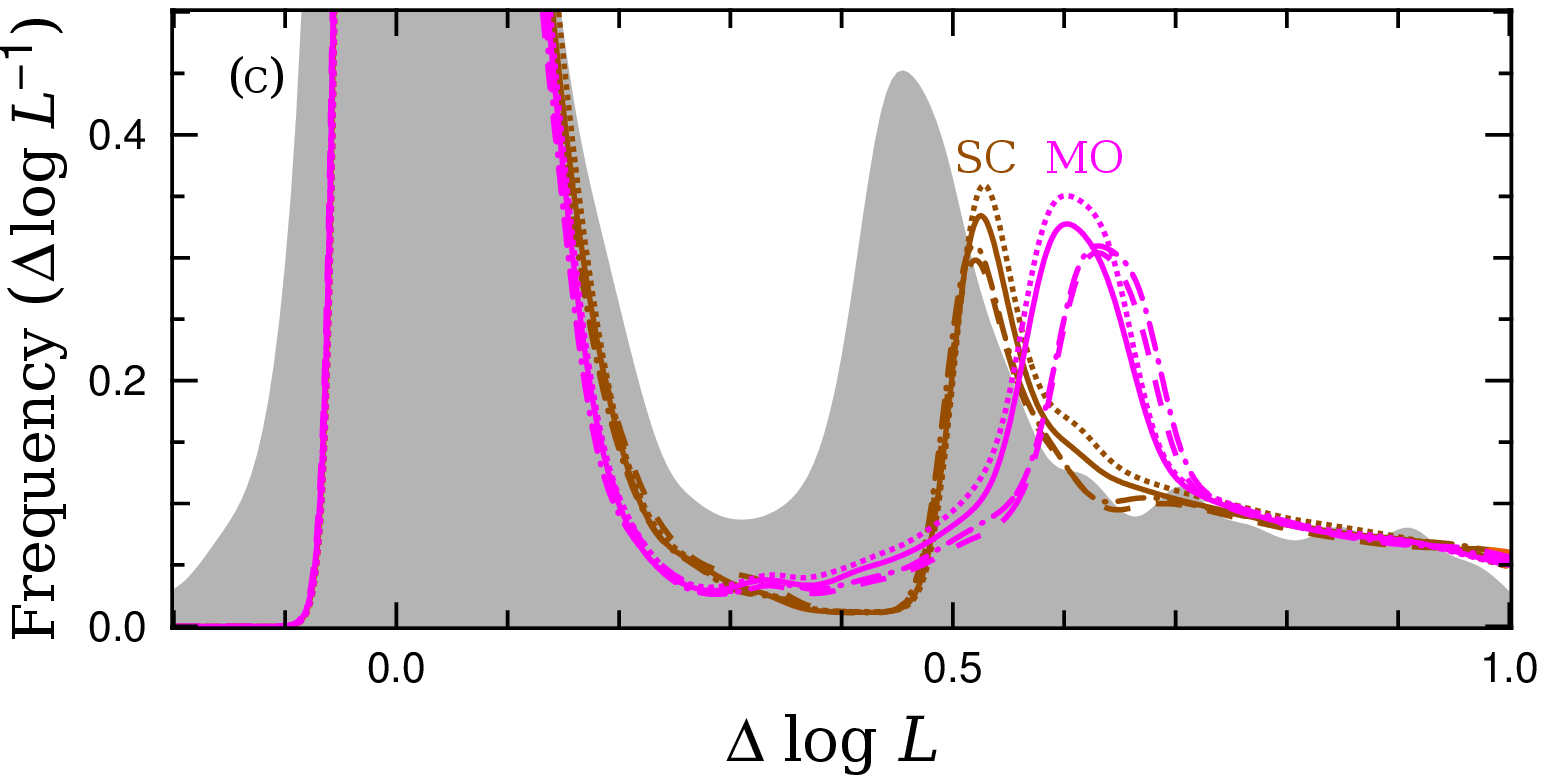}
  \caption{Comparison of CHeB and early-AGB evolution for models with semiconvection (orange) and maximal overshoot (magenta) and different helium burning reaction rates.  The models have standard reaction rates (solid lines), double the triple-$\alpha$ rate (dotted lines), double the triple-$\alpha$ and ${}^{12}\text{C}(\alpha,\gamma)^{16}\text{O}$ rates (dotted dashed lines),  and double the ${}^{12}\text{C}(\alpha,\gamma)^{16}\text{O}$ rate (dashed lines).  The panels are the same as Figure~\ref{figure_mixing_comparison}.}
  \label{figure_different_reaction_rates1}
\end{figure}

\begin{figure}
\includegraphics[width=\linewidth]{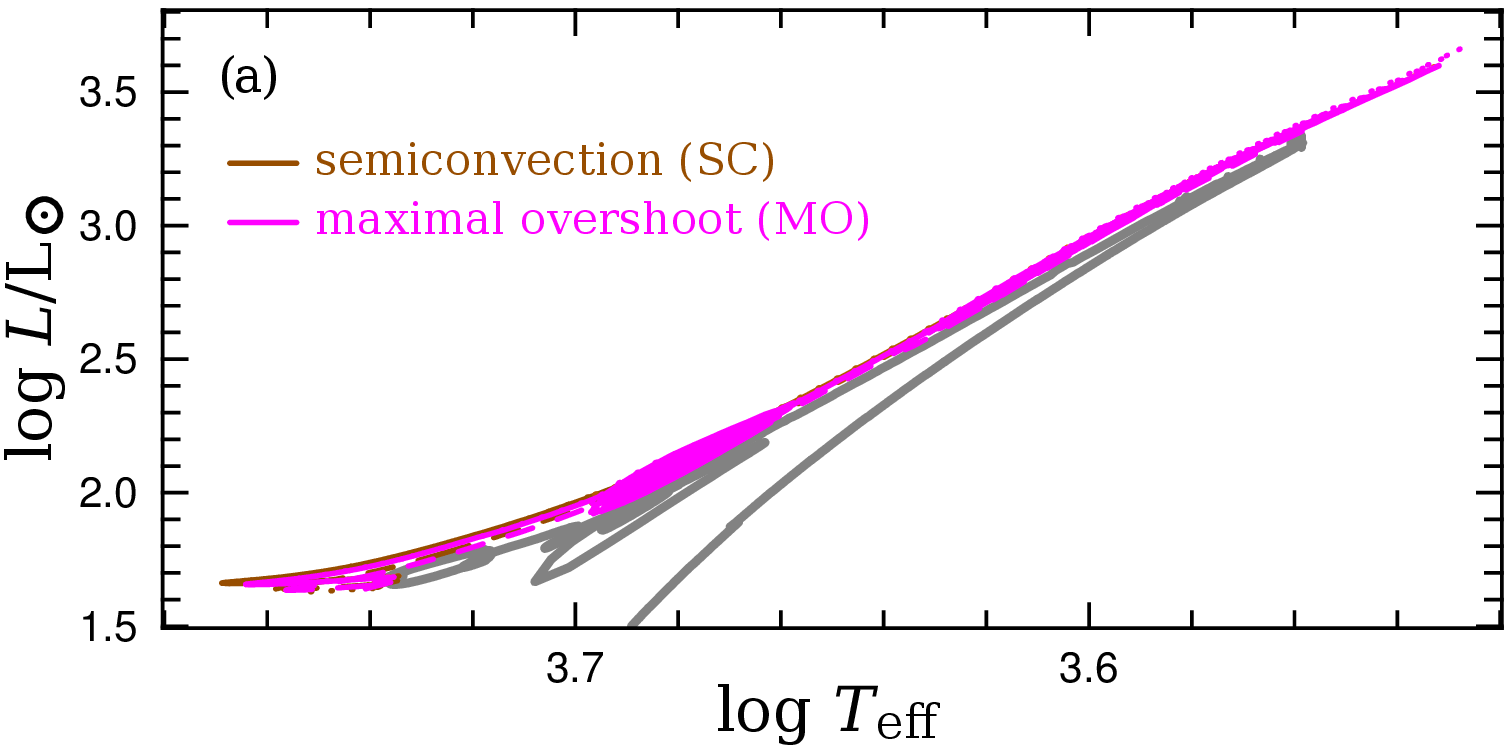}
\par
\vspace{0.35cm}
\includegraphics[width=\linewidth]{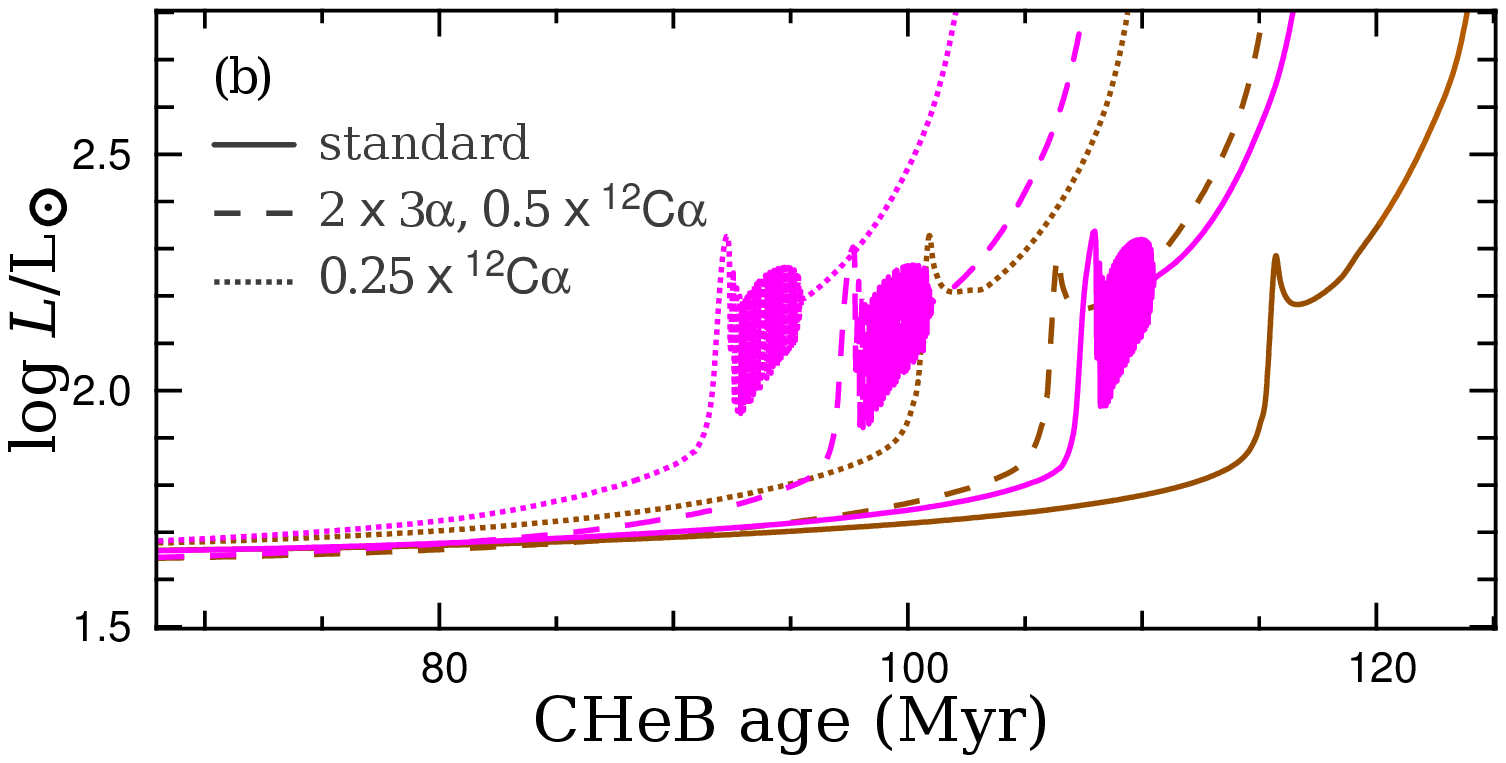}
\par
\vspace{0.35cm}
\includegraphics[width=\linewidth]{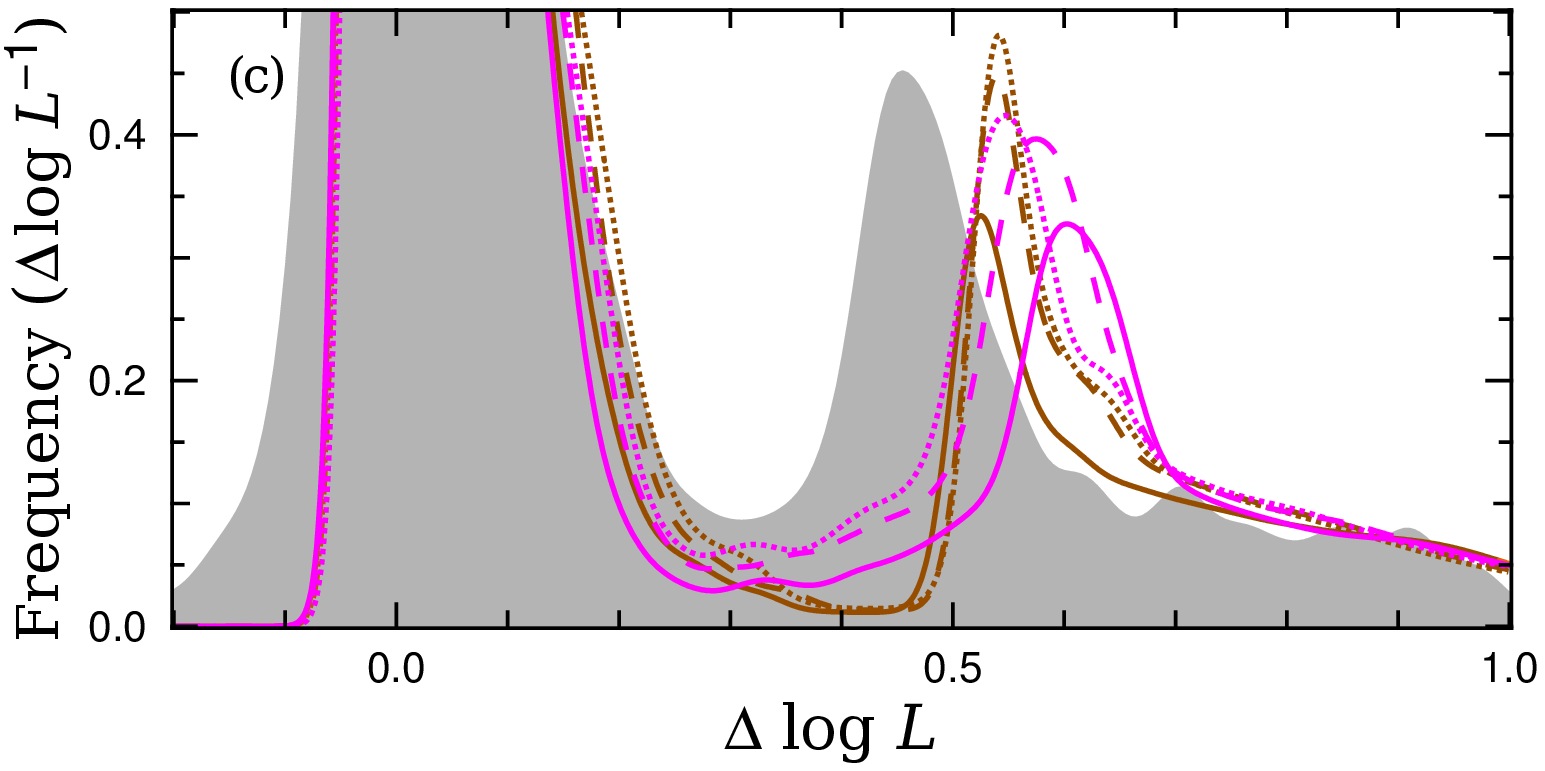}
  \caption{Comparison of CHeB and early-AGB evolution for models with semiconvection (orange) and maximal overshoot (magenta) and different helium burning reaction rates.  The models have standard reaction rates (solid lines), the ${}^{12}\text{C}(\alpha,\gamma)^{16}\text{O}$ rate multiplied by a factor of a quarter (dotted lines), and double the triple-$\alpha$ and half the ${}^{12}\text{C}(\alpha,\gamma)^{16}\text{O}$ rate (dashed lines).  The panels are the same as Figure~\ref{figure_mixing_comparison}.}
  \label{figure_different_reaction_rates2}
\end{figure}

Uncertainties in helium burning reaction rates are an important complication for efforts to constrain the mixing in CHeB models.  We have examined the effects of changing the triple-$\alpha$ and ${}^{12}\text{C}(\alpha,\gamma)^{16}\text{O}$ reaction rates, both separately and concurrently.  We change these reaction rates by up to factors of two and four, which is larger than their respective uncertainties of 15 per cent and 40 per cent \citep{1999NuPhA.656....3A}.   Examples of these tests are presented in Figures~\ref{figure_different_reaction_rates1} and \ref{figure_different_reaction_rates2}.  

Increasing the triple-$\alpha$ rate reduces the H-exhausted core mass at the flash and therefore also the mass of the convective core during subsequent CHeB (see Paper~I).  Later in CHeB, the triple-$\alpha$ reaction is favoured at the expense of the ${}^{12}\text{C}(\alpha,\gamma)^{16}\text{O}$, reducing the total energy that can be released from helium burning.  Both of these consequences contribute to the shortening of the CHeB phase.  In contrast, increasing the ${}^{12}\text{C}(\alpha,\gamma)^{16}\text{O}$ rate obviously favours that reaction, releasing more energy, which then causes an increase in the fuel supply by expanding the mass enclosed by the convective core.  These effects lead to an increase in the CHeB lifetime.

Doubling the triple-$\alpha$ rate decreases the absolute luminosity of the HB and AGB clump equally and therefore has little effect on $\Delta \log{L_\text{HB}^\text{AGB}}$.  Increasing the ${}^{12}\text{C}(\alpha,\gamma)^{16}\text{O}$ rate by the same factor has only about one fifth of the effect on the HB luminosity compared with an equal change of the triple-$\alpha$ rate.  In the semiconvection models the early-AGB luminosity (and therefore also $\Delta \log{L_\text{HB}^\text{AGB}}$) is relatively unaffected.  This contrasts with the maximal-overshoot models, where the increased ${}^{12}\text{C}(\alpha,\gamma)^{16}\text{O}$ rate, and consequently larger convective core at the end of CHeB, pushes the position of subsequent He-burning shell outward during the early-AGB and increases the luminosity, and thus also $\Delta \log{L_\text{HB}^\text{AGB}}$.  In these maximal-overshoot models we find that 
\begin{equation}
\frac{\partial \Delta \log{L_\text{HB}^\text{AGB}}}{\partial \log{r_{\text{C}\alpha}}} = 0.09.
\end{equation}

In models with either semiconvection or maximal overshoot, increasing the triple-$\alpha$ rate increases $R_2$ whereas increasing the  ${}^{12}\text{C}(\alpha,\gamma)^{16}\text{O}$ rate decreases $R_2$.  The strengths of the effects, however, depend on the mixing scheme.  In the maximal-overshoot case we find that
\begin{equation}
\frac{\partial R_2}{\partial \log{r_{3\alpha}}} = 0.025,
\end{equation}
and
\begin{equation}
\frac{\partial R_2}{\partial \log{r_{\text{C}\alpha}}} = -0.04,
\end{equation}
which are both around double that for the semiconvection models.  We have also confirmed that these partial derivatives hold when the two reaction rates are changed simultaneously.

\subsubsection{Effect of neutrino emission rate}
\label{sec:neutrino}
The neutrino production mechanism most important to the evolution of CHeB stars is the plasma process, which is an efficient cooling mechanism for the degenerate core prior to the ignition of helium.   Observations of globular cluster stars provide some of the best constraints for non-standard neutrino electromagnetic coupling \citep[e.g.][]{1999ARNPS..49..163R,2013PhRvL.111w1301V}.  This is because additional cooling from plasma neutrino emission would delay the core flash and allow the core mass to grow further, thereby increasing the luminosity of the RGB-tip stars.  In Paper~I we showed that an increased H-exhausted core mass at the flash could help resolve the discrepancy between the predicted and observationally inferred asymptotic g-mode period spacing $\Delta\Pi_1$ in \textit{Kepler} field stars.  Other exotic processes could also affect CHeB evolution, such as axion production via the Primakoff effect.  This would be most significant during the CHeB phase, when the core is non-degenerate, and would shorten the HB lifetime \citep{2012arXiv1201.1637R}.

In this section we test the results of an ad hoc increase to the neutrino emission rate by a factor of four.  Because this prolongs the RGB evolution we also halt mass loss when the total mass reaches that of the standard run at the RGB-tip, ensuring that the comparisons are between models of the same total mass.  We have also computed sequences in which the neutrino emission rate is returned to the standard rate after the core flash.  This serves as a proxy for other physical uncertainties whose main effect is to alter the core mass at the flash.

The modification to the neutrino emission rate $\epsilon_\nu$ throughout the evolution of the semiconvection and maximal-overshoot runs causes a decrease in $R_2$ according to 
\begin{equation}
\frac{\partial R_2}{\partial \log{\epsilon_\nu}} = -0.03.
\end{equation}
This dependence is due to the additional early-AGB neutrino losses rather than any effect on the preceding RGB evolution.  Given that $R_2$ in standard models is already lower than the observed range, this appears to be another strong restriction on the magnetic dipole moments of neutrinos, which would increase the emission from plasmon decay.

If the standard neutrino emission rate is restored after the flash, the models both show
\begin{equation}
 \label{eq:R2_neutrino}
 \left\lvert \frac{\partial R_2}{\partial \log{\epsilon_\nu}} \right\rvert < 0.008.
\end{equation}
The main structural change in those models is an increase of the H-exhausted core mass at the flash.  Multiplying the neutrino emission rate by a factor of four increases the H-exhausted core mass at the flash $M_\text{He}$ by $0.029\,\text{M}_\odot$.  The dependence in Equation~\ref{eq:R2_neutrino} can therefore be expressed as
\begin{equation}
 \left\lvert \frac{\partial R_2}{\partial M_\text{He}/\text{M}_\odot} \right\rvert < 0.15.
\end{equation}
In the maximal overshoot sequence for instance, the 0.029\,M$_\odot$ increase in $M_\text{He}$, which is larger than permitted by other constraints \citep{1996ApJ...461..231C}, decreases $R_2$ by $0.001$ and increases $\Delta \log{L_\text{HB}^\text{AGB}}$ by just $0.015$.  This demonstrates that reasonable uncertainty in the core mass at the flash does not significantly affect $R_2$ or $\Delta \log{L_\text{HB}^\text{AGB}}$.

The ratio $R_2$ could form an even tighter constraint on novel particle emission than the previously used ratio $R = n_\text{HB}/n_\text{RGB}$, because any `dark channel' that is more active during the early-AGB than CHeB would further lower $R_2$ and worsen the agreement between observations and standard models.  This should be an obligatory consideration when comparing stellar models with globular cluster observations to determine these constraints.  Although this is beyond the scope of this study, we note that unlike the earlier RGB evolution in which the core is degenerate, the burning shell that surrounds the degenerate core during the early-AGB is hot enough to support helium burning (and is hotter than CHeB).  This would have implications for temperature sensitive effects, such as Primakoff conversion from axion-photon coupling which has a specific energy loss rate that goes as $T^7/\rho$ \citep{2013PhRvL.110f1101F,2015arXiv150202357A}.  Indeed, \citet{1999MNRAS.306L...1D} have already shown that this can significantly truncate the early-AGB lifetime in more massive models ($M \ge 3\,\text{M}_\odot$).

When the altered neutrino loss rate is applied only before the core flash we find
\begin{equation}
\frac{\partial \Delta \log{L}_\text{HB}^\text{AGB}}{\partial \log{\epsilon_\nu}} \approx 0.025,
\end{equation}
or equivalently, a weak dependence on H-exhausted core mass:
\begin{equation}
\frac{\partial \Delta \log{L}_\text{HB}^\text{AGB}}{\partial M_\text{He}/\text{M}_\odot} \approx 0.50.
\end{equation}
Retaining the higher neutrino emission during the CHeB and early-AGB phases further increases $\Delta \log{L}_\text{HB}^\text{AGB}$ in models with maximal overshoot.  In models with the altered neutrino emission rate for the whole evolution, we find
\begin{equation}
\frac{\partial \Delta \log{L}_\text{HB}^\text{AGB}}{\partial \log{\epsilon_\nu}} \approx 0.08,
\end{equation}
which is a rate of change more than three times that for models with altered neutrino losses only during the RGB.  In contrast, $\epsilon_\nu$ has a negligible effect on $\Delta \log{L}_\text{HB}^\text{AGB}$ for the semiconvection models.  We further discuss the early-AGB evolution of the maximal overshoot models in Section~\ref{sec:gravonuclear_loops}.

In Paper~I we showed that the increased $M_\text{He}$ from enhanced neutrino emission during the RGB improves the agreement between CHeB models and asteroseismology. Interestingly though, increasing neutrino emission throughout the evolution reduces $R_2$ and worsens the agreement with the observations.  Standard models are also not improved when the excess neutrino losses are stopped after the core flash (emulating any process whose main effect is to increase $M_\text{He}$).  Even a substantial core mass increase of $\Delta M_\text{HB} = 0.029\,\text{M}_\odot$ produces only small changes to $R_2$ and $\Delta \log{L_\text{HB}^\text{AGB}}$.  The insensitivity of $R_2$ (according to the definition in this study) to uncertainties in $M_\text{He}$ increases its diagnostic power for CHeB mixing.

\subsubsection{Stellar initial mass / age}
\label{sec:initial_mass}
We have tested the effect of increasing the initial stellar mass for $\text{[Fe/H]} = -0.5$ models while keeping other parameters unchanged.  In this case, increasing the initial mass of the $M = 0.84\,\text{M}_\odot$ model by 0.05\,M$_\odot$ reduces the age at the beginning of CHeB by 2.4\,Gyr.  In sequences with either semiconvection or maximal overshoot, increasing the initial mass decreases both $R_2$ and $\Delta \log{L}_\text{HB}^\text{AGB}$.  We can quantify the changes (with respect to either initial mass or age) by
\begin{equation}
\frac{\partial \Delta \log{L}_\text{HB}^\text{AGB}}{\partial M_\text{i}/\text{M}_\odot} \approx -0.30,
\end{equation}
or
\begin{equation}
\frac{\partial \Delta \log{L}_\text{HB}^\text{AGB}}{\partial t_\text{ZAHB}} \approx 0.007,
\end{equation}
where $t_\text{ZAHB}$ is the age in Gyr.  This dependence is not significant given the size of the uncertainty in globular cluster stellar mass and age.  $R_2$ is similarly insensitive to initial mass, although in that case the effect is not consistent between the two mixing schemes.  In both cases we find that
\begin{equation}
-0.07 < \frac{\partial R_2}{\partial M_\text{i}/\text{M}_\odot}, 
\end{equation}
and
\begin{equation}
\frac{\partial R_2}{\partial t_\text{ZAHB}} < 0.0015.
\end{equation}

We have also isolated the dependence of $R_2$ and $ \Delta \log{L}_\text{HB}^\text{AGB}$ on the zero-age HB mass by halting mass loss before the RGB tip. We find for the $\text{[Fe/H]} = -0.5$ models that
\begin{equation}
\frac{\partial \Delta \log{L}_\text{HB}^\text{AGB}}{\partial M_\text{ZAHB}/\text{M}_\odot} = -0.23,
\end{equation}
and
\begin{equation}
\frac{\partial R_2}{\partial M_\text{ZAHB}/\text{M}_\odot} = -0.019.
\end{equation}
In contrast, there are only negligible differences for models that differ in age by 2.4\,Gyr but have the same total ZAHB mass.  This conclusively shows that the dissimilar CHeB evolution in models with a different initial mass stems from how it affects the zero-age HB mass.  This is important because there is some degeneracy between mass loss rate and initial mass/cluster age.  Moreover, the change in $R_2$ from increasing the initial mass but keeping helium constant is opposite to that found when helium is also adjusted so that the age is kept constant when the mass is increased (see Section~\ref{sec:helium}).  The weak dependence of $R_2$ and $\Delta \log{L}_\text{HB}^\text{AGB}$ on stellar age validates our assumption in Section~\ref{sec:stellar_models} that it is unnecessary to make specific models to match the age of each cluster.

\subsubsection{Summary of these effects}
\label{sec:summary_of_effects}
We can use the dependencies identified in this section to estimate the overall uncertainties for predictions of $R_2$ and $\Delta \log{L_\text{HB}^\text{AGB}}$ from factors other than the mixing scheme.  In order to do this we use the following approximate uncertainties in the models: $\Delta r_{3\alpha}/r_{3\alpha} = 0.15$, $\Delta r_{\text{C}\alpha}/r_{\text{C}\alpha} = 0.40$ \citep{1999NuPhA.656....3A}, $\Delta Y = 0.04$, $\Delta \text{[Fe/H]} = 1.0$, $\Delta t_\text{ZAHB} = 2\,\text{Gyr}$, $\Delta M_\text{HB} = 0.01\,\text{M}_\odot$, and $\Delta M_\text{ZAHB} = 0.05\,\text{M}_\odot$.  By adding each of the consequential contributions to the uncertainties of $R_2$ and $\Delta \log{L_\text{HB}^\text{AGB}}$ in quadrature, we find that the 1-$\sigma$ uncertainty in $R_2$ is 0.009 and in $\Delta \log{L_\text{HB}^\text{AGB}}$ it is 0.04.  The dominant source of uncertainty for $R_2$ is the $^{12}\text{C}(\alpha,\gamma)^{16}\text{O}$ reaction rate.  The most important factor for $\Delta \log{L_\text{HB}^\text{AGB}}$ is composition, including both helium abundance and metallicity.  The uncertainty of $R_2$ in models is comparable to the statistical uncertainty in the observations ($\sigma_{R_2,\text{obs}} = 0.005$) but for $\Delta \log{L_\text{HB}^\text{AGB}}$ the uncertainty in models is many times larger than the statistical uncertainty in the observations($\sigma_{\Delta \log{L},\text{obs}} = 0.012$).  Both of these uncertainties are smaller than the changes resulting from the use of different mixing schemes (see e.g. Figure~\ref{figure_mixing_comparison}c).  This confirms that $R_2$ and $\Delta \log{L_\text{HB}^\text{AGB}}$ are powerful constraints for the mixing in the core.

\subsection{Numerical effects}
\label{sec:numerical_dependence}

\subsubsection{Dependence on the overshooting parameter}
\label{sec:overshoot_dependence}

\begin{figure}
\includegraphics[width=\linewidth]{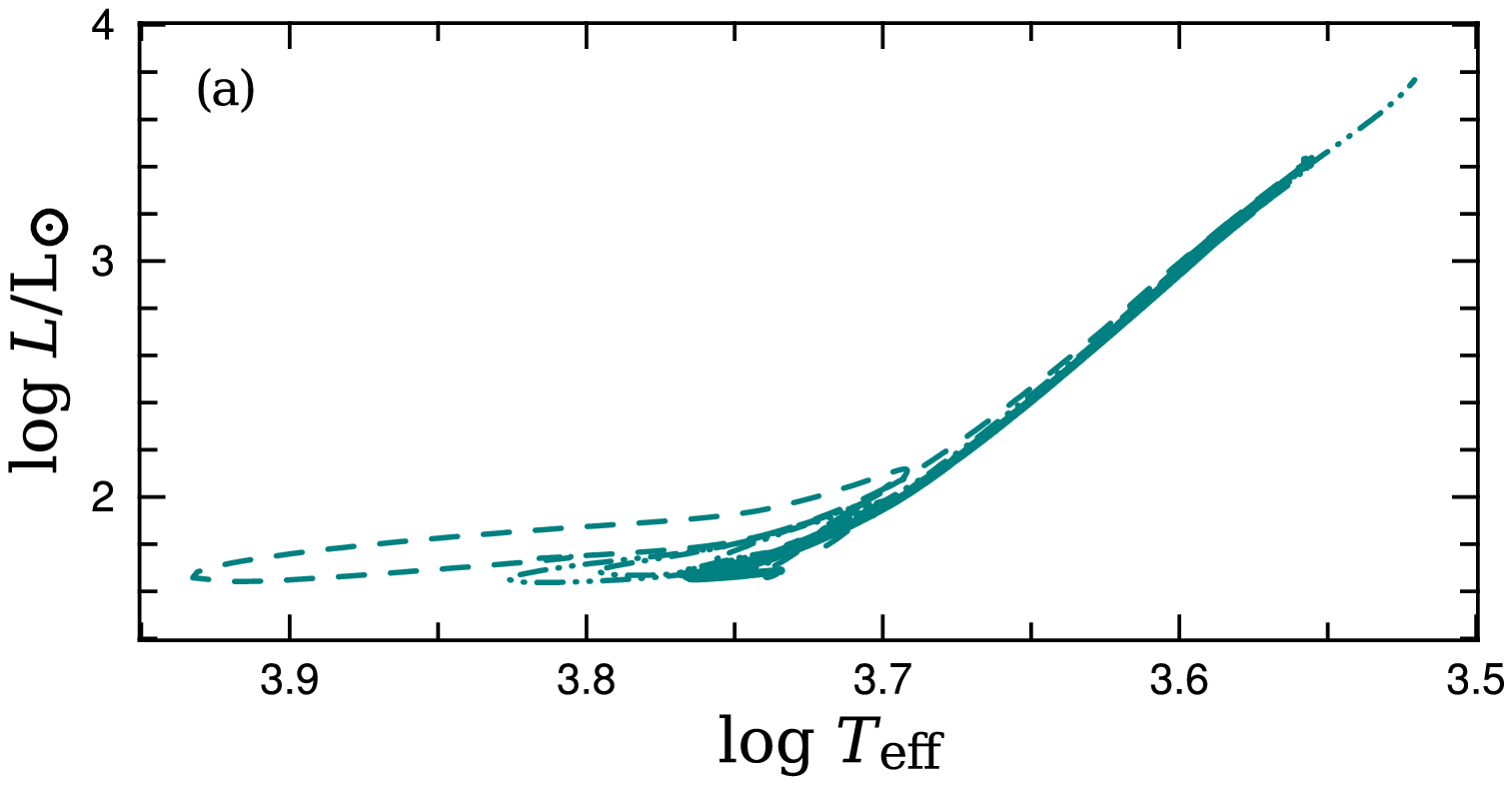}
\par
\vspace{0.35cm}
\includegraphics[width=\linewidth]{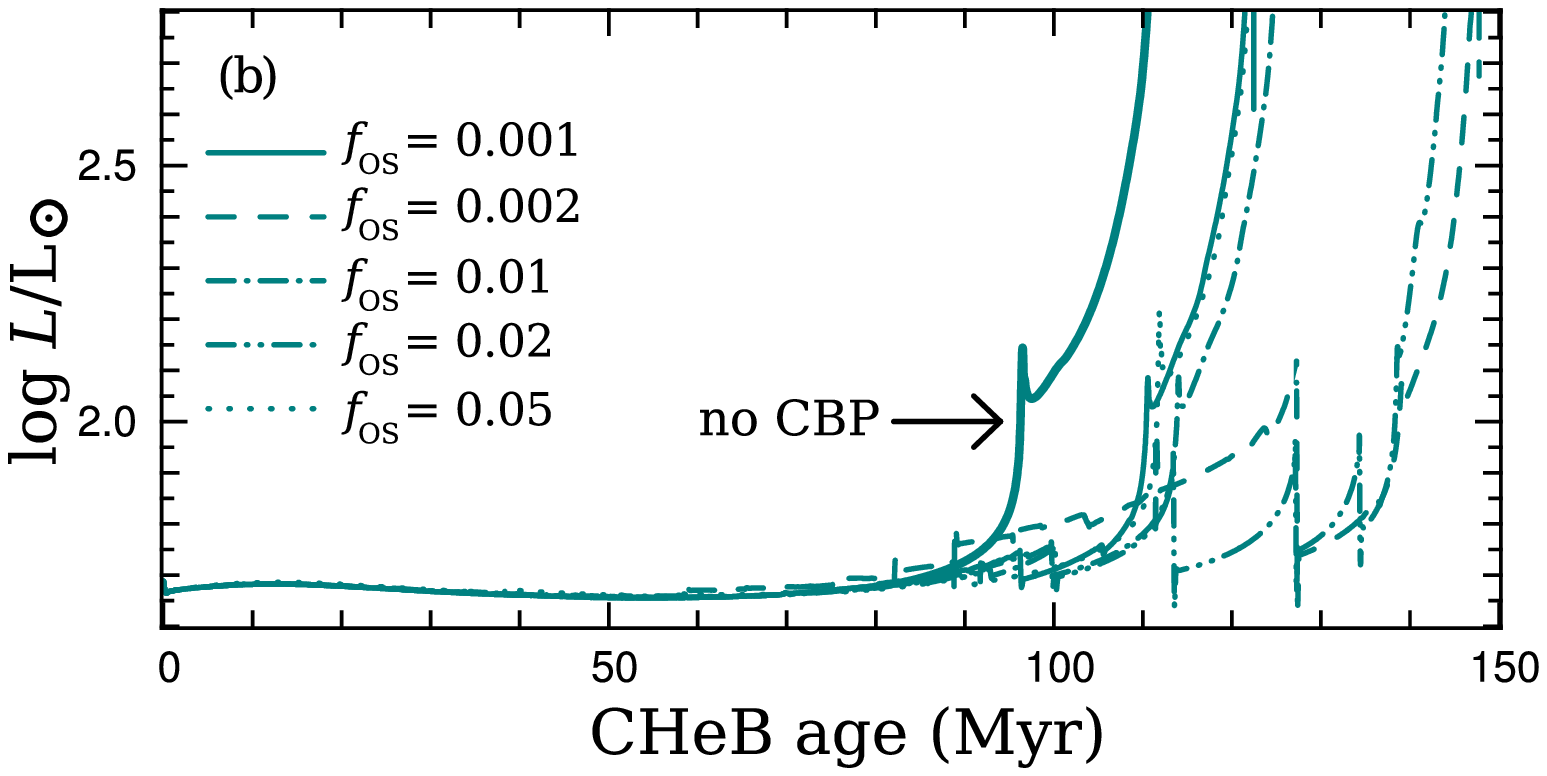}
\par
\vspace{0.35cm}
\includegraphics[width=\linewidth]{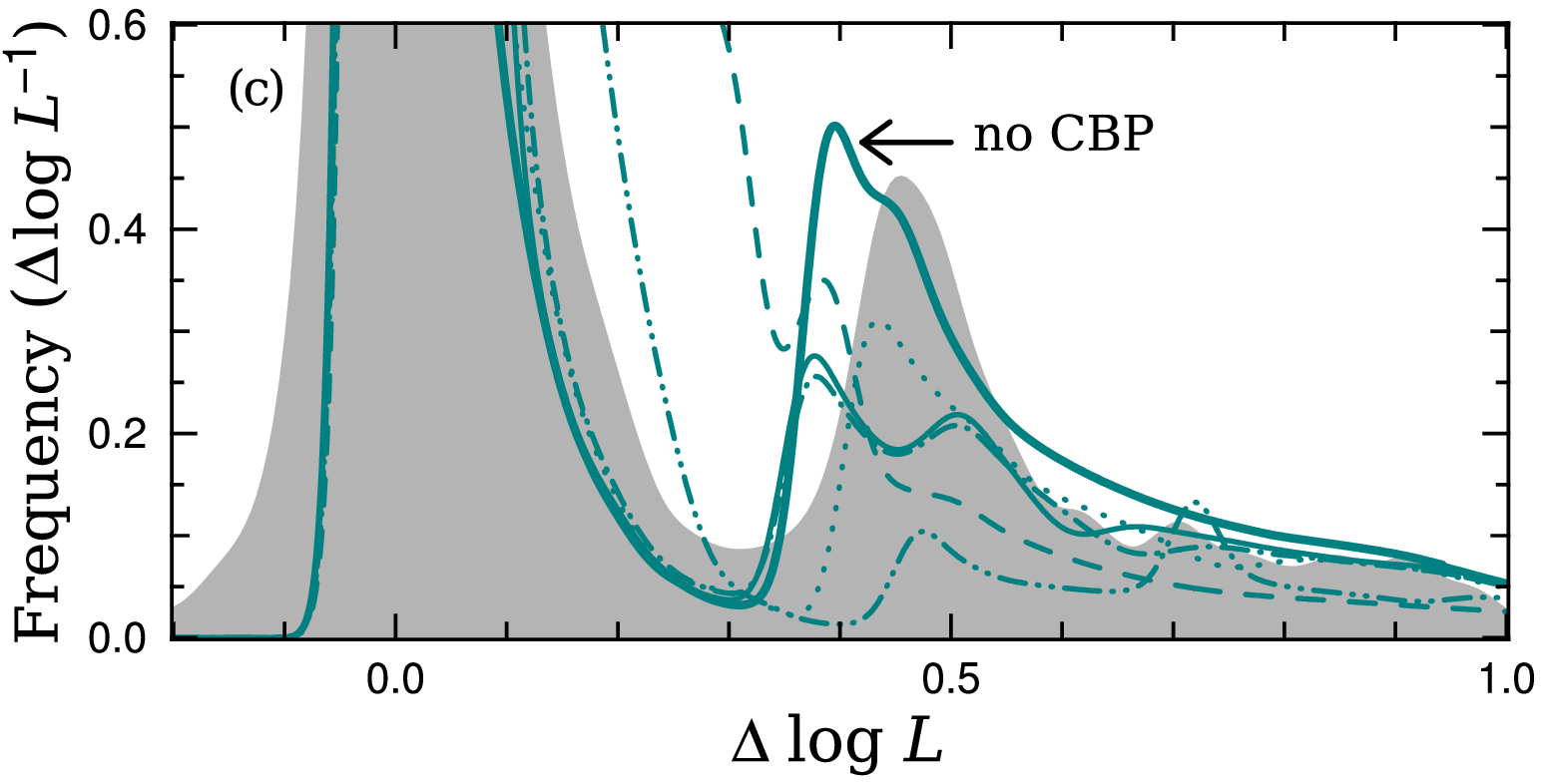}
  \caption{Comparison of standard overshoot evolution sequences with different overshooting parameter $f_\text{OS}$.  The additional thick lines (not listed in the key) are for a model with $f_\text{OS}=0.001$ initially but $f_\text{OS}=0.0$ when the mass fraction of helium in the centre drops below 0.17, to emulate the suppression of breathing pulses \citep[see e.g.][]{1989ApJ...340..241C}.  These models have average time step $10^4\,\text{yr} < \overline{\Delta t} < 2 \times 10^4\,\text{yr}$.  The panels are the same as Figure~\ref{figure_mixing_comparison}.}
  \label{figure_standardovershoot_comparison}
\end{figure}

In our standard overshoot models we apply the overshooting scheme proposed by \citet{1997A&A...324L..81H} where there is an exponential decay in diffusion coefficient according to 
\begin{equation}
D_\text{OS}(z)=D_\text{0} e^{\frac{-2z}{H_\text{v}}},
\end{equation}
where $D_\text{OS}(z)$ is the diffusion coefficient at distance $z$ from the convective boundary and $D_\text{0}$ is the diffusion coefficient just inside the boundary.  $H_\text{v}$ is the `velocity scale height' defined as
\begin{equation}
H_\text{v}=f_\text{OS} H_\text{p},
\end{equation}
where $H_\text{p}$ is the pressure scale height, and we have chosen $f_\text{OS}=0.001$ for this study. In Figure~\ref{figure_standardovershoot_comparison} we show the consequences of altering this value and an example of suppressing core breathing pulses (thick line) by stopping overshooting when the central helium abundance is low (this is analysed in Section~\ref{sec:cbp}).

It is clear from Figure~\ref{figure_standardovershoot_comparison}(b) that the extent of overshoot does not significantly alter the luminosity until late in core helium burning.  This is not surprising, given that the luminosity evolution during CHeB is scarcely affected by the choice of mixing prescription (Figure~\ref{figure_mixing_comparison}).  Near core helium exhaustion, however, the range of variation between standard overshoot models with different values of $f_\text{OS}$ is greater than it is for models with entirely different mixing schemes (compare e.g., the standard overshoot and semiconvection runs in Figures~\ref{figure_mixing_comparison}b and \ref{figure_standardovershoot_comparison}b).  In these tests, modifying $f_\text{OS}$ can change the CHeB lifetime by up to 30\,Myr (more than 20 per cent of the CHeB lifetime). This also leads to extremely large variations in $R_2$ and $\Delta \log{L_\text{HB}^\text{AGB}}$.  In the five models shown in Figure~\ref{figure_standardovershoot_comparison} with $0.001 \leq f_\text{OS} \leq 0.05$, we find ranges of $0.036 \leq R_2 \leq 0.091$ and $ 0.38 \leq \Delta \log{L_\text{HB}^\text{AGB}} \leq 0.72$.  During CHeB, these models have average time step $10^4\,\text{yr} < \overline{\Delta t} < 2 \times 10^4\,\text{yr}$.

Two of the five evolution sequences in Figure~\ref{figure_standardovershoot_comparison}(a) show blueward excursions, or `blue loops', in the HR diagram.  In this example they belong to the two longest lived CHeB sequences.  They immediately follow the ingestion of helium into the core during CBP and last for about 200\,kyr (which is less than 0.2 per cent of the CHeB lifetime).  If real, these would be sufficiently short-lived to make it unlikely that a star in this phase would be observed and thus provide evidence for the existence of CBP.  Even if they were observed, they could also be interpreted as less massive stars because they share the same position in the HR diagram.

Each of the models in Figure~\ref{figure_standardovershoot_comparison}(c) shows multiple peaks in AGB region of the luminosity PDF.  These are not seen in models with the other mixing schemes.  The first (lowest $\log{L}$) peak is caused by the drop and subsequent slow increase in luminosity immediately after core helium exhaustion.  The subsequent peaks are caused by the helium burning shell encountering a region richer in helium when it moves through a composition discontinuity in the partially mixed region.  This temporarily speeds up the rate of increase of the surface luminosity.  These episodes are analogous to the RGB luminosity function bump which is caused by the advance of the hydrogen-burning shell through the composition discontinuity left by first dredge-up.  This explanation makes it clear why none of the other mixing schemes show this phenomenon: the no-overshoot and maximal-overshoot models do not have a partially mixed zone and the semiconvection models do not leave behind any composition discontinuities.

The clarity of the subsequent peaks in the AGB luminosity PDF depends on both the difference in composition across the discontinuities and the mass enclosed by them.  If two discontinuities are close together in mass (or one is near to the earlier boundary of the convective core) then it can be hard to distinguish the two peaks.  In the case where CBP are prevented, for example, the first two peaks are separated by $\Delta \log{L} = 0.07$ (thick line in Figure~\ref{figure_standardovershoot_comparison}).  If the partially mixed region is very large it is similarly difficult to detect the second peak because the burning front moves through the edge of the partially mixed region at higher luminosity, when the evolution is fast (there are examples of this in the range $0.6 \leq \Delta \log{L} \leq 0.9$ shown in Figure~\ref{figure_standardovershoot_comparison}c).  The differences in the luminosity PDFs that arise from the suite of standard overshoot sequences reflects the broad variation in the structure of their partially mixed regions by the end of CHeB.

In general, there is a stochastic dependence of $R_2$ and $\Delta \log{L_\text{HB}^\text{AGB}}$ on $f_\text{OS}$.  An exception to this is when a very large $f_\text{OS}$ is used.  In the example run with $f_\text{OS} = 0.05$ the overshoot penetrates so far that the partially mixed region consists of a single zone with a homogeneous helium abundance between that of the convective core and the helium-rich shell surrounding it.  This means that after core helium exhaustion there is no composition discontinuity to burn through until the front moves to the edge of the partially mixed zone, which occurs when the luminosity is much higher (when $\log{L/L_\odot}>2.5$ in that example).  That model also shows the earliest instability in the core boundary, after only about 40\,Myr.  There are four small CBP throughout CHeB, each separated by 16\,Myr, rather than the typical large CBP late in CHeB (when $Y<0.1$ in the core).  Despite the unusual evolution during CHeB, the sequence has $R_2 = 0.086$ and $\Delta \log{L_\text{HB}^\text{AGB}}=0.44$ which are both unremarkable for standard overshoot.

\subsubsection{Time step dependence}
\label{sec:numerical_dependence}

\begin{figure}
\includegraphics[width=\linewidth]{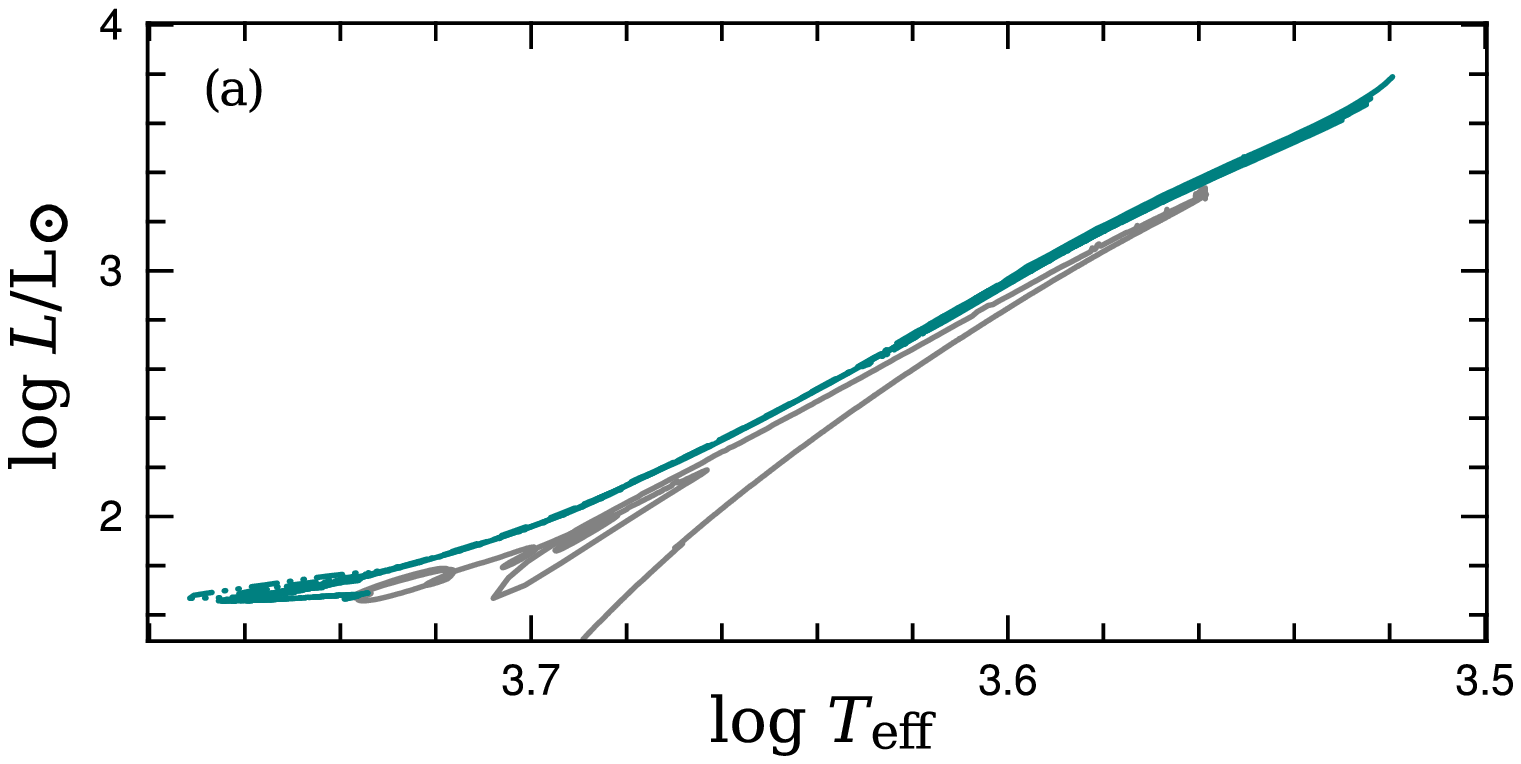}
\par
\vspace{0.35cm}
\includegraphics[width=\linewidth]{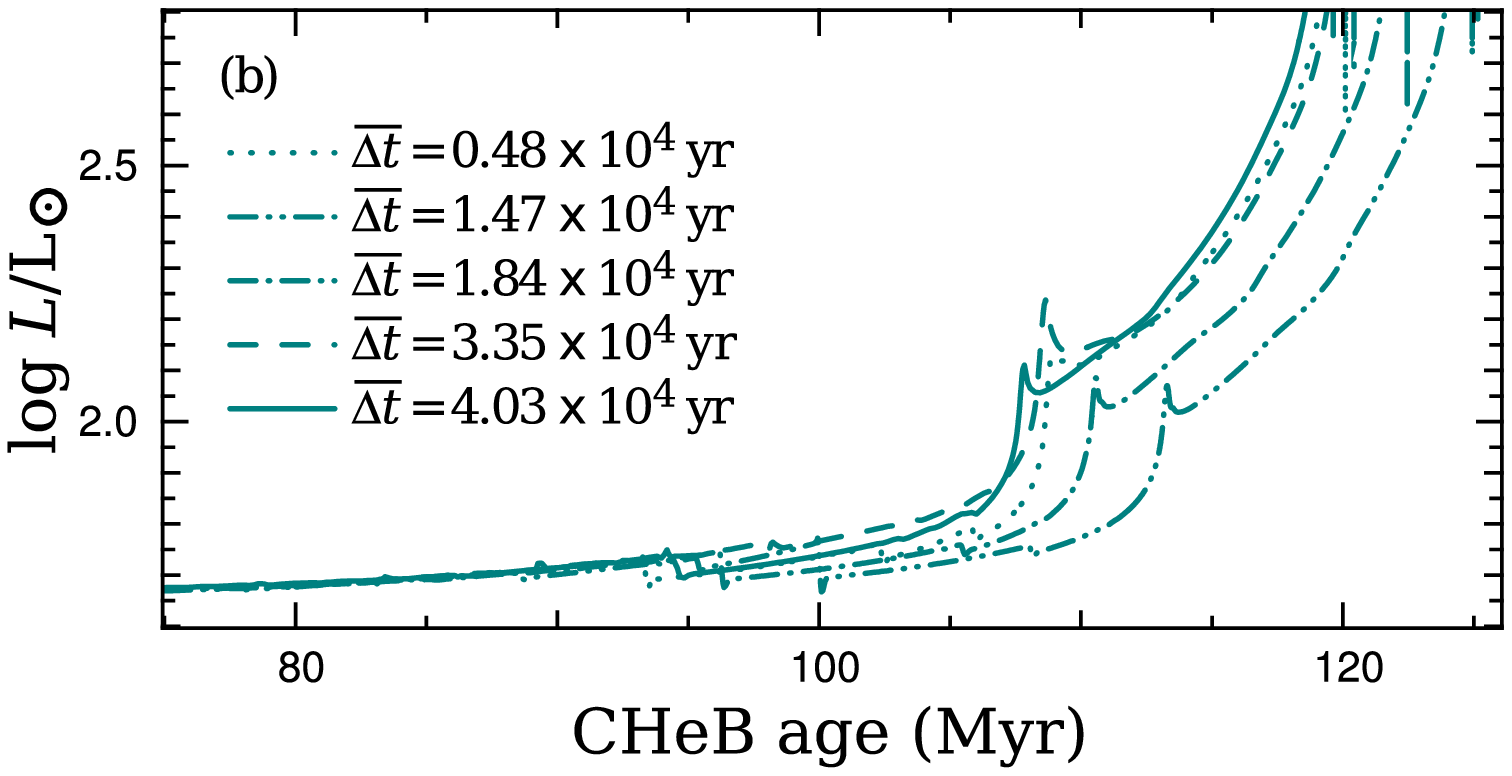}
\par
\vspace{0.35cm}
\includegraphics[width=\linewidth]{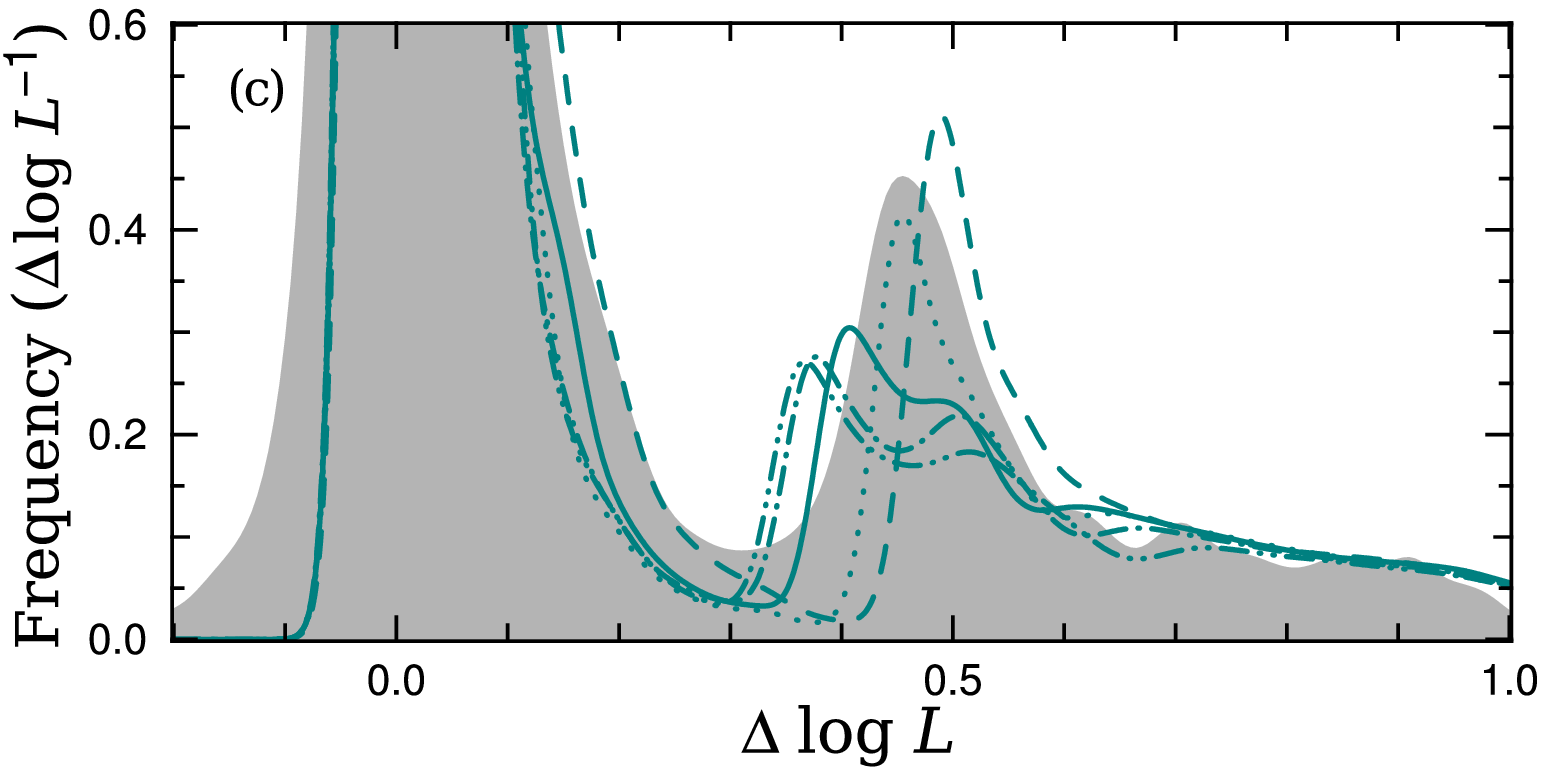}
  \caption{Comparison of standard overshoot runs with different time step constraints.  Each run is identified by the resulting average time step $\overline{\Delta t}$ during CHeB.  Each model has overshooting parameter $f_\text{OS} = 0.001$.  The panels are the same as Figure~\ref{figure_mixing_comparison}.}
  \label{figure_standardovershoot_dt}
\end{figure}

In this section we investigate how different time step constraints affect the evolution of our standard-overshoot runs.  In Figure~\ref{figure_standardovershoot_dt} we show five standard-overshoot models with $f_\text{OS}=0.001$ that differ only in the number of time steps taken during the CHeB phase.  Among the different sequences we find $0.36 \leq \Delta \log{L_\text{HB}^\text{AGB}} \leq 0.51$ and $0.080 \leq R_2 \leq 0.096$.  Overall, the evolution of the suite of standard-overshoot models with different average time step is more consistent than it is among the group with different $f_\text{OS}$ shown in Section~\ref{sec:overshoot_dependence}.  The ranges of $\Delta \log{L_\text{HB}^\text{AGB}}$ and $R_2$ are both smaller, but there is still a significant spread.  The 10\,Myr ($\sim 9$ per cent of the CHeB lifetime; Figure~\ref{figure_standardovershoot_dt}b) variation in the CHeB lifetime resulting from different time steps is also smaller than it is for the suite of models with different $f_\text{OS}$ (Figure~\ref{figure_standardovershoot_comparison}b).  We do not find a correlation between the size of the time steps and the properties of the evolution, mirroring our finding for modifications to $f_\text{OS}$.

We have shown that time step constraints and the overshoot prescription can both have unpredictable and severe effects on the evolution of standard-overshoot models late in CHeB, importantly including predictions for $\Delta \log{L_\text{HB}^\text{AGB}}$ and $R_2$.  The stochasticity of the evolution is the reason we did not use standard overshoot in the earlier sections to quantify the effects of altering the input physics.  Moreover, we note that the evolution of a single model produces AGB clump peaks in the luminosity PDF (e.g. Figure~\ref{figure_standardovershoot_comparison}c and~\ref{figure_standardovershoot_dt}c) that are broader than both of those resulting from the addition of observed data for multiple clusters (Figure~\ref{figure_all_red_hb}).  Considering that combining data from different clusters, photometric errors, and contamination are all likely to \textit{widen} the AGB clump peak, this disagreement provides further strong evidence against the credibility of the late CHeB evolution that arises from the use of standard overshoot (i.e. CBP).  Instead, it points towards the existence of a smooth or entirely flat composition profile outside the convective core.

\subsection{Core breathing pulses and their suppression}
\label{sec:cbp}
Among the four treatments of convective boundaries, only the runs with standard overshoot exhibit CBP (e.g. near 96\,Myr in Figure~\ref{figure_mixing_comparison}b).  The other sequences have a monotonic decrease in central helium abundance and a stability in the size of the convective core.

We have performed an experiment to separate the immediate effects of each mixing prescription from their cumulative effect on the stellar structure.  We began by selecting three late-CHeB 1\,M$_\odot$ models from Paper~I with no overshoot, semiconvection, and maximal overshoot -- the three schemes that avoid CBP.  Beginning from models with the same central helium abundance ($Y = 0.1$), we then continued the evolution using the standard-overshoot prescription.  Each of the three then displayed CPB, but to different extents.  The no-overshoot model showed the largest breathing pulse.  This led to an increase in central helium abundance of $\Delta Y=0.58$, compared with $\Delta Y = 0.17$ for the largest core breathing pulse in the original standard overshoot run.  This demonstrates how unstable this small core is late in CHeB.  The next largest core breathing pulse was seen in the maximal-overshoot model which had $\Delta Y=0.04$.  The substantial difference between this and the no-overshoot model may be attributed to the already large convective core.  The smallest CBP seen in the semiconvection model which had $\Delta Y = 0.02$.  In that case, the radiative region immediately outside the convective core was only marginally richer in helium, limiting the potential for feedback when it was mixed into the burning region by overshoot.  

CBP can be prevented in models by omitting the gravitational energy term \citep{1993ApJ...409..387D} or by halting the enlargement of the convective core if it will lead to an increase in the central helium abundance \citep[e.g.][]{1989ApJ...340..241C,1997ApJ...489..822B,2001A&A...366..578C}. Our semiconvection model is unusual in that CBP are avoided without explicitly altering the physics for the end of CHeB.  We have shown that in this phase the structure is such that if the Schwarzschild criterion for convection is used, any overshoot will trigger CBP.  Like the models from \citet{1986MmSAI..57..411B} without CBP, our two methods that do have mixing beyond the Schwarzschild boundary but avoid CBP also have a non-local treatment of convection.  The semiconvection method has a limit on how steeply the diffusion coefficient can change through the structure.  In the maximal overshoot method, the extent of overshooting at the boundary is determined by $\nabla_\text{rad} / \nabla_\text{ad}$ further inside the convection zone.

The thick line in Figure~\ref{figure_standardovershoot_comparison} shows the result of suppressing CBP by turning off convective overshoot before the size of the convective core becomes unstable (in this case when the central helium abundance is $Y = 0.17$).  It is clear from panel~(c) that this model spends relatively more time on the AGB (i.e. with $\Delta \log{L}> 0.3$), and therefore that this method of preventing CBP increases $R_2$.  This is caused by both a reduction in the CHeB lifetime and an increase in the early-AGB lifetime (Figure~\ref{figure_standardovershoot_comparison}b).  

The drastic effect on the evolution of standard-overshoot sequences from small changes in $f_\text{OS}$ and time step (Section~\ref{sec:overshoot_dependence} and~\ref{sec:numerical_dependence}) is primarily due to how these influence the development of CBP.  The luminosity of the AGB clump for the model with CBP suppressed (thick line in Figure~\ref{figure_standardovershoot_comparison}c) is consistent with that in models with CBP.  The peak in the luminosity PDF is noticeably thinner, however, for reasons explained in Section~\ref{sec:overshoot_dependence}, and it better matches the observations.  In contrast with $\Delta \log{L_\text{HB}^\text{AGB}}$, $R_2$ is strongly affected by the prevention of CBP.  In the test model, $R_2$ is increased to 0.143, compared with a range of $0.036 \leq R_2 \leq 0.091$ (with median $R_2 = 0.086$) for the other sequences shown.  This consequence of the suppression of CBP (by different means) has been shown before \citep[e.g.][]{1989ApJ...340..241C,2001A&A...366..578C}.  

\subsection{Post-CHeB maximal overshoot evolution}
\label{sec:gravonuclear_loops}

In Paper~I we showed that the maximal-overshoot mixing scheme was the only one of the four that could match the high asymptotic g-mode period spacing inferred from asteroseismic observations (i.e., it did not rely on the effects of mode trapping).  The models with this scheme that we have shown so far, however, predict $R_2$ much lower, and $\Delta \log{L}_\text{HB}^\text{AGB}$ much higher, than observed (Figure~\ref{figure_mixing_comparison}).

The no-overshoot and maximal-overshoot sequences display a phenomenon known as gravonuclear loops \citep[see e.g.][]{1986ApJ...304..217I,1997ApJ...479..279B,1997ApJ...489..822B,2000LIACo..35..529S,2009A&A...507.1575P}.  These are evident from the oscillation in surface luminosity near 55\,Myr and 108\,Myr for the respective sequences in Figure~\ref{figure_mixing_comparison}(b).  There is also an oscillation in the effective temperature during this period, thereby giving rise to 'loops' within the AGB clump in the HR diagram.  This phenomenon occurs when the He-burning shell encounters a large composition discontinuity at the former boundary of the convective core, causing discrete episodes of strong helium burning \citep[see e.g.][]{2000LIACo..35..529S}.  The energy generation is high enough to trigger convection temporarily.  The gravonuclear loops finally end when the convective mixing has smoothed the composition gradient that originally induced them.  The standard overshoot and semiconvection models have (relatively) smooth composition gradients at the end of CHeB, and hence avoid the gravonuclear instability.

\begin{figure}
\includegraphics[width=\linewidth]{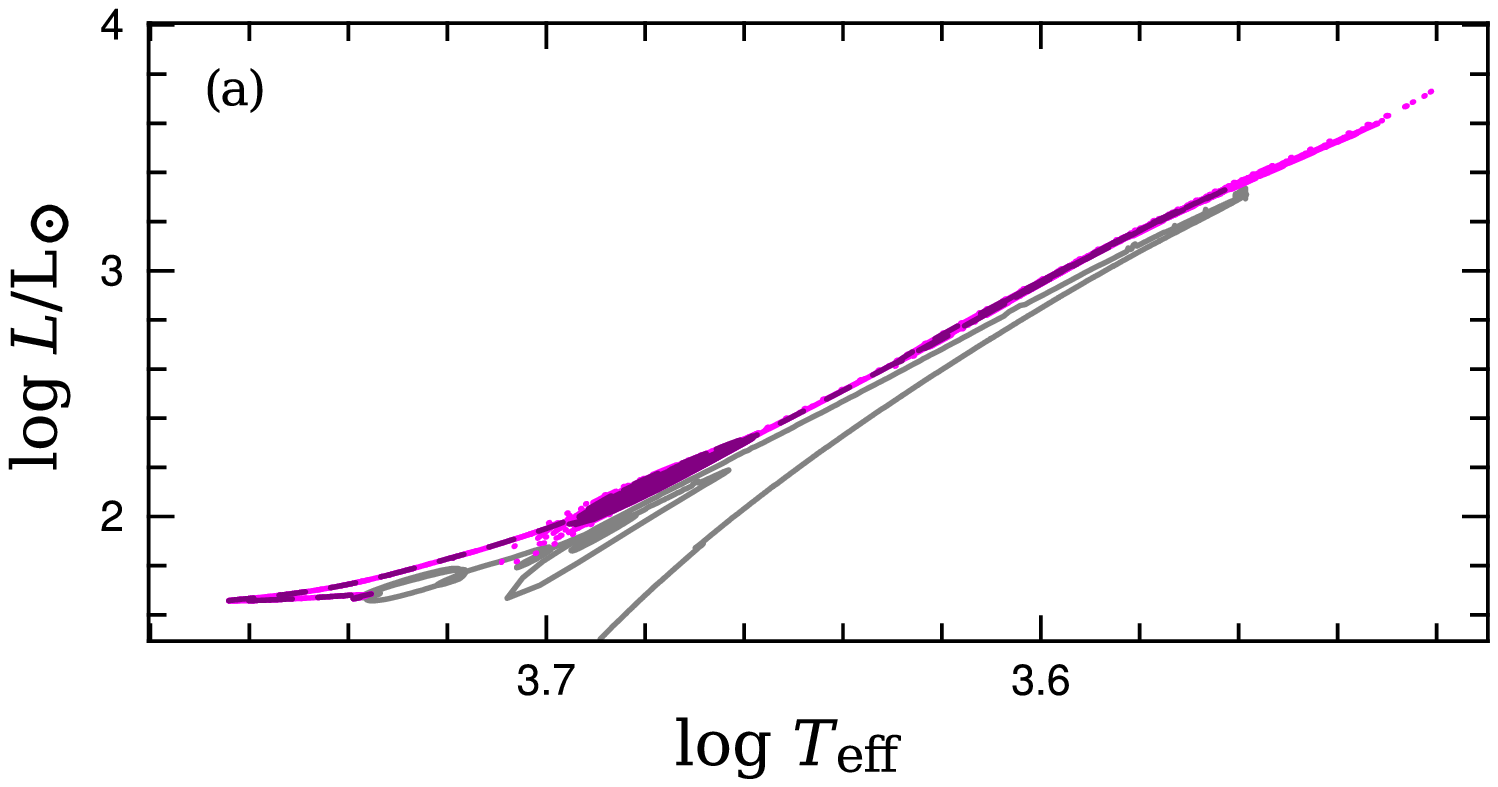}
\par
\vspace{0.35cm}
\includegraphics[width=\linewidth]{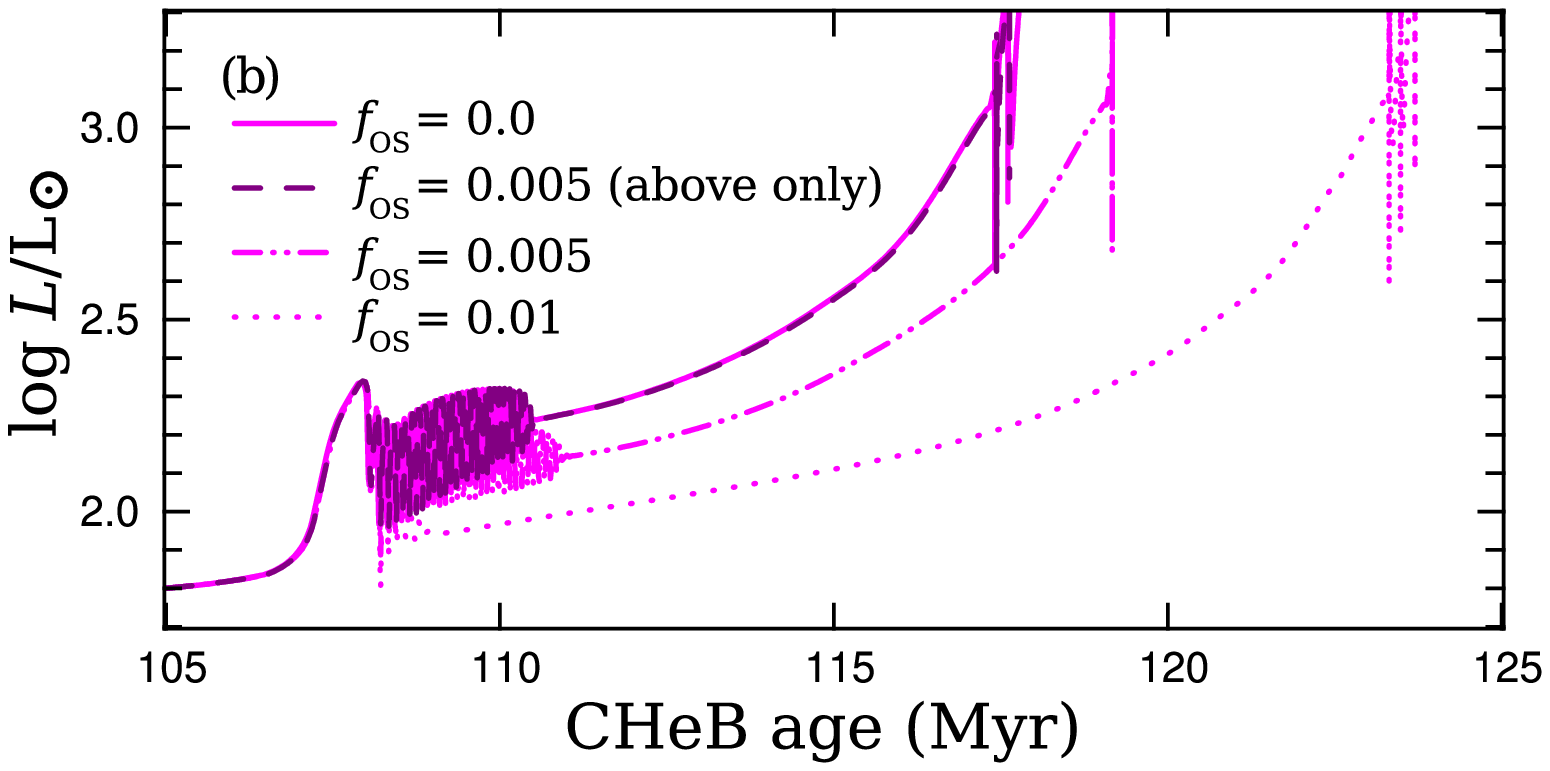}
\par
\vspace{0.35cm}
\includegraphics[width=\linewidth]{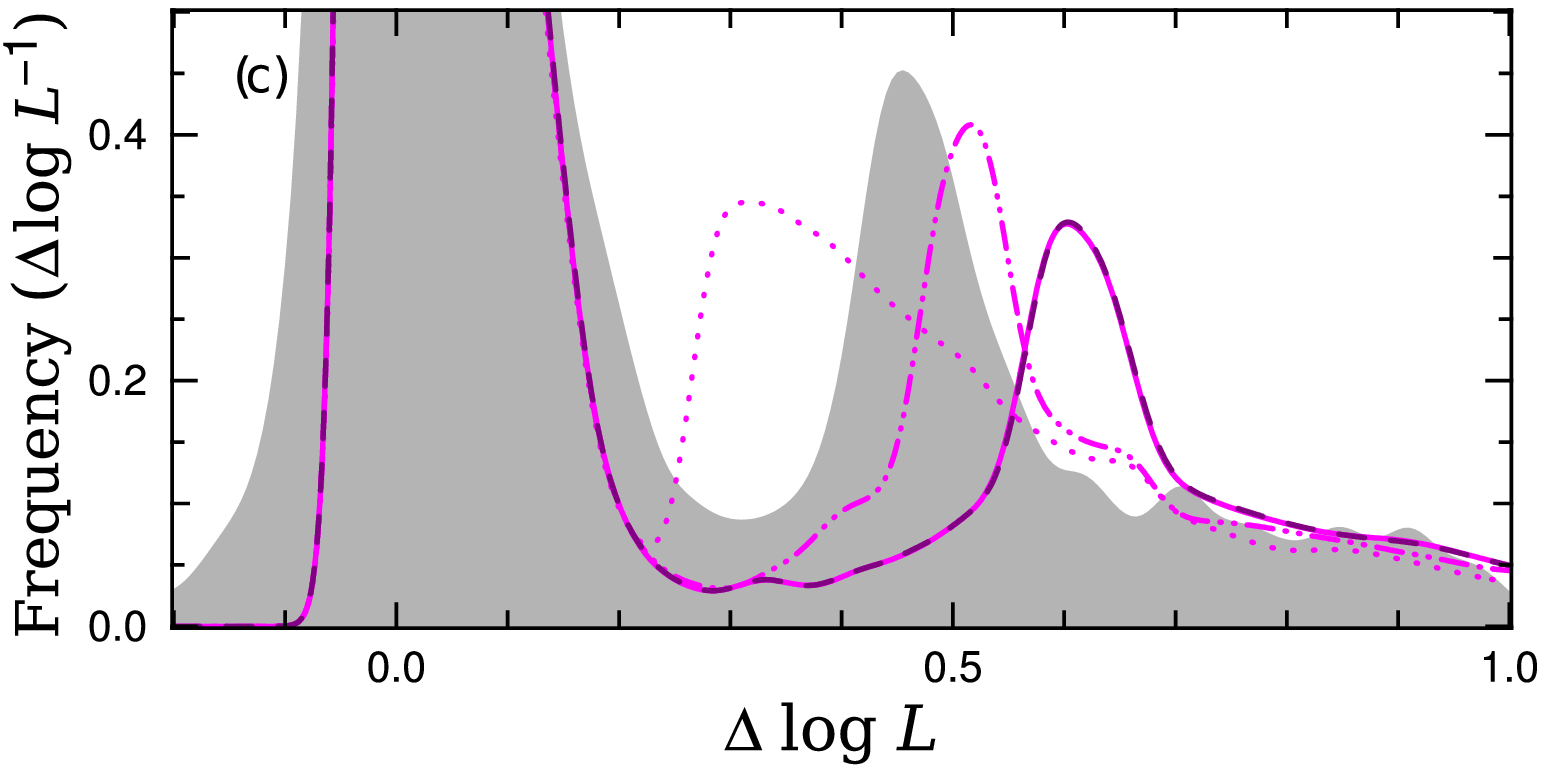}
  \caption{Comparison of early-AGB (post maximal-overshoot) evolution sequences with different treatments of convective overshoot.  The models have no overshoot (solid magenta line), overshoot with $f_\text{OS} = 0.01$ (magenta dots), overshoot with $f_\text{OS} = 0.005$ (magenta dashed dotted line), and overshoot only at outer boundaries with $f_\text{OS} = 0.005$ (dark magenta dashes).  The panels are the same as Figure~\ref{figure_mixing_comparison}.}
  \label{figure_mo_os}
\end{figure}

In the maximal-overshoot sequences shown in Figures~\ref{figure_mixing_comparison}-\ref{figure_different_reaction_rates2} there is no overshooting at convective boundaries after core helium burning finishes.  In Figure~\ref{figure_mo_os} we show the effect of including overshooting at the boundaries of the convection zones that emerge during helium shell burning.  Those three sequences show that the existence and extent of overshooting has a substantial impact on the early-AGB evolution.  In the run with $f_\text{OS} = 0.01$, for example, the early-AGB lifetime is extended by 4\,Myr, increasing $R_2$ from 0.082 to 0.139, i.e. to above the observed value of $R_2 = 0.117 \pm 0.005$.  

\begin{figure}
\includegraphics[width=\linewidth]{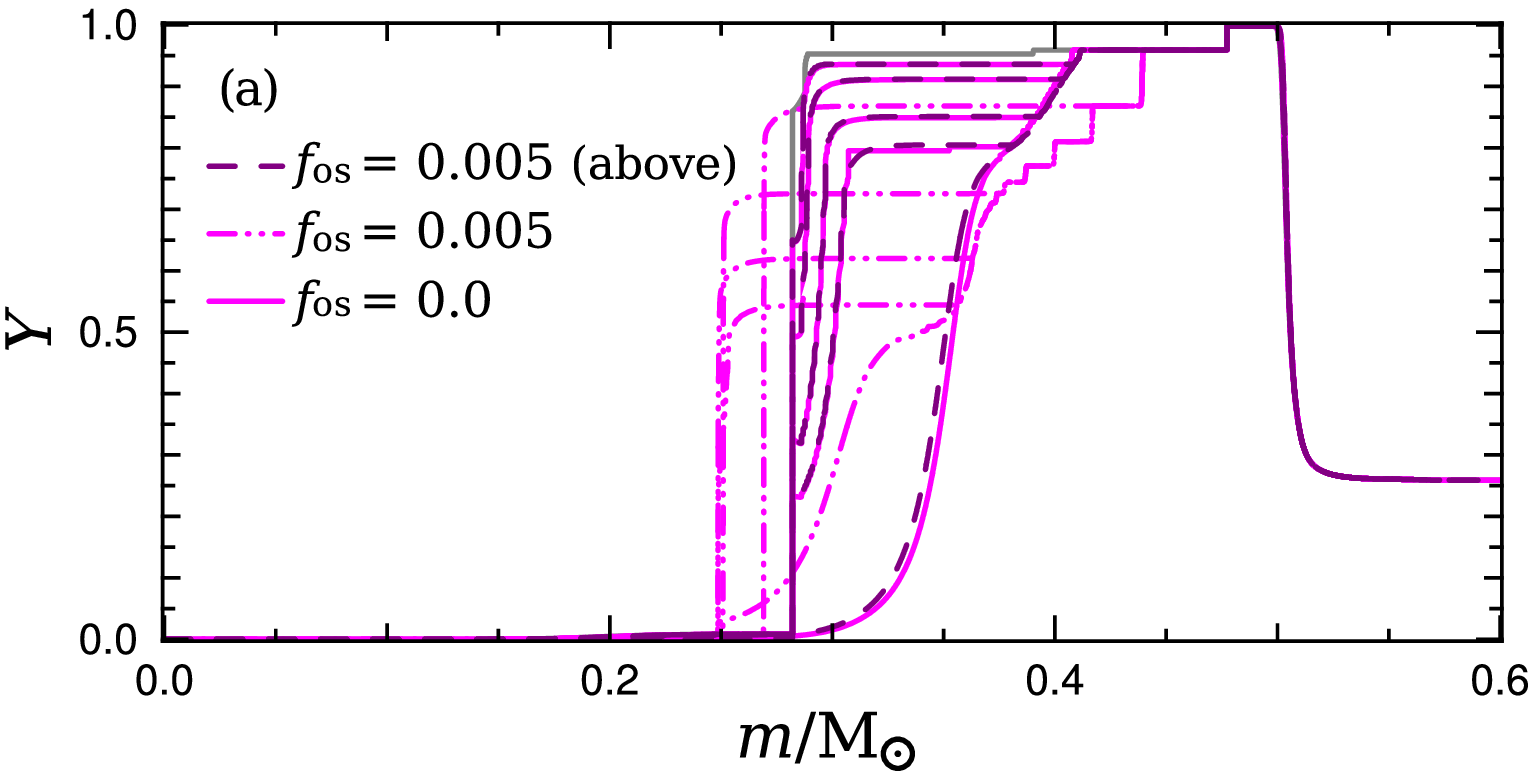}
\par
\vspace{0.35cm}
\includegraphics[width=\linewidth]{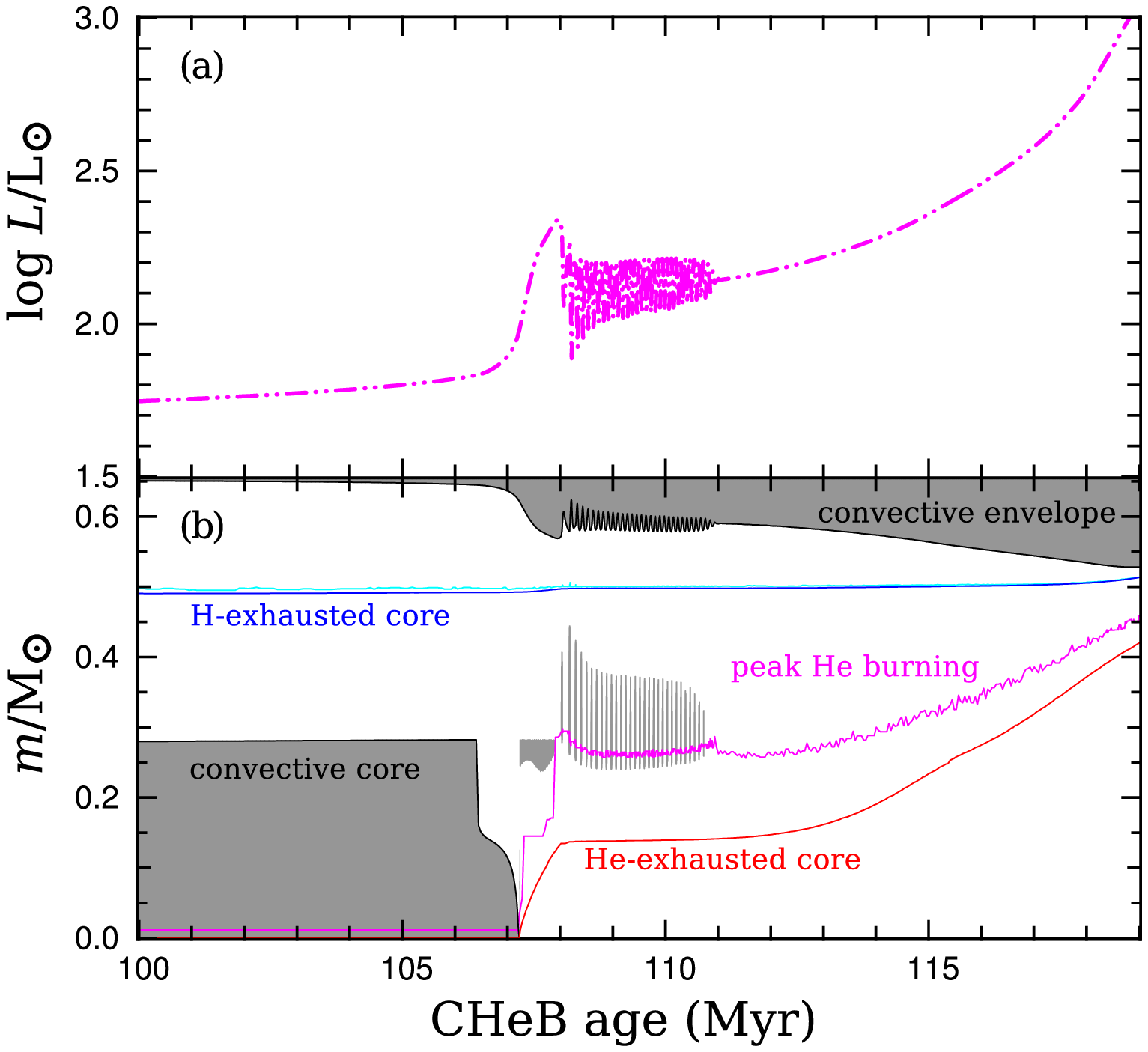}
  \caption{\textbf{Upper panel:} evolution of internal helium mass fraction $Y$ of early-AGB, post maximal-overshoot CHeB models with different treatments of convective overshoot.  The models have no overshoot (solid magenta line), overshoot with $f_\text{OS} = 0.005$ (magenta dashed dotted line), and overshoot only at outer boundaries with $f_\text{OS} = 0.005$ (dark magenta dashes).  The composition at the end of CHeB is shown in grey.  \textbf{Lower panels:} evolution of surface luminosity and Kippenhahn plot of the sequence with overshoot at all boundaries shown in the upper panel.  The positions of peak H burning, H exhaustion, peak He burning, and He exhaustion are shown by cyan, blue, magenta, and red lines, respectively.}
  \label{figure_maximal_overshoot_eagb}
\end{figure}

Figure~\ref{figure_maximal_overshoot_eagb} shows the internal evolution for three cases: no overshooting, overshooting only at the outer convective boundaries, and overshooting at all convective boundaries.  The two runs that only differ by their treatment of overshoot at the outer boundaries of convection zones are nearly identical.  In contrast, overshooting beneath the helium burning convection zones has profound consequences.  The inclusion of convective overshooting with $f_\text{OS} = 0.005$ beneath convection zones has a significant effect.  The position of peak helium burning moves inward by about 0.04\,M$_\odot$.  The gravonuclear loop phase is extended by about 0.5\,Myr, but this only accounts for part of the almost 2\,Myr increase in early-AGB lifetime.  The model also finishes the gravonuclear loop phase with lower luminosity and having burnt less helium.  Once the luminosity increases, the rates of change of the H-burning and He-burning luminosity are the same as for models without overshooting beneath the helium burning convection zones (only it is offset by 1.8\,Myr).

If more penetrating overshoot is used, such as in the run with $f_\text{OS} = 0.01$, the gravonuclear loop convection zones rapidly reach the ashes of helium burning.  This puts an end to the gravonuclear loop phase and quiescent shell helium burning begins.  During this period the peak of nuclear burning moves inward by 0.12\,M$_\odot$ and the surface luminosity drops almost to the HB level (Figure~\ref{figure_mo_os}b).  Once the helium burning shell advances far enough to enclose the same mass as it does in the model without overshoot after the gravonuclear loop phase, the rates of change of the H-burning and He-burning luminosity are again the same as that model (this time offset by 4\,Myr).  The subsequent slow luminosity increase during quiescent helium burning produces a diffuse peak in the luminosity PDF (Figure~\ref{figure_mo_os}c) that is at odds with observations (Figure~\ref{figure_mo_os}c).

Analogous with the case of hydrogen shell burning, when helium burning occurs with a less massive helium exhausted core beneath it, the burning occurs at a lower temperature.  This happens at the beginning of shell helium burning as a consequence of overshoot eroding the He-exhausted core.  Introducing overshoot of $f_\text{OS}=0.005$ and $f_\text{OS}=0.01$ reduces the shell temperature immediately after the cessation of gravonuclear loops from 128\,MK to 99\,MK and 90\,MK, respectively.  As previously mentioned, this slows the evolution, but it also favours the $^{12}\text{C}(\alpha,\gamma)^{16}\text{O}$ reaction, decreasing the final C/O ratio in the degenerate CO core.

Figure~\ref{figure_mo_os}(c) shows that increasing $f_\text{OS}$ decreases $\Delta \log{L_\text{HB}^\text{AGB}}$ and increases $R_2$.  Although the values of $f_\text{OS}$ were arbitrary, if we take the average from the models with $f_\text{OS} = 0.005$ and $f_\text{OS} = 0.01$ we find $R_2 = 0.118$ and $\Delta \log{L_\text{HB}^\text{AGB}} = 0.42$, which is a reasonable match to the observations.  This demonstrates that our models with a large convective core and a single composition discontinuity at the end of CHeB can also be consistent with the globular cluster observations, but only if convective overshoot moves the helium burning front inward (in mass) by a particular amount.

The maximal overshoot runs with subsequent convective overshoot have a core helium burning lifetime more than 10\,Myr longer than the standard overshoot models without core breathing pulses.  Both of these models can fit the $R_2$ and $\Delta \log{L_\text{HB}^\text{AGB}}$ constraints and neither can yet be ruled out by asteroseismology (see Paper~I).  The lifetime disparity is therefore an important uncertainty for various other constraints that are derived from counting HB stars, such as globular cluster initial helium abundance \citep[e.g.][]{2003ApJ...588..862C,2004A&A...420..911S} and bounds on axion-photon coupling \citep[e.g.][]{2014PhRvL.113s1302A}.  The difference in CHeB lifetime between these two runs, and therefore predictions for the ratio $R = n_\text{HB}/n_\text{RGB}$, equates to a change in inferred initial helium abundance of $\Delta Y \approx 0.02$.

\subsection{The AGB luminosity limit}

In this section we examine how the $R_2$ comparison between models and observations is affected by the choice of the luminosity cut-off for the AGB that we introduced in Section~\ref{sec:red_hb_clusters}.  We tested reducing the maximum AGB luminosity from $\log{L_\text{AGB}} = \log{L_\text{HB}}+1.0$ to $\log{L_\text{AGB}}=\log{L_\text{HB}}+0.7$, which is high enough for the AGB clump to still be included in the luminosity PDF for clusters without a blue extension to the HB (Figure~\ref{figure_cluster_properties_z}) as well as (most of) the computed sequences.  This change reduces the observed $R_2$ from 0.114 and 0.127 to 0.091 and 0.095 for the \citet{2002A&A...391..945P} and \citet{2007AJ....133.1658S} samples, respectively, and interestingly improves their consistency.

Lowering the AGB luminosity limit has variable consequences for predictions of $R_2$.  Unsurprisingly, the decrease in the predicted R$_2$ is largest for models with a high $\Delta \log{L_\text{HB}^\text{AGB}}$, because this truncation excludes part of the AGB clump peak in the luminosity PDF (but almost never more than half).  The luminosity PDFs of the models with different mixing schemes are generally quite similar for $\Delta \log{L} > 0.7$ (and especially so for $\Delta \log{L} > 0.8$) so reducing the cut-off has a uniform effect on $R_2$ for most cases (see e.g. Figure~\ref{figure_mixing_comparison}c, \ref{figure_different_reaction_rates2}c, \ref{figure_standardovershoot_dt}c, \ref{figure_mo_os}c).  The sequences shown in Figure~\ref{figure_mixing_comparison} are typical: by decreasing the cut-off to $\Delta \log{L} < 0.7$ $R_2$ is reduced from 0.096, 0.068, and 0.082 to 0.070, 0.043, and 0.057 for the standard-overshoot, semiconvection, and maximal overshoot models, respectively.  This $\Delta R_2 \approx -0.025$ in all three cases, compared with $\Delta R_2 \approx -0.028$ for the observations.  The equivalent standard-overshoot model with CBP suppressed (thick line in Figure~\ref{figure_standardovershoot_comparison}) shows a larger reduction of $R_2$, from 0.143 to 0.099.  Reducing the luminosity cut-off for the $R_2$ calculation for the two maximal-overshoot sequences that include overshooting after CHeB ceases (with $f_\text{OS} = 0.005$ and $f_\text{OS} = 0.01$ respectively; described in Section~\ref{sec:gravonuclear_loops}) has an effect similar to that on the observations.  Those respective models show reductions of $R_2$ from 0.097 and 0.139 to 0.075 and 0.119.

Overall, it appears that reducing the luminosity limit for the AGB does not significantly alter the (dis)agreement between models and observations: those which can match with $\log{L_\text{AGB}}<\log{L_\text{HB}}+1.0$ can also match with $\log{L_\text{AGB}}<\log{L_\text{HB}}+0.7$, and vice versa, although in some cases the disagreement is exacerbated.  The insensitivity to the luminosity limit is not unexpected considering the predominance of AGB stars near the clump in both observations and theoretical predictions (e.g., near $\Delta \log{L_\text{HB}^\text{AGB}} \approx 0.5$ in Figure~\ref{figure_mixing_comparison}c).

\section{Summary and conclusions}

In this paper we extended our study into the mixing in core helium burning stars in Paper~I by confronting models with observations of globular clusters.  The particular observational probes of core helium burning we used were (i)  $R_2$, the ratio of AGB to HB stars, and (ii) $\Delta \log{L_\text{HB}^\text{AGB}}$, the luminosity difference between the AGB clump and the HB, for 48 Galactic globular clusters with suitable HST photometry \citep{2002A&A...391..945P,2007AJ....133.1658S}.  We compared these data to a suite of stellar models that includes four different mixing prescriptions, variations in the initial composition, and an exploration of their numerical dependence and physical uncertainties.

In Section~\ref{sec:photometry_comparison} we showed there is a considerable spread in $R_2$ determined from observations.  The scatter is apparent for distinct photometry of a given cluster and for homogeneous photometry of different clusters.  This casts doubt on inferences about stellar evolution from $R_2$ derived from the photometry of a small number of clusters, or even a single cluster, which are common in the literature.  By combining data for 48 clusters from two HST surveys, we minimized the dominant statistical uncertainty.  Encouragingly, the cluster to cluster variation in $R_2$ is also smaller for the newer HST photometry than for inhomogeneous photometry in the literature.

By combining photometry for the 15 clusters common to the three data sets, we found $R_2 = 0.121 \pm 0.006$ \citep{2002A&A...391..945P}, $R_2 = 0.125 \pm 0.005$ \citep{2007AJ....133.1658S}, and $R_2 = 0.152 \pm 0.007$ \citep{2000MNRAS.313..571S}, where the 1-sigma errors are calculated from Equation~\ref{eq:R2_sigma}.  The two new determinations of a lower $R_2$ lessen the disagreement with standard models, and bring $R_2$ into alignment with models in the literature where core breathing pulses have been suppressed \citep[e.g.][]{2003ApJ...588..862C}.

We investigated the sources of the discrepancies between star counts from different photometry of the same clusters and found two main causes: (i) photometry can be incomplete, especially for blue HB stars, and (ii) it can be impossible to distinguish between the more luminous RGB and AGB stars.  To minimize errors from the latter problem we restricted the counts to stars less than 10 times as luminous as HB stars.  When this method was used we did not detect any dependence of $R_2$ on metallicity (see Section~\ref{sec:z_and_hb_morph}).  Furthermore, we showed in Section~\ref{sec:photometry_comparison} that the statistics of finite sampling can explain the majority of the scatter in $R_2$.  We also found evidence that clusters that host the bluest HB stars have abnormally low $R_2$, supporting conjecture that a significant proportion of those stars do not evolve to the AGB.

In order to better compare our models with observations we further limited the star counts in Section~\ref{sec:red_hb_clusters} to clusters without a blue extension of the HB.  This yielded a sample of 21 CMDs (of 14 unique clusters) comprising 6366 stars.  Models with each of the four different mixing schemes that were tested (standard overshoot, no overshoot, semiconvection, and maximal overshoot; described in Section~\ref{sec:mixing_schemes}) typically cannot simultaneously match $R_2 = 0.117 \pm 0.005$ and $\Delta \log{L_\text{HB}^\text{AGB}} = 0.455 \pm 0.012$ from the observations.  

Compared with observations, the evolution sequences without convective overshoot have $\Delta \log{L_\text{HB}^\text{AGB}}$ far too low and $R_2$ far too high.  This is consistent with previous findings that models without overshoot disagree with globular cluster observations and asteroseismology (e.g. Paper~I).  In contrast, our initial tests with each of the other schemes predicted $R_2$ well below that derived from observations.  The semiconvection models have $\Delta \log{L_\text{HB}^\text{AGB}}$ slightly too large whereas the standard overshoot models typically have about the observed value.  In Section~\ref{sec:numerical_dependence} we showed that the predicted luminosity probability density function from standard-overshoot models with core breathing pulses is not strongly peaked enough near the AGB clump (Figure~\ref{figure_standardovershoot_comparison}c).  Suppressing core breathing pulses removes this discrepancy and also increases $R_2$ (to even higher than the observed value in our ad hoc test).  These two factors are strong arguments against the validity of standard-overshoot runs with core breathing pulses that produce multiple large composition discontinuities in the partially mixed region.  Furthermore, in Section~\ref{sec:gravonuclear_loops} we demonstrated that models with the maximal-overshoot prescription can simultaneously match the observed $R_2$ and $\Delta \log{L_\text{HB}^\text{AGB}}$, but only if there is a particular amount of convective overshoot beneath the shell helium burning `gravonuclear' convection zones that appear during the early-AGB phase.

In Section~\ref{composition_and_input_physics} we quantified the effect that stellar mass and composition, and various physical uncertainties (other than mixing), have on predictions of $R_2$ and $\Delta \log{L_\text{HB}^\text{AGB}}$.  By adding each effect in quadrature we found that the respective uncertainties are $\sigma_{R_2} \approx 0.009$ and $\sigma_{\Delta \log{L}} \approx 0.04$.  Initial composition and HB stellar mass can each account for small changes in $R_2$, but not enough to resolve the difference between models and observations.  The most important uncertainty for $R_2$ is the $^{12}\text{C}(\alpha,\gamma)^{16}\text{O}$ reaction rate, which dominates late in core helium burning.  A reduction of this rate tends to decrease the HB lifetime and increase $R_2$.  We found that uncertainty in the H-exhausted core mass, which we showed in Paper~I could potentially account for some of the disagreement with asteroseismology, makes no appreciable change to either $R_2$ or $\Delta \log{L_\text{HB}^\text{AGB}}$.  The treatment of mixing is the dominant source of uncertainty in the models (see Section~\ref{sec:summary_of_effects}).  Even among models with mixing beyond the Schwarzschild boundary, uncertainties in the treatment of convective boundaries and other physics can cause more than a 40\,Myr (roughly $30$ per cent) variation in the HB lifetime, which would significantly affect any inferences from counts of HB stars.  Moreover, using photometry to constrain the treatment of convective boundaries in models is also made more difficult by the stochastic effect that the numerical treatment can have on the evolution during this phase (e.g. models with standard overshoot in this study).

In our asteroseismology study in Paper~I we found that mixing schemes that produce either a large partially mixed region that can trap modes or a very large convective core, predict an $\ell = 1$ mixed-mode period spacing that can be consistent with the observations.  This work is ongoing, but it should be pointed out that in our models, these mixing schemes are only consistent with globular cluster observations if (i) core breathing pulses do not significantly extend the core helium burning lifetime or create large composition discontinuities within the partially mixed region (see Section~\ref{sec:numerical_dependence}), or (ii) they develop very large convective cores (e.g. maximal overshoot) and there is a particular extent of convective overshoot beneath the convection zones that appear in the subsequent `gravonuclear loop' phase during the early-AGB (Section~\ref{sec:gravonuclear_loops}).  The physics behind these different possibilities may perhaps now best be addressed with multi-dimensional fluid-dynamics simulations or improved theories of convective boundary mixing \citep[see e.g.][]{2015ApJ...809...30A,2015A&A...582L...2S}.  Finally, we emphasise that this work could be complemented by ground-based photometry that can better differentiate the AGB and RGB sequences, e.g. with the $UBV$ bands. 

\section*{Acknowledgements}

This work was supported in part by computational resources provided by the Australian Government through the National Computational Merit Allocation Scheme and the Pawsey Supercomputing Centre (projects g61 and ew6).

\clearpage
\footnotesize{
 \bibliographystyle{mnras}
  \bibliography{refs2}
}

\end{document}